\documentclass[toc=bibliography]{scrreprt}
\usepackage[utf8]{inputenc}
\usepackage[T1]{fontenc}
\usepackage[english,french]{babel}

\usepackage{graphicx}
\graphicspath{{figures/}}
\usepackage{amsmath,amssymb,amsthm}
\usepackage{esint} 
\usepackage{bbm}
\usepackage{enumitem,color}
\usepackage{hyperref}
\usepackage{pdfpages} 

\usepackage{multibib}
\newcites{my}{List of publications}
\usepackage{cite}

\usepackage{xstring}

\newtheorem{thm}{Theorem}
\numberwithin{thm}{chapter}

\theoremstyle{definition}
\newtheorem{Def}{Definition}
\numberwithin{Def}{chapter}

\theoremstyle{remark}
\newtheorem{rem}{Remark}
\numberwithin{rem}{chapter}
\newtheorem{exmp}[rem]{Example}

\newcommand{\fig}[2]{\includegraphics[width=#1\textwidth]{#2}}

\DeclareMathOperator{\Cat}{Cat}
\newcommand{\paths}[3]{p^{(#1)}_{#2,#3}}

\newcommand{\Z}{\mathbb{Z}}
\newcommand{\R}{\mathbb{R}}
\newcommand{\C}{\mathbb{C}}
\DeclareMathOperator{\Prob}{Prob}
\DeclareMathOperator{\E}{\mathbb{E}}

\DeclareMathOperator{\Ai}{Ai}

\DeclareMathOperator{\Tr}{Tr}
\DeclareMathOperator{\pf}{pf}

\newcommand{\norm}[1]{\lvert #1 \rvert}

\DeclareMathOperator{\ex}{ex}
\DeclareMathOperator{\arccot}{arccot}

\begin{document}
\includepdf{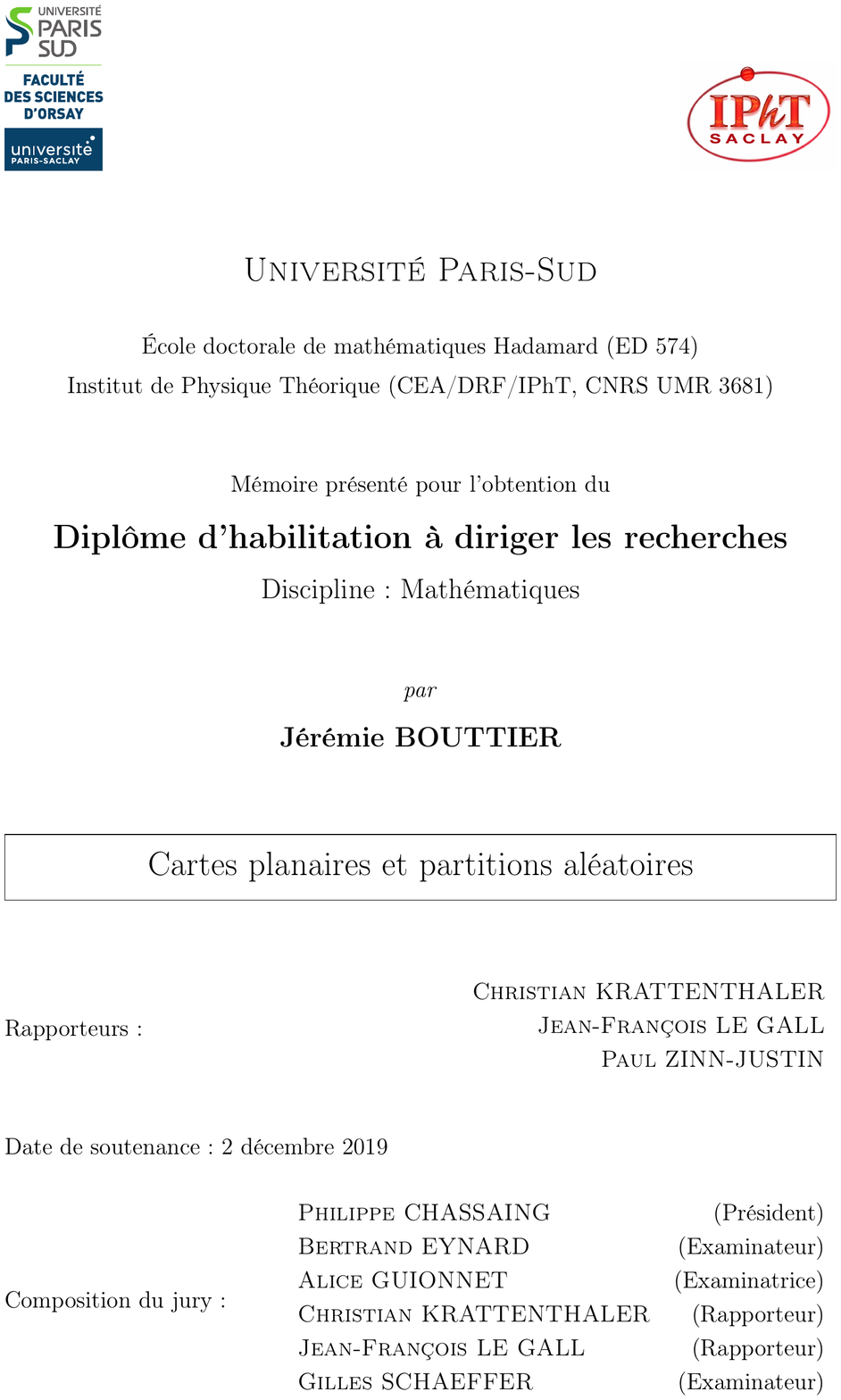}

\selectlanguage{french}
\section*{Cartes planaires et partitions aléatoires}

\paragraph{Résumé} Ce mémoire d'habilitation est une synthèse des
travaux de recherche que j'ai effectués entre 2005 et 2019. Il
s'organise en quatre chapitres. Les trois premiers portent sur les
cartes planaires aléatoires. Le chapitre~\ref{chap:dist} s'intéresse à
leurs propriétés métriques: à partir d'une bijection générale entre
cartes et mobiles, on calcule la fonction à trois points des
quadrangulations, avant d'évoquer le lien avec les fractions
continues. Le chapitre~\ref{chap:slices} présente la décomposition en
tranches, une approche bijective unifiée qui s'applique notamment aux
cartes irréductibles. Le chapitre~\ref{chap:on} a pour sujet le modèle
de boucles $O(n)$ sur les cartes planaires: par une décomposition
combinatoire récursive on obtient le diagramme de phase, puis on
étudie les statistiques d'emboîtements entre boucles. Le
chapitre~\ref{chap:dominos} porte quant à lui sur les partitions
aléatoires et processus de Schur, en allant des pavages de dominos
pentus aux systèmes fermioniques.

\bigskip

\noindent Hormis un avant-propos en français, le document est écrit en
anglais. La numérotation des pages de cette version électronique
diffère de celle des exemplaires imprimés distribués lors de la
soutenance.

\bigskip

\selectlanguage{english}
\section*{Planar maps and random partitions}

\paragraph{Abstract} This habilitation thesis summarizes the research
that I have carried out from 2005 to 2019. It is organized in four
chapters. The first three deal with random planar
maps. Chapter~\ref{chap:dist} is about their metric properties: from a
general map-mobile bijection, we compute the three-point function of
quadrangulations, before discussing the connection with continued
fractions. Chapter~\ref{chap:slices} presents the slice decomposition,
a unified bijective approach that applies notably to irreducible
maps. Chapter~\ref{chap:on} concerns the $O(n)$ loop model on planar
maps: by a combinatorial decomposition, we obtain the phase diagram
before studying loop nesting statistics. Chapter~\ref{chap:dominos}
deals with random partitions and Schur processes, from steep domino
tilings to fermionic systems.

\bigskip

\noindent The document is written in English, except an introduction
in French. The page numbers in this electronic version differ from
those of the printed copies given at the defence.

\vspace{\stretch{1}}

\selectlanguage{french}

{\small
  \noindent Email: \href{mailto:jeremie.bouttier@ipht.fr}{\texttt{jeremie.bouttier@ipht.fr}}
  \medskip
  
  \noindent Affiliations:\\
  Institut de Physique Théorique, Université Paris-Saclay, CEA, CNRS, F-91191 Gif-sur-Yvette\\
  Univ Lyon, Ens de Lyon, Univ Claude Bernard, CNRS, Laboratoire de Physique, F-69342 Lyon
  \medskip
  
  \noindent Support: \\
  Agence Nationale de la Recherche
  (ANR-12-JS02-001-01/Cartaplus, ANR-14-CE25-0014/Graal,
  ANR-18-CE40-0033/Dimers), Ville de Paris (programme Émergences
  \emph{Combinatoire à Paris}), CNRS INSMI (PEPS CARMA)}


\pagestyle{headings}

\selectlanguage{english}

\tableofcontents


\addchap{Preface}

This aim of this document is to summarize the research that I have
carried out after the completion of my doctoral thesis in 2005. This
research has been published in the articles
\citemy{HObipar,mcrt,fomap,vacancy,statgeod,threepoint,loops,pseudoquad,quadwithnoME,hankel,OHRMT,recuron,constfpsac,moreloops,pottsloop,irredmaps,irredsuite,wings,gen2p,juggling,pyramids,sampling,multispecies,dimerstat,treeloop,freeboundaries,cylindricschur,BBNVfpsac2019}\footnote{A separate bibliography is provided for these articles,
  with numbered citations such as~\citemy{HObipar}. References to
  works by other authors, or to works done during my doctoral thesis, are given
  in the general bibliography with ``alpha-style'' citations such
  as~\cite{these}.} \footnote{Due to the important fluctuations of the
  duration of the refereeing process, these articles are sorted
  according to the date of their initial posting on arXiv, which is a
  slightly more accurate indication of the period at which they were
  conceived.}. Of course I shall not attempt to be exhaustive, but I
will rather try to present how these different papers articulate
with one another.

My presentation consists of four chapters, which roughly correspond to
four different ``threads'' of my research. The first three chapters
deal with planar maps. Chapter~\ref{chap:dist} describes results about
the metric properties of random planar maps, which form the
continuation of the work done during my doctoral
thesis. Chapter~\ref{chap:slices} is devoted to the slice
decomposition, an alternate bijective approach for studying planar
maps. Chapter~\ref{chap:on} deals with the $O(n)$ loop model on random
planar maps, a statistical physics model with a rich critical
behavior. In the last Chapter~\ref{chap:dominos}, we leave the realm
of planar maps and enter that of Schur processes, which have been the
main topic of my recent research.

Each chapter aims at being as self-contained as possible: it starts
with some reminders of the context, then consists of two or three core
sections where I highlight one or two main results, and ends with a
conclusion discussing perspectives for future research. For convenience
I have gathered some common definitions in the mathematical
preliminaries at the beginning of the document.

Note that this document is deliberately \emph{not} written in the
style of a mathematical paper, as my goal is to narrate stories about
my research. Even if I follow the tradition of emphasizing some
theorems, definitions and other remarks, I avoided using the proof
environment: where relevant I provide some elements of justification
in the main text, and in any case the reader is invited to consult the
original papers where full proofs are given.

For brevity, I had to exclude several papers from my presentation. The
most notable exclusion are the three
papers~\citemy{juggling,multispecies,wings} which deal with certain
Markov chains whose study was motivated by juggling. But two out of
these papers were written with my former student François Nunzi (whom
I supervised with Sylvie Corteel), so I may simply point to his
thesis~\cite{theseNunzi} for a nice account of this topic.


\selectlanguage{french}

\addchap{Avant-propos}

\begin{quote}
  \emph{Mathématiques discrètes et continues se rencontrent et se
    complètent volontiers harmonieusement} \cite{ChFl03}.
\end{quote}

\bigskip

Ce mémoire a pour but de présenter les travaux de recherche que j'ai
effectués depuis ma soutenance de thèse de doctorat en 2005. Il
s'appuie sur les
publications~\citemy{HObipar,mcrt,fomap,vacancy,statgeod,threepoint,loops,pseudoquad,quadwithnoME,hankel,OHRMT,recuron,constfpsac,moreloops,pottsloop,irredmaps,irredsuite,wings,gen2p,juggling,pyramids,sampling,multispecies,dimerstat,treeloop,freeboundaries,cylindricschur,BBNVfpsac2019}
\footnote{Afin de différencier ces publications des citations de
  travaux d'autres auteurs, ou bien de travaux effectués pendant ma
  thèse de doctorat, j'utiliserai des références numérotées
  comme~\citemy{HObipar} pour les premières, et des références dans le
  style «alpha», comme~\cite{these}, pour les secondes qui sont
  listées dans la bibliographie générale à la fin du mémoire.}  dont
la liste\footnote{Cette liste est ordonnée selon la date de
  publication sur arXiv, qui reflète mieux que celle du journal la
  période de conception des articles.} est donnée un peu plus loin. Il
ne s'agit bien sûr pas de donner une présentation exhaustive mais
plutôt, au travers de quatre chapitres, de présenter quelques idées et
le cheminement qui m'a conduit de l'une à l'autre.

\bigskip

Commençons tout d'abord par tenter d'identifier la démarche générale
dans laquelle s'inscrivent mes recherches, qu'on peut situer à
l'interface entre la physique statistique, la combinatoire et la
théorie des probabilités. Le lien entre physique statistique et
combinatoire remonte à Boltzmann lui-même, et à sa célèbre formule
$S=k_B \ln W$ qui définit l'entropie microcanonique $S$ d'un système à
l'équilibre macroscopique en fonction de son nombre $W$ de micro-états
($k_B$ étant la constante de Boltzmann). Ce lien s'est ensuite
particulièrement développé autour des \emph{modèles exactement
  solubles}\footnote{Voir par exemple le livre de
  Baxter~\cite{BaxterBook} qui traite des modèles sur réseaux
  bidimensionnels solubles par l'ansatz de Bethe et méthodes liées.}:
la physique statistique fait grand usage de tels modèles dans
lesquelles les grandeurs physiques pertinentes (fonction de partition,
fonctions de corrélations...)  peuvent être déterminées de manière
exacte, au sens où on en a une expression suffisamment explicite pour
pouvoir passer à la limite thermodynamique, calculer des exposants
critiques, etc. Ce phénomène résulte d'une structure mathématique
sous-jacente, appelée généralement \emph{intégrabilité}\footnote{Même
  si cette notion est difficile à définir précisément au-delà du cadre
  de la mécanique classique.}. L'étude des modèles intégrables
discrets amène de nombreuses questions de nature combinatoire.  Nous
en verrons deux illustrations au cours de ce mémoire, au travers des
cartes planaires d'une part, des processus de Schur d'autre part. La
théorie des probabilités apporte quant à elle de nouveaux outils,
notamment pour l'étude des \emph{limites d'échelle}\footnote{Un autre
  apport intéressant des probabilités, peu discutée dans ce mémoire,
  est celle des «limites locales» que l'on peut rapprocher du
  formalisme des mesures de Gibbs en mécanique statistique. L'idée est
  alors de décrire directement un système infini sans passer par les
  systèmes de taille finie. Par bien des aspects, et paradoxalement,
  cette description peut s'avérer plus simple.}. En effet, dès lors
qu'on s'intéresse aux propriétés macroscopiques des modèles, il est
possible d'oublier certains de leurs détails microscopiques, et les
probabilités fournissent un cadre rigoureux pour ce faire. On arrive
parfois ainsi à donner une définition mathématique d'un objet-limite
continu capturant l'essence de la physique à grande échelle. Par
exemple, la carte brownienne\footnote{Voir les références données à la
  section~\ref{sec:distcontext}.} correspond à la limite d'échelle
générique des cartes planaires aléatoires non décorées, et le
processus d'Airy à celle des processus de Schur autour d'un point
générique de la courbe arctique\footnote{Voir par exemple l'article de
  Prähofer et Spohn~\cite{PrSp02} qui traite du modèle de croissance
  polynucléaire (\textsl{PNG}), qui se ramène à un processus de
  Schur. Nous rencontrerons brièvement le noyau d'Airy et son analogue
  à température finie au chapitre~\ref{chap:dominos}.}. Cette démarche
donne un sens mathématique au phénomène d'\emph{universalité}, puisque
différents modèles discrets peuvent avoir une même limite d'échelle
continue et donc les mêmes propriétés macroscopiques. En outre, cela
justifie la pertinence des modèles exactement solubles: en effet on
s'appuie typiquement sur leur intégrabilité afin de les étudier,
analyse qui pourra être ensuite étendue au continu et à d'autres
modèles par universalité. Comme aboutissement, on arrive à des
\emph{résultats rigoureux} qui, au côté des validations
expérimentales, permettent de confirmer ou réfuter les prédictions de
la physique théorique.

\bigskip

Entrons à présent dans le contenu de ce mémoire.  Il est constitué de
quatre chapitres principaux, qui sont essentiellement indépendants les
uns des autres: chaque chapitre commence par un bref rappel du
contexte, comprend ensuite deux ou trois sections présentant certains
de mes résultats, et se conclut sur une discussion de quelques pistes
de recherche future. Par commodité, quelques définitions communes sont
regroupées dans les préliminaires mathématiques en introduction du
mémoire.

Les trois premiers chapitres portent sur les cartes planaires, et sont
consacrés respectivement aux propriétés métriques des cartes
aléatoires, un sujet auquel j'ai continué à m'intéresser après ma
thèse de doctorat, à la décomposition en tranches, une nouvelle
approche bijective pour l'étude des cartes, et enfin au modèle de
boucles $O(n)$ sur les cartes planaires, un modèle de physique
statistique au riche comportement critique. Je ne donnerai pas ici de
présentation générale du vaste sujet de recherche que sont les cartes,
ce qu'on pourra trouver dans d'autres références. Je mentionnerai,
entre autres, la revue récente de Gilles Schaeffer~\cite{Schaeffer15}
pour les aspects relevant de la combinatoire énumérative et bijective,
les notes de cours successives de Jean-François Le Gall, Grégory
Miermont et Nicolas Curien~\cite{LGMi12,MiermontSFnotes,CurienSFnotes}
pour les aspects probabilistes, le mémoire d'habilitation de Guillaume
Chapuy~\cite{ChapuyHDR} pour une évocation du lien avec les
hiérarchies intégrables, et enfin le premier chapitre de ma thèse de
doctorat~\cite{these} ou la première partie de celle d'Ariane
Carrance~\cite{theseCarrance} pour une discussion des liens entre
cartes et physique théorique.

Dans le quatrième et dernier chapitre, je quitte le monde des cartes
planaires pour évoquer mes travaux récents autour des partitions
aléatoires, et plus précisément des processus de Schur. Le domaine
général est celui des probabilités intégrables auxquelles les notes de
cours d'Alexei Borodin et Vadim Gorin~\cite{BoGo16} donnent une bonne
introduction. Ma présentation suivra la manière dont j'ai été amené à
m'intéresser à ce domaine, en allant des pavages de dominos aux
systèmes de fermions à température finie.

Dans un souci de concision, j'ai été amené à exclure plusieurs travaux
de la présentation qui est faite dans ce mémoire. L'omission la plus
importante concerne les trois
articles~\citemy{juggling,multispecies,wings} qui portent sur des
modèles de jonglage aléatoire. Sur ce sujet, le lecteur pourra
consulter la thèse de François Nunzi~\cite{theseNunzi}.

\addsec{Résumé détaillé du mémoire}

Le chapitre~\ref{chap:dist} est consacré aux propriétés métriques des
cartes planaires aléatoires. J'y présente quelques résultats issus de
travaux poursuivant ceux effectués pendant ma thèse de doctorat. La
présentation n'est pas chronologique puisque, après un rappel du
contexte dans la section~\ref{sec:distcontext}, je commence par
présenter en section~\ref{sec:distmobiles} une bijection obtenue en
2013 avec Éric Fusy et Emmanuel Guitter dans
l'article~\citemy{gen2p}. Il s'agit d'une correspondance entre
\emph{cartes convenablement étiquetées} (cartes planaires dont les
sommets sont étiquetés par des entiers, de telle sorte que deux
sommets adjacents ont des étiquettes consécutives) et \emph{mobiles}
(cartes planaires biparties, où les sommets d'un type sont étiquetés
par des entiers, et les autres ne sont pas étiquetés mais imposent
certaines contraintes sur leurs étiquettes adjacentes). Elle unifie
plusieurs constructions antérieures: la célèbre bijection de
Cori-Vauquelin-Schaeffer~\cite{CoVa81,theseSchaeffer} entre
quadrangulations enracinées et arbres bien étiquetés, une première
généralisation que nous avions obtenue avec Philippe Di Francesco et
Emmanuel Guitter~\cite{mobiles} et qui relie cartes biparties
enracinées et mobiles à une face, et une seconde généralisation due à
Miermont, Ambj{\o}rn et Budd~\cite{Miermont09,AmBu13}\footnote{Voir la
  remarque 1 dans~\citemy{gen2p} pour une discussion de l'équivalence
  entre la formulation de Miermont et celle d'Ambj{\o}rn et Budd.}
qui relie quadrangulations convenablement étiquetées et cartes bien
étiquetées.

Cette \emph{approche bijective} est particulièrement intéressante pour
étudier les distances dans les cartes planaires. En effet, par des
choix d'étiquetages judicieux, on peut lire certaines propriétés
métriques des cartes dans les mobiles correspondants, qui se trouvent
être des arbres ou des cartes à un petit nombre de faces. L'analyse de
ces derniers objets peut alors être menée à bien par les différentes
méthodes connues pour étudier les arbres: décompositions récursives
donnant des équations pour les séries génératrices, codage par des
marches permettant de passer à la limite d'échelle, etc.

Dans la section~\ref{sec:dist3P}, je donne un exemple d'une telle
analyse pour le calcul de la \emph{fonction à trois points} des
quadrangulations, d'après l'article~\citemy{threepoint} écrit avec
Emmanuel Guitter. Le problème est de trouver la série génératrice des
quadrangulations ayant trois points marqués à distances
prescrites. Par une spécialisation de la bijection présentée en
section~\ref{sec:distmobiles}, le problème est ramené à celui du
comptage de cartes bien étiquetées à trois faces, avec des valeurs
prescrites pour les étiquettes minimales incidentes à chaque face. De
façon surprenante, ce calcul peut être mené à bien de manière
explicite, au sens où on peut donner une formule close pour les séries
génératrices correspondantes. Il s'agit d'une manifestation d'un
phénomène d'\emph{intégrabilité discrète} que nous avions déjà observé
dans le calcul de la fonction à deux points~\cite{geod}. À partir de
l'expression exacte discrète, il est facile de passer à la limite
d'échelle et calculer ainsi la fonction à trois points de la carte
brownienne, c'est-à-dire la loi jointe des distances entre trois
points tirés au hasard uniformément.

La section~\ref{sec:distCF} revient sur le phénomène d'intégrabilité
discrète mentionné ci-dessus, dans un cadre légèrement différent: on
considère la fonction à deux points seulement, mais dans le contexte
plus général des \emph{cartes de Boltzmann}. Avec Philippe Di
Francesco et Emmanuel Guitter, nous avions conjecturé une expression
exacte pour cette fonction à deux points dans
l'article~\cite{geod}. Cette expression a été prouvée quelques années
plus tard, dans l'article~\citemy{hankel} écrit avec Emmanuel
Guitter. Le point de départ de ce travail est l'observation, faite
auparavant dans l'article~\citemy{pseudoquad}, que la fonction à deux
points intervient dans le développement en \emph{fraction continue} de
la série génératrice des cartes à un bord. Par la théorie générale des
fractions continues, on peut donc déduire la première de la seconde, à
l'aide de \emph{déterminants de Hankel} qu'on peut identifier à des
caractères de groupes orthogonaux ou symplectiques.

Je conclus ce chapitre par la section~\ref{sec:distconc} qui
évoque quelques questions ouvertes.

\bigskip

Le chapitre~\ref{chap:slices} est consacré à la \emph{décomposition en
  tranches} des cartes planaires. Il s'agissait initialement d'une
reformulation de la bijection de~\cite{mobiles} évoquée dans les
articles~\citemy{pseudoquad,hankel,constfpsac}, mais sa portée plus
générale a été comprise dans l'article~\citemy{irredmaps} écrit avec
Emmanuel Guitter. En effet, nous y expliquons comment adapter la
construction au cas des cartes \emph{irréductibles}, c'est-à-dire des
cartes dans lesquelles tout cycle a une longueur au moins égale à un
entier $d\geq 0$ fixé, et tout cycle de longueur $d$ est le bord d'une
face. Une telle contrainte semble difficile à concilier avec la
bijection de la section~\ref{sec:distmobiles}.

La section~\ref{sec:slicecontext} rappelle le contexte et notamment le
développement du \emph{canevas bijectif} qui vise à unifier les
différentes bijections cartes-arbres connues et à en découvrir de
nouvelles, en s'appuyant sur la théorie des orientations.

La section~\ref{sec:slicegen} donne la définition générale des
tranches. Il s'agit de cartes avec un bord et trois points marqués
$A,B,C$ sur celui-ci, tels que le côté $AB$ est une géodésique entre
$A$ et $B$, et le côté $AC$ est l'unique géodésique entre $A$ et
$C$. Je décris ensuite deux opérations importantes:
\begin{itemize}
\item la décomposition d'une tranche en tranches \emph{élémentaires},
  qui consiste à découper le long de toutes les \emph{géodésiques
    gauches} allant des sommets du côté $BC$ vers $A$,
\item l'\emph{enroulement}, qui consiste à identifier les côtés $AB$
  et $AC$, en laissant éventuellement quelques arêtes non identifiées
  si les longueurs sont différentes: cette opération produit alors des
  cartes \emph{annulaires} (ayant deux faces marquées).
\end{itemize}
À l'aide de ces deux opérations, on parvient à exprimer la série
génératrice des cartes annulaires en fonction de celle des tranches
élémentaires, qui est elle-même égale à celle bien connue des cartes
pointées enracinées.

Dans la section~\ref{sec:sliceirr}, j'explique comment adapter la
décomposition en tranches au cas des cartes irréductibles. La
difficulté principale est d'énumérer les tranches élémentaires, qui ne
sont plus en bijection avec les cartes pointées enracinées. On est
amené à introduire la notion de \emph{quasi-tranche}, en modifiant
légèrement la définition des tranches élémentaires: le côté $AB$ n'est
plus nécessairement une géodésique entre $A$ et $B$, mais seulement un
chemin de longueur minimale parmi tous les chemins évitant l'arête
$BC$. Par une nouvelle méthode de décomposition des quasi-tranches, on
obtient un système d'équations algébriques qui détermine leurs séries
génératrices. On peut alors énumérer les cartes annulaires
irréductibles en appliquant l'opération d'enroulement. Celle-ci ne
pose pas de difficulté particulière, une fois observé qu'il convient
de traiter à part les cycles séparant les deux faces marquées.

Quelques pistes de recherche sont mentionnées en
section~\ref{sec:sliceconc}.

\bigskip

Au chapitre~\ref{chap:on}, nous considérons un modèle de physique
statistique sur les cartes planaires, le \emph{modèle de boucles
  $O(n)$}. Les configurations sont des cartes planaires décorées de
boucles (cycles disjoints sur la carte duale), chacune recevant un
poids $n$ qui joue le rôle d'un paramètre non local.

La section~\ref{sec:oncont} passe brièvement en revue la littérature
au sujet des modèles de physique statistique sur les cartes. Beaucoup
de travaux ont porté sur le modèle de Potts à $q$ états, ses
différentes reformulations et ses spécialisations. En particulier, le
modèle de boucles $O(n)$ est lié aux points «auto-duaux» du modèle de
Potts, avec $q=n^2$. Lorsque $n$ varie dans l'intervalle $[0,2]$, il
existe une famille de points critiques dont les exposants varient
continument avec $n$. On s'attend à ce que la géométrie des cartes
sous-jacentes soit affectée par les corrélations à longue échelle aux
points critiques, mais cela reste une question largement ouverte.

Les sections~\ref{sec:ongasket} et~\ref{sec:onsol} présentent
principalement les résultats de l'article~\citemy{recuron} écrit avec
Gaëtan Borot et Emmanuel Guitter. Je me concentrerai sur le cas des
boucles~\emph{rigides} sur les quadrangulations qui est plus simple à
analyser. Dans la section~\ref{sec:ongasket}, je présente la
décomposition combinatoire dite des \emph{poupées russes}\footnote{Ce
  terme me semble préférable à celui de \emph{joint de culasse} qui
  serait la traduction littérale de \textsl{gasket}.}: à une
configuration de boucles sur une quadrangulation à un bord, on associe
la carte externe formée par les arêtes extérieures à toutes les
boucles. Il s'agit d'une carte sans boucles mais pouvant avoir des
faces de degrés arbitrairement grands. Nous verrons que cette carte
suit la loi d'une carte de Boltzmann, dont les paramètres sont donnés
par une équation de point fixe qui donne également la fonction de
partition du modèle.  J'esquisse ensuite dans la
section~\ref{sec:onsol} la façon dont l'équation de point fixe peut
être résolue, en réécrivant celle-ci sous forme d'une équation
fonctionnelle pour la~\emph{résolvante}. Cette équation fonctionnelle
possède une solution explicite particulièrement simple aux
\emph{points critiques non génériques} du modèle. En particulier, on
peut voir que, à ces points, la loi du degré d'une face typique de la
face externe a une queue lourde. Par des résultats de Le Gall et
Miermont \cite{LGMi11}, la limite d'échelle diffère de la carte
brownienne.

La section~\ref{sec:onnesting} s'inspire de
l'article~\citemy{treeloop} écrit avec Gaëtan Borot et Bertrand
Duplantier. On s'y intéresse à la structure des emboîtements entre
boucles et, plus précisément, à la \emph{profondeur} d'un sommet
marqué, qui est défini comme le nombre de boucles séparant ce sommet
du bord. On montre que, à un point critique non générique, la
profondeur d'un sommet tiré au hasard croît logarithmiquement avec la
longueur du bord (plus précisément, on établit un théorème central
limite et un principe de grandes déviations pour cette profondeur). La
preuve repose sur le calcul d'une série génératrice raffinée, qu'on
parvient à mener à bien en adaptant la méthode utilisée à la
section~\ref{sec:onsol}.

La section~\ref{sec:onconc} présente quelques perspectives.

\bigskip

Dans le chapitre~\ref{chap:dominos}, je quitte le monde des cartes
planaires pour évoquer mes travaux récents autour des \emph{processus
  de Schur}. La section~\ref{sec:dominoscont} présente brièvement le
contexte, qui est celui des \emph{probabilités intégrables}. J'y
raconte également de quelle manière j'ai été amené à m'intéresser
personnellement à ce sujet, par un groupe de travail sur les
hiérarchies intégrables où je m'étais familiarisé avec le formalisme
des \emph{fermions libres}.

La section~\ref{sec:dominosschur} présente mes premières contributions
au domaine, et reprend les articles~\citemy{pyramids,dimerstat} écrits
tous deux avec Guillaume Chapuy et Sylvie Corteel, ainsi que Cédric
Boutillier et Sanjay Ramassamy pour le second. Nous définissons la
famille des \emph{pavages pentus}, qui sont des pavages du plan par
des dominos (rectangles $2 \times 1$) obtenus à partir d'un certain
pavage fondamental en effectuant une séquence de mouvements
élémentaires. Ces pavages sont en bijection avec des configurations de
dimères sur un graphe dit en \emph{gare de triage}. Par
spécialisation, nous retrouvons plusieurs familles de pavages connues:
partitions planes, partitions-pyramides et pavages du diamant
aztèque. Nous donnons des expressions explicites pour la série
génératrice des pavages pentus, ainsi que pour les \emph{fonctions de
  corrélation} qui sont les probabilités d'occurrence conjointe de
certains types de dominos à des positions prescrites. Nous faisons
également un lien avec la théorie de Kasteleyn pour les modèles de
dimères.

Dans la section~\ref{sec:dominosbc}, je m'intéresse aux pavages pentus
avec des \emph{conditions aux bords périodiques ou libres}. Leurs
séries génératrices avaient été calculées dans~\citemy{pyramids}, mais
nous n'avions pas été alors capables de calculer leurs fonctions de
corrélation. C'est un problème sur lequel je suis revenu dans les
articles~\citemy{freeboundaries,cylindricschur} écrits tous deux avec
Dan Betea, ainsi que Peter Nejjar et Mirjana Vuleti\'c pour le
premier. Nous y calculons les fonctions de corrélation des processus
de Schur avec des conditions aux bords respectivement libres et
périodiques, via le formalisme des fermions libres\footnote{Dans le
  cas périodique, les fonctions de corrélations avaient précédemment
  été calculées par Borodin~\cite{Borodin07} via une autre méthode. La
  nôtre semble plus simple.}. J'esquisse les idées principales de
notre approche, qui consiste à considérer des fermions \emph{à
  température finie} dans le cas périodique, et effectuer une
\emph{transformation de Bogolioubov} dans le cas libre.

La section~\ref{sec:dominosasymp} porte sur les applications des
considérations précédentes. En effet, on s'attend à ce que la
modification des conditions aux bords permette d'observer de nouvelles
limites d'échelles. Je considère principalement le cas périodique,
d'après~\citemy{cylindricschur} (le cas libre étant l'objet de travaux
en cours). On s'intéresse au cas le plus simple de processus de Schur
périodique, qui est une mesure sur les partitions d'entiers appelée
\emph{mesure de Plancherel cylindrique}. Cette mesure interpole entre
la mesure canonique et la mesure de Plancherel usuelle sur les
partitions. On considère la loi de la plus grande part et on montre
que, après recentrage et normalisation, celle-ci admet plusieurs
limites possibles:
\begin{itemize}
\item la loi de Gumbel dans un régime dit de haute température
  (incluant la mesure canonique),
\item la loi de Tracy-Widom de l'ensemble gaussien unitaire (\textsl{GUE}) dans
  un régime dit de basse température (incluant la mesure de Plancherel
  usuelle),
\item et enfin, dans un régime intermédiaire, une loi interpolant
  entre les deux lois précédentes, qui avait précédemment été
  rencontrée par Johansson dans le cadre des matrices aléatoires.
\end{itemize}
Du point de vue physique, cette transition de phase peut être
expliquée par une compétition entre les fluctuations d'origine
thermique et celles d'origine quantique.

Enfin, la section~\ref{sec:dominosconc} indique quelques
directions pour le futur.

\addsec{Remerciements}

Je voudrais rendre ici hommage à toutes les personnes qui m'auront
amené à présenter ce mémoire d'habilitation, et plus généralement
d'avoir la chance et le plaisir d'exercer le métier de chercheur. La
liste de ces personnes est longue, et je prie d'avance celles que
j'oublierais de citer ici nommément de bien vouloir me pardonner.

\bigskip

Comme d'usage je commencerai par le jury. Merci tout d'abord à
Jean-François Le Gall qui a bien voulu être mon rapporteur interne
pour l'école doctorale, et m'a permis de passer la première étape de
la Commission de la recherche. Merci ensuite à Christian Krattenthaler
et Paul Zinn-Justin qui ont accepté d'être mes rapporteurs externes,
et ont recommandé de m'autoriser à soutenir. Merci enfin à Philippe
Chassaing, Bertrand Eynard, Alice Guionnet et Gilles Schaeffer qui me
font l'honneur d'être mes examinateurs.

J'ajoute au passage des remerciements pour Frédéric Paulin, directeur
de l'école doctorale, pour ses nombreux conseils tout au cours de la
procédure, et au comité \emph{ad hoc} qui a donné tous les avis
favorables nécessaires.

\bigskip

Je voudrais ensuite témoigner de ma profonde gratitude envers les
institutions de recherche françaises, et envers les personnes qui les
font vivre.

En premier lieu vient mon employeur, le Commissariat à l'énergie
atomique et aux énergies alternatives, et au sein de celui-ci
l'Institut de physique théorique où j'ai le privilège d'être chercheur
permanent depuis 2006. Je dois beaucoup à mes chefs d'institut
successifs: Henri Orland, Michel Bauer et François David. Henri a pris
la décision capitale de m'offrir un poste, Michel et François m'ont
soutenu avec énergie dans la suite de ma carrière, y compris pour mes
demandes de mobilité hors de l'IPhT. Je remercie également tout le
personnel de support: Sylvie Zaffanella qui a traité avec efficacité
et gentillesse toutes mes missions et autres demandes, Laure Sauboy
qui m'a notamment aidé pour l'organisation de la conférence Itzykson
2015, Anne Angles à qui je souhaite beaucoup de succès dans ses
nouvelles fonctions, Emmanuelle de Laborderie, Loïc Bervas, Patrick
Berthelot et Laurent Sengmanivanh... Merci enfin à tous mes collègues,
parmi lesquels mes commensaux réguliers Jean-Marc Luck, Marc
Barthélémy, Cécile Monthus, Jérôme Houdayer et Lenka Zdeborová, pour
les nombreuses discussions intéressantes que nous avons eues.

J'ai passé l'année 2011-2012 au Laboratoire d'informatique
algorithmique: fondements et applications (LIAFA, devenu depuis
IRIF). Je souhaite y remercier notamment son directeur d'alors, Pierre
Fraigniaud, qui a bien voulu m'accueillir, Noëlle Delgado qui a traité
mes demandes administratives y compris après mon séjour en raison de
l'ANR Cartaplus, et tous les membres de l'équipe Combinatoire et du
groupe de lecture.

De 2012 à 2016, j'ai été professeur associé à temps partiel au
Département de mathématiques et applications de l'École normale
supérieure. Merci à Wendelin Werner et Thierry Bodineau qui ont pensé
à moi pour ce poste, à Olivier Debarre et Claude Viterbo qui ont été
mes directeurs successifs au DMA, à Bénédicte Auffray et Zaïna Elmir
qui m'ont offert leur précieux soutien administratif au laboratoire, à
Laurence Vincent puis Albane Trémeau qui ont fait de même côté
enseignement, aux collègues de l'équipe Probabilités et statistiques,
à Guilhem Semerjian avec qui j'ai eu la chance d'enseigner... sans
oublier les élèves avec qui j'ai eu de nombreux échanges
enrichissants.

Depuis 2016, je suis mis à disposition du Laboratoire de physique à
l'École normale supérieure de Lyon. Je remercie Thierry Dauxois, son
directeur, et Jean-Michel Maillet, chef de l'équipe de physique
théorique, qui ont appuyé mon «transfert» et plusieurs de mes projets
lyonnais. Je salue Grégory Miermont que j'ai eu le plaisir de
retrouver à Lyon, ne perdons pas nos bonnes habitudes d'organiser des
évènements scientifiques! Merci également à Laurence Mauduit, Nadine
Clervaux et Fatiha Bouchneb pour leur soutien administratif, au
secrétariat de l'UMPA qui fait de même lorsque je m'aventure au
quatrième étage, à Krzysztof Gawędzki et Giuliano Niccoli qui ont
partagé ou partagent leur bureau avec moi, et à Laurent Chevillard
avec qui j'échange toujours volontiers au détour d'un couloir.

Je dois ajouter à ma liste d'institutions celles qui financent mes
recherches, notamment l'Agence nationale de la recherche et la Ville
de Paris. Merci surtout aux porteurs de projets ---Cartaplus, Graal,
Combinatoire à Paris, Carma et Dimers--- pour leurs efforts qui
s'avèrent de plus en plus nécessaires pour que nous puissions
effectuer nos missions et nous équiper sans contraintes excessives.

Enfin, j'ai débuté ma carrière postdoctorale hors de France, à
l'Institut de physique théorique de l'Université d'Amsterdam. Merci à
Bernard Nienhuis qui m'a offert un contrat de deux ans, et à tous les
collègues que j'ai côtoyés là-bas et que j'ai encore plaisir à
recroiser aujourd'hui.

\bigskip

Je me tourne à présent vers mes collaborateurs, compagnons de route
qui me sont indispensables comme ma liste de publications peut en
témoigner.

Tout d'abord vient Emmanuel Guitter, qui de directeur de thèse est
devenu mon principal coauteur et le reste encore à ce jour. Ce fut une
grande chance de pouvoir débuter ma carrière avec lui, et d'obtenir
tous les résultats qui peuplent l'essentiel des trois premiers
chapitres de ce mémoire. Merci également à Philippe Di Francesco, mon
autre directeur de thèse, à qui je suis toujours fier de pouvoir
raconter mes recherches, lorsque nous parvenons à nous croiser à
Saclay.

Je dois beaucoup à Sylvie Corteel, qui m'a proposé de venir passer une
année au LIAFA. Cela a eu un grand impact sur ma carrière, et m'a
permis m'intéresser à de nouvelles questions qui ont notamment abouti
au quatrième chapitre de ce mémoire.

Je suis reconnaissant envers Dan Betea, avec qui j'ai cheminé ces
dernières années dans le monde des probabilités intégrables. C'est
bien souvent vers lui que je me tourne en premier pour vérifier l'état
de l'art ou tester une idée.

Marie Albenque tient une place particulière puisqu'elle a été ma
première collaboratrice. J'espère que nous arriverons enfin à «plier»
les tranches eulériennes et autres fractions multicontinues! Merci
également à Gaëtan Borot qui m'aura tout appris sur les subtilités du
modèle $O(n)$ et qui aura été si patient, à Guillaume Chapuy qui est
non seulement un chercheur accompli mais aussi un porteur
bienveillant, à Arvind Ayyer qui m'a transmis le goût du jonglage
aléatoire, et aussi à Chikashi Arita, Cédric Boutillier, Mark Bowick,
Bertrand Duplantier, Éric Fusy, Momo Jeng, Paul Krapivsky, Svante
Linusson, Kirone Mallick, Peter Nejjar, Sanjay Ramassamy et Mirjana
Vuleti\'c.  Je suis heureux d'avoir coencadré les thèses de François
Nunzi et Linxiao Chen, avec comme codirecteurs respectifs Sylvie
Corteel et Nicolas Curien.

\bigskip

Je terminerai par ma famille qui m'entoure de son affection au
quotidien. J'ai le bonheur de pouvoir toujours compter sur le soutien
inconditionnel de mes parents, après plus de quarante ans. Merci aussi
à Monique et Bernard qui viennent parfois prêter main forte à la
maison quand je suis en déplacement, à Mathieu et Mélanie qui sont
d'une grande complicité, et à Olivier et Kirsten chez qui j'ai terminé
la rédaction de ce mémoire. J'ai une pensée émue pour mes
grands-parents, qui m'ont tant donné et que j'aurais aimé pouvoir
inviter à ma soutenance. Merci enfin à Sophie qui partage ma vie avec
amour. Et, pour conclure, je souhaite beaucoup de joie à mes petits
chercheurs en herbe, Chloé et Raphaël, dans leurs découvertes de tous
les jours.


\selectlanguage{english}

\bibliographystylemy{hdr_myhunsrt}
\bibliographymy{HDRmy}

\addchap{Mathematical preliminaries}

We gather here for convenience some general terminology, definitions
or notations concerning the objects that we encounter in this
document. We believe that most of it is either standard or
self-explanatory.

\paragraph{Maps} \label{page:defmaps}

\begin{figure}
  \centering
  \fig{}{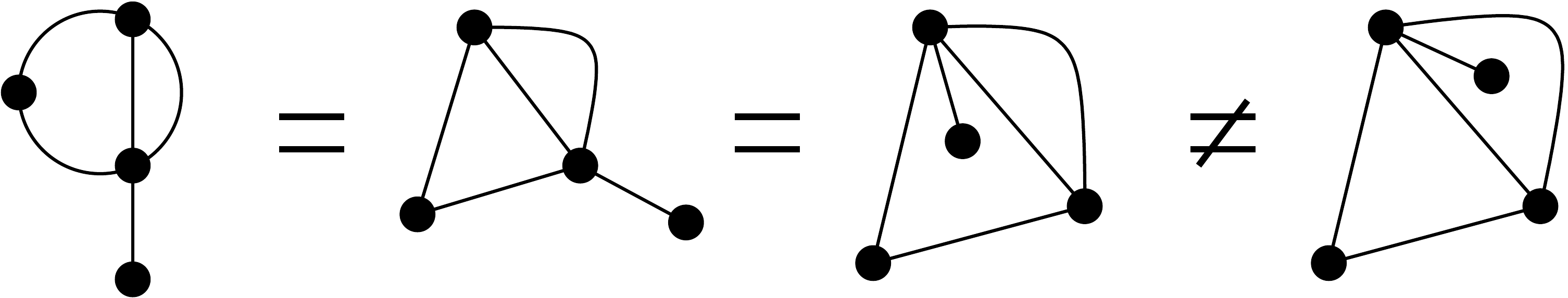}
  \caption{Different drawings of the same graph in the plane. The first
    three correspond to the same planar map: the first two differ by a
    simple deformation, and the third by a different choice for the
    outer face. The fourth corresponds to a different planar map, as
    may be seen for instance by considering the face degrees.}
  \label{fig:cellmap}
\end{figure}

A \emph{planar map} is a connected graph drawn on the sphere without
edge crossings, considered up to continuous deformation. It consists
of \emph{vertices}, \emph{edges}, \emph{faces} and \emph{corners}. In
general we allow the graph to have loops and multiple edges; otherwise
the map is said \emph{simple}. The \emph{degree} of a vertex or
face is its number of incident corners. One usually represents a
planar map via a stereographic projection of the sphere into the
plane, which amounts to distinguishing an \emph{outer} face containing
the ``point at infinity'', see Figure~\ref{fig:cellmap}.  The outer
vertices, edges and corners are those incident to the outer face. All
the rest are said \emph{inner}.
A planar map having only one face is nothing but a (plane)
\emph{tree}.

One may consider maps on surfaces other than the sphere, such as
closed orientable surfaces of higher genus or even nonorientable
surfaces. We will however not do so in this document, hence unless
stated otherwise all maps will be assumed to be planar. We might
sometimes encounter maps with \emph{boundaries}, which simply means
that some faces are distinguished. A (planar) map with one boundary is
sometimes called a \emph{disk} or \emph{plane map}---the distinguished
face is then taken as the outer face---and a map with two boundaries a
\emph{cylinder} or \emph{annular map}.

When enumerating maps, as well as graphs, trees, etc., a slight
subtlety concerns the proper treatment of ``symmetries''. This problem
is usually circumvented by considering \emph{rooted} maps. Rooting is
traditionally done by marking and orienting an edge, which amounts to
distinguishing a \emph{half-edge}; a more recent trend is to do it by
marking a corner. Of course, the two ways are equivalent as there is a
one-to-one correspondence between half-edges and corners in a
map. Rooting also allows to distinguish one \emph{root vertex} and one
\emph{root face}\footnote{This is an advantage of corner-rooting where
  the choice is canonical. For edge-rooting one needs a convention,
  for instance that the root face is the one on the right of the root
  edge.}, which is usually chosen as the outer face.
A \emph{pointed} map has a distinguished vertex called the
\emph{origin}. A rooted map is naturally pointed, but in a
\emph{pointed rooted} map the origin and the root vertex may differ.

\begin{figure}
  \centering
  \fig{.5}{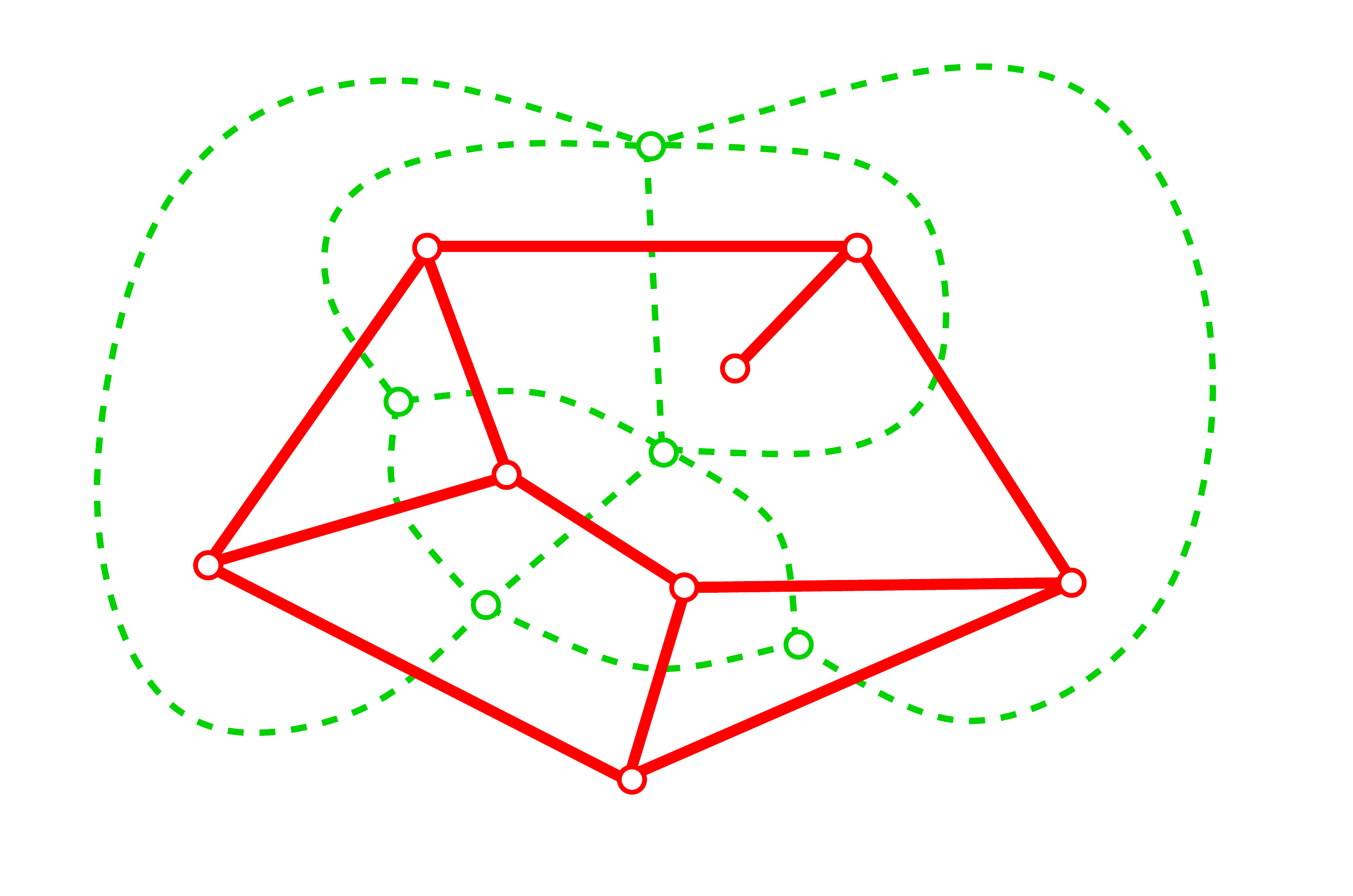}
  \caption{A map and its dual map (drawn with red solid and green
    dashed edges respectively).}
  \label{fig:dualmap}
\end{figure}

To each map we may associate its \emph{dual map} which, abstractly
speaking, is obtained by exchanging the role of vertices and
faces. More concretely, to construct the dual map, we add a new dual
vertex inside each face of the original (primal) map and, for each
primal edge, we consider its two incident faces and draw a dual edge
connecting the two dual vertices inside them. See
Figure~\ref{fig:dualmap} for an example. Note that the dual of a
simple map is not necessarily simple.

A map is said \emph{bipartite} if its vertex set can be partitioned in
two subsets so that no edge has its two endpoints in the same
subset. This implies that all the faces have even degrees and, in the
planar case, the converse is also true. Through duality we obtain an
\emph{Eulerian} map, i.e.\ a map whose vertices all have even degree.
A common combinatorial trick consists in transforming an arbitrary map
into a bipartite map by adding a \emph{dummy} vertex in the middle of
each edge. The dummy vertices have degree two and form one of the two
subsets of the bipartition. In this sense, bipartite (or dually
Eulerian) maps appear as ``more general'' than ordinary maps.  A
closely related concept is that of \emph{hypermap}, see for instance
\cite[p.~43]{LaZv04}.

\begin{figure}
  \centering
  \fig{.5}{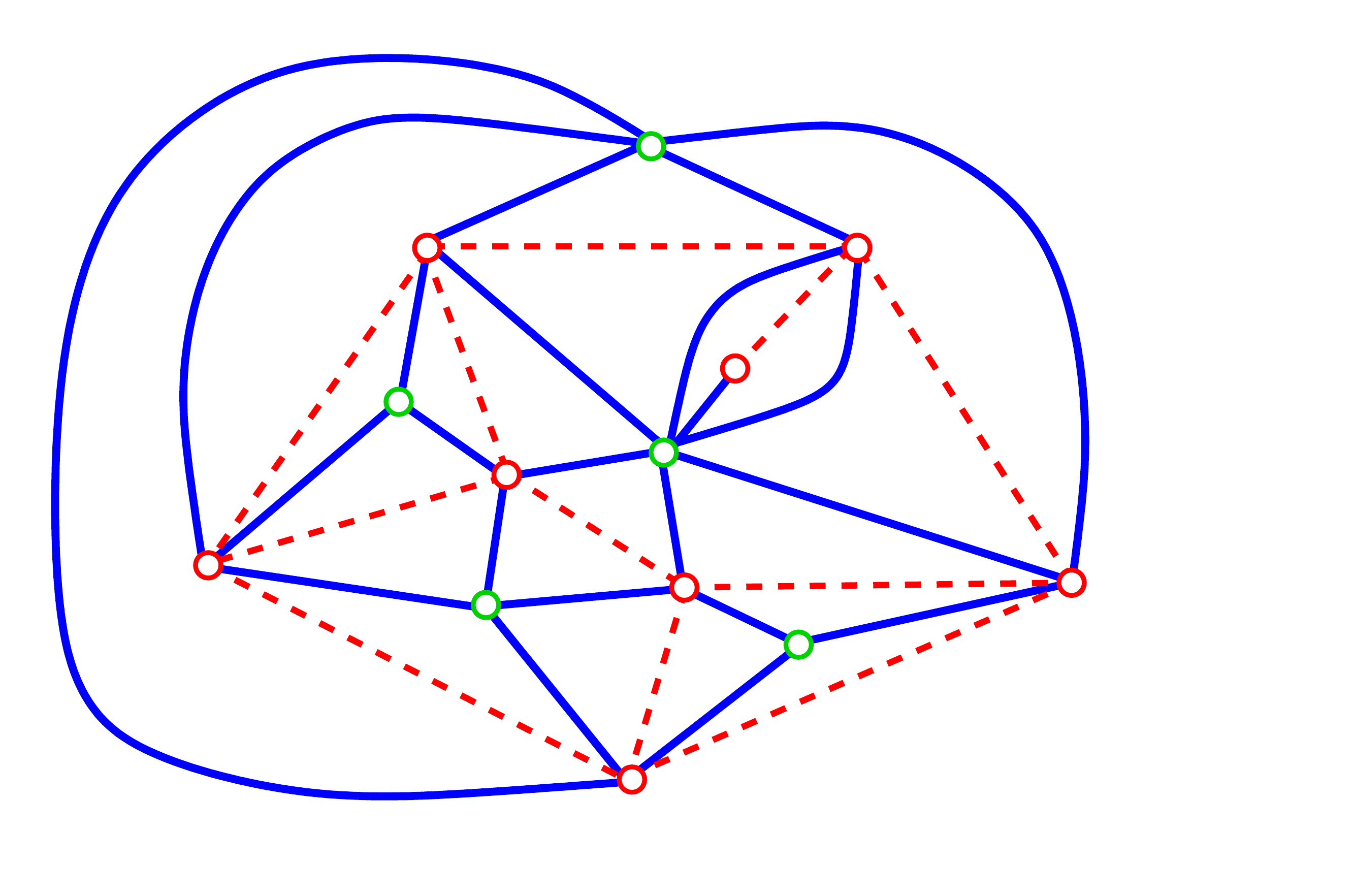}
  \caption{The quadrangulation (blue solid edges) associated with the
    map of Figure~\ref{fig:dualmap} (here shown with red dashed
    edges).}
  \label{fig:quadmap}
\end{figure}

A \emph{triangulation} is a map all of whose faces have degree
three. Similarly a \emph{quadrangulation} has all its faces of degree
four. Triangulations and quadrangulations are respectively dual to
\emph{cubic} and \emph{quartic} maps, where the degree constraint is
on the vertices. In a triangulation or quadrangulation \emph{with a
  boundary}, we relax the degree constraint on the outer face. A
planar quadrangulation (possibly with one boundary) is always
bipartite.  There is a well-known correspondence between arbitrary
maps with $n$ edges and quadrangulations with $n$ faces, see
Figure~\ref{fig:quadmap}. This correspondence is two-to-one, since the
dual map is sent to the same quadrangulation. It becomes one-to-one
after rooting, since the quadrangulation has twice as many corners as
the original map. The dual of the quadrangulation is sometimes called
the \emph{medial graph} (though it is actually a map). A celebrated result
of Tutte~\cite{Tutte63} states that the number of rooted planar maps
with $n$ edges, or equivalently the number of rooted planar
quadrangulations with $n$ faces, is equal to
\begin{equation}
  \label{eq:prelmn}
  m_n = \frac{2 \cdot 3^n}{(n+1)(n+2)} \binom{2n}{n} \sim \frac{2}{\sqrt{\pi}} \frac{12^n}{n^{5/2}} \quad (n \to \infty).
\end{equation}

A \emph{well-labeled map} (resp.\ \emph{suitably labeled
  map}\footnote{Or ``very well-labeled'' map~\cite{Arques86}}) is a
map whose vertices are labeled by integers, in such a way that the
labels of adjacent vertices differ by at most $1$ (resp.\ exactly
$1$). Note that a suitably labeled map is necessarily bipartite, since
vertices with odd labels can only be adjacent to vertices with even
labels, and vice versa.

For us, the notions of \emph{path} and \emph{walk} on a map (or graph)
are synonymous. The~\emph{length} of a path is its number of edges. A
path whose two endpoints are equal is said~\emph{closed}. A path which
never visits the same vertex twice, except possibly at the endpoints,
is said~\emph{simple} or \emph{self-avoiding}. A simple closed path is
called a~\emph{cycle}.  The \emph{contour} of a face is the closed
path formed by its incident edges. A face is said simple if its
contour is a cycle (the contour is not a cycle when the face is
incident to bridges or ``pinch points''). A~\emph{geodesic} between
two vertices is a path of minimal length connecting them. It is
necessarily simple. The~\emph{girth} of a map is the length of a
shortest cycle. It is infinite in the case of a tree.

\paragraph{Partitions} \label{page:defpar}

An \emph{integer partition}, hereafter called a \emph{partition} for
short, is a way to decompose a nonnegative integer, called the
\emph{size} of the partition, as a sum of positive integers. The order
of the terms in the sum is disregarded\footnote{If we do care about
  the order, we have a \emph{composition}.}, therefore we may order
them in weakly decreasing order, and view a partition as a finite
nonincreasing sequence of positive integers called~\emph{parts}.  The
number of parts is the \emph{length} of the partition.

The tradition is to denote partitions with lower greek letters:
$\lambda$, $\mu$, $\nu$... Given a partition $\lambda$, its size is
denoted $\norm{\lambda}$, its length is denoted $\ell(\lambda)$, and
its parts are denoted $\lambda_1,\ldots,\lambda_{\ell(\lambda)}$ with
$\lambda_1 \geq \cdots \geq \lambda_{\ell(\lambda)}$. It is convenient
to set $\lambda_i=0$ for $i>\ell(\lambda)$. The notation
$\lambda \vdash n$ means that $\lambda$ is a partition of size
$n$. The \emph{empty partition}, denoted $\emptyset$, is the unique
partition of size zero.

\begin{figure}
  \centering
  \includegraphics[scale=.7]{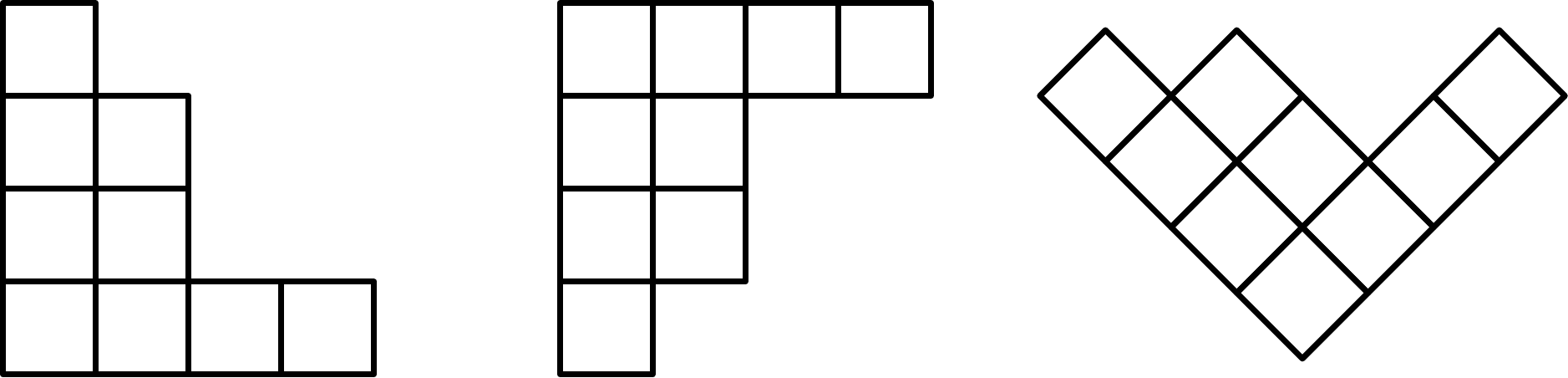}
  \caption{The Young diagram of the partition $(4,2,2,1)$ shown in
    French (left), English (middle) and Russian (right)
    conventions. The conjugate partition is $(4,3,1,1)$.}
  \label{fig:YoungDiagramFER}
\end{figure}

A well-known graphical representation of partitions is via \emph{Young
  diagrams}. Trusting that a picture is worth a thousand words, we
refer to Figure~\ref{fig:YoungDiagramFER} for an example. The small
squares are called \emph{cells}. If we reflect the Young diagram of a
partition $\lambda$ along its main diagonal, we obtain the Young
diagram of its \emph{conjugate} partition, denoted $\lambda'$. In
mathematical notation, we have $\lambda'_i=\#\{j:\lambda_j\geq i\}$.

\begin{figure}
  \centering
  \includegraphics[scale=.7]{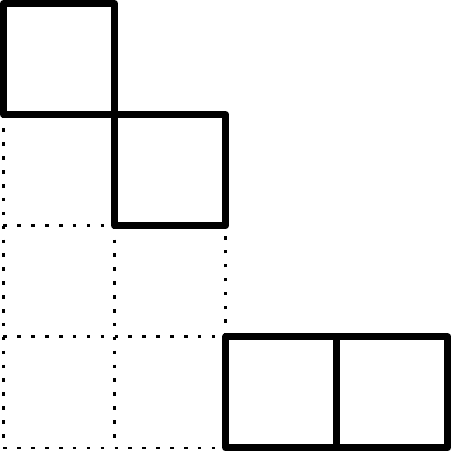}
  \caption{The skew Young diagram of shape $(4,2,2,1)/(2,2,1)$, shown
    in French convention. It is a horizontal strip since no two cells
    are in the same column.}
  \label{fig:YoungDiagramSkew}
\end{figure}

For two partitions $\lambda,\mu$, we write $\lambda \supset \mu$ if
the Young diagram of $\lambda$ contains that of $\mu$, in other words
if we have $\lambda_i \geq \mu_i$ for all $i$. Such pair
$(\lambda,\mu)$ is called a \emph{skew shape}, and is denoted
$\lambda/\mu$. It is represented as a \emph{skew Young diagram}, see
Figure~\ref{fig:YoungDiagramSkew}. Its size is
$\norm{\lambda/\mu}:=\norm{\lambda}-\norm{\mu}$.

The skew shape $\lambda/\mu$ is called a \emph{horizontal strip} if we have
\begin{equation}
  \lambda_1 \geq \mu_1 \geq \lambda_2 \geq \mu_2 \geq \cdots
\end{equation}
If so, we write
\begin{equation}
  \lambda \succ \mu
  \label{eq:succnot}
\end{equation}
and we say that the partitions $\lambda$ and $\mu$ are (horizontally)
\emph{interlaced}. Graphically, this says that the cells of the skew
Young diagram of shape $\lambda/\mu$ lie in different columns, see
again Figure~\ref{fig:YoungDiagramSkew}.

If $\lambda$ and $\mu$ are such that their conjugates satisfy
$\lambda' \succ \mu'$, then we say that the skew shape $\lambda/\mu$
is a \emph{vertical strip}, that the two partitions are
\emph{vertically interlaced}, and we use the alternate notation
\begin{equation}
  \lambda \succ' \mu.
\end{equation}
Note that vertical interlacing amounts to the condition that
\begin{equation}
  \label{eq:vertint}
  \lambda_i-\mu_i \in \{0,1\}, \qquad i=1,2,\ldots.
\end{equation}

Partitions are classically related with symmetric
functions~\cite{Macdonald95}. In Section~\ref{chap:dominos}, we will
encounter Schur functions. To a skew shape $\lambda/\mu$ we associate
the \emph{skew Schur function} $s_{\lambda/\mu}$ that may be defined
through the Jacobi-Trudi formula
\begin{equation}
  \label{eq:jacobitrudi}
  s_{\lambda/\mu}(x_1,\ldots,x_n) = \det_{1 \leq i,j \leq \ell(\lambda)} h_{\lambda_i-i-\mu_j+j}(x_1,\ldots,x_n)
\end{equation}
where $h_k(x_1,\ldots,x_n)$ is the complete homogeneous symmetric
polynomial\footnote{For simplicity we will ignore the distinction
  between symmetric polynomials and symmetric functions.} of degree
$k$, which is conveniently defined by the generating function
\begin{equation}
  \sum_{k=0}^\infty h_k(x_1,\ldots,x_n) z^k = \prod_{i=1}^n \frac{1}{1-x_i z}
\end{equation}
(with $h_k=0$ for $k<0$). The classical Schur function $s_\lambda$ is
obtained by taking $\mu$ equal to the empty partition $\emptyset$. It
may also be expressed as a ratio of alternants
\begin{equation}
  s_\lambda(x_1,\ldots,x_n) = \frac{\det_{1 \leq i,j \leq n} x_j^{\lambda_i+n-i}}{
    \det_{1 \leq i,j \leq n} x_j^{n-i}}
\end{equation}
which may be seen as an instance of the Weyl character formula for the
general linear group. In Section~\ref{sec:distCF} we will also
encounter symplectic and orthogonal Schur functions, corresponding to
characters of the respective groups (but we do not give the formulas
here for brevity).

It is an elementary exercise to check that the skew Schur function of
a single variable $x$ is given by
\begin{equation}
  \label{eq:schursingle}
  s_{\lambda/\mu}(x) =
  \begin{cases}
    x^{\norm{\lambda/\mu}} & \text{if $\lambda \succ \mu$,}\\
    0 & \text{otherwise.}
  \end{cases}
\end{equation}
This relation, together with the ``branching'' relation
\begin{equation}
  s_{\lambda/\mu}(x_1,\ldots,x_n) = \sum_\nu s_{\lambda/\nu}(x_1,\ldots,x_m) s_{\nu/\mu}(x_{m+1},\ldots,x_n) \qquad (m \leq n)
\end{equation}
which results from the Jacobi-Trudi formula~\eqref{eq:jacobitrudi} and
the Cauchy-Binet formula, yields the expression
\begin{equation}
  s_{\lambda/\mu}(x_1,\ldots,x_n) = \sum_{\substack{\lambda^{(0)} \prec \lambda^{(1)} \prec \cdots \prec \lambda^{(n)} \\ \lambda^{(0)}=\mu,\ \lambda^{(n)}=\lambda}} x_1^{\norm{\lambda^{(1)}/\lambda^{(0)}}} \cdots
  x_n^{\norm{\lambda^{(n)}/\lambda^{(n-1)}}}
\end{equation}
which amounts to the combinatorial definition of the Schur functions
in terms of semistandard Young tableaux (that are in bijection with
the sequences of interlaced partitions on which we sum). For $y$ a
variable, the~\emph{exponential specialization}
$\ex_y$~\cite[p.~304]{Stanley99} is formally obtained by taking
$x_1=\cdots=x_n=\frac{y}{n}$ and letting $n$ tend to infinity. When
applied to the skew Schur function $s_{\lambda/\mu}$, it gives
\begin{equation}
  \label{eq:expspec}
  s_{\lambda/\mu}(\ex_y) = \lim_{n \to \infty} s_{\lambda/\mu}\Big(
  \underbrace{\frac{y}{n},\ldots,\frac{y}{n}}_{n \text{ times}}\Big) =
  \frac{\dim(\lambda/\mu)}{\norm{\lambda/\mu}!} y^{\norm{\lambda/\mu}}
\end{equation}
where $\dim(\lambda/\mu)$ is the number of standard Young tableaux of
shape $\lambda/\mu$, i.e.\ the number of ways to fill the cells of the
Young diagram of $\lambda/\mu$ with the numbers
$1,\ldots,\norm{\lambda/\mu}$ in such a way that cell numbers increase
strictly along rows and colums (from left to right and from bottom to
top in French convention).


\chapter{Mobiles and distance statistics of random planar maps}
\label{chap:dist}

The purpose of this chapter is to present some of my contributions to
the study of the metric properties of random planar maps. The
publications closest to this topic are
\citemy{statgeod,threepoint,loops,pseudoquad,quadwithnoME,hankel,constfpsac,gen2p}. The
general strategy is the following: using a suitable bijection we may
translate the problem of determining the distribution of a certain
``metric observable'' of a map into the problem of counting
``mobiles'' subject to certain constraints. In several favorable
cases, this counting problem may often be solved exactly, and we may
then take asymptotics to deduce properties of large random maps.

My presentation will neither be exhaustive nor chronological. After
recalling the context in Section~\ref{sec:distcontext}, I will start
in Section~\ref{sec:distmobiles} with a general ``map-mobile''
correspondence taken from~\citemy{gen2p}. I will then apply two
particular cases of this correspondence: in Section~\ref{sec:dist3P}
to compute the three-point function of
quadrangulations~\citemy{threepoint}, and in Section~\ref{sec:distCF}
to study the two-point function of Boltzmann maps and its connection
with continued fractions~\citemy{hankel}. Perspectives are discussed
in Section~\ref{sec:distconc}.

\section{Context}
\label{sec:distcontext}

The study of the metric properties of random maps has been an
exploding topic of research over the last 20 years. To my knowledge,
considering the \emph{distance} in 2D quantum gravity was first
advocated by Watabiki and his collaborators, see e.g.~\cite{AmWa95}
and references therein. In particular, this paper successfully
predicted\footnote{It has recently been realized by Timothy Budd that
  the derivation of~\cite{AmWa95}, which is usually regarded as
  nonrigorous, can be made rigorous if one understands that it
  actually describes the first passage percolation distance in cubic
  maps with exponential edge weights~\cite{AmBu16}.} the asymptotic
form of the so-called \emph{two-point function} of pure
gravity. Coincidentally, Schaeffer introduced in his
thesis~\cite[Chapter~6]{theseSchaeffer} one of the most important
tools for a rigorous study of distances, namely the bijection between
quadrangulations and well-labeled trees. His construction is actually
a clever reformulation ---which makes the role of distance manifest---
of an earlier bijection due Cori and Vauquelin~\cite{CoVa81}, hence it
is nowadays known as the CVS bijection.  It was exploited
in~\cite{ChSc04} to prove the first convergence result for a
metric-related observable, the radius.

This is the time where I entered into this subject, during my own
thesis.
This led to several papers
written with Philippe Di Francesco and Emmanuel Guitter: in a
nutshell we brought the following two main contributions to the field.
\begin{enumerate}
\item First, we gave a rigorous derivation of the Ambj{\o}rn-Watabiki
  expression for the two-point function~\cite{geod}, relying on an
  unexpected property of \emph{discrete integrability}. At the time
  this was a mere observation, but understanding the combinatorial
  origin of this integrability, as well as working out its
  applications, has been a major topic of my research in the early
  period covered by this memoir. This research will be summarized in
  Sections~\ref{sec:dist3P} and~\ref{sec:distCF}.
\item Second, we gave a generalization of the CVS bijection to maps
  with arbitrary face degrees~\cite{mobiles}.
  It is now often called the ``BDG'' bijection but I shall rather use
  the term~\emph{mobile} which we introduced to designate the type of
  trees produced by our construction. While more elementary than the
  previous item, it turned out to have important applications as it
  allows to establish \emph{universality} results: the CVS bijection
  allows to establish convergence results for distance in
  quadrangulations, our bijection allows to extend them to
  triangulations, pentagulations, etc. In
  Section~\ref{sec:distmobiles}, I will give a further extension of
  the mobile bijection, obtained jointly with Éric Fusy and Emmanuel
  Guitter~\citemy{gen2p}.
\end{enumerate}

After that, the subject became a major topic in probability
theory. Let me highlight some works, without the ambition of being
exhaustive.  First, Marckert and Mokkadem~\cite{MaMo06} introduced the
\emph{Brownian map} describing the scaling limit of
quadrangulations. They proved the convergence in a certain weak sense
and, in his ICM2006 lecture, Schramm~\cite[Problem~4.1]{Schr07} asked
whether the convergence also holds in the stronger Gromov-Hausdorff
sense. This question was addressed in a series of papers mostly by Le
Gall and
Miermont~\cite{LeGall07,LGPa08,Miermont08,Miermont09,LeGall10,Miermont13,LeGall13}
which ended up with an affirmative answer to Schramm's question. See
for instance~\cite{MiermontSFnotes} for a pedagogical
presentation. Further developments went in several directions, among
which the case of maps of higher genus~\cite{CMS09,Bet10,Bet12,Cha17}
or on nonorientable surfaces~\cite{ChDo17,Bet15}, and proofs of
convergence to the Brownian map for various families of
maps~\cite{BeLG13,BJM14,Abr16,AdAl17,Mar18} (some of them not
analyzable using the CVS/mobiles bijections). A direction which is
still very open is the investigation of families of maps which do
\emph{not} converge to the Brownian map, but for instance to the
so-called stable maps \cite{LGMi11,Mar18b}. We shall return to this
question in Chapter~\ref{chap:on}, where we discuss its connection
with the $O(n)$ loop model on random maps.

So far I have only mentioned works that rely on a bijective approach,
but further away from the considerations developed in this chapter
there are at least two other sides to the story (that were mostly
developed after the publications presented in this chapter):
\begin{itemize}
\item local limits of planar maps, whose study was initiated by Angel
  and Schramm~\cite{AnSc03}, and which recently brought new insights
  in the study of distances, see e.g.~\cite{CuLG15,BuCu17} and
  references therein,
\item Liouville quantum gravity, whose metric structure and relation
  with random maps have been recently investigated by Miller,
  Sheffield and their collaborators in an imposing series of papers
  including~\cite{MiSh15a,MiSh15b,MiSh16a,MiSh16b,GMS17}. Interestingly,
  it seems that this approach is able to provide information about
  families of maps not converging to the Brownian
  map~\cite{GHS17,DiGw18}.
\end{itemize}

\section{From maps to mobiles}
\label{sec:distmobiles}

In this section we present a bijection obtained with Éric Fusy and
Emmanuel Guitter in~\citemy{gen2p}. It may be seen as the ``merger''
of two generalizations of the CVS bijection: the bijection
of~\cite{mobiles} mapping bipartite maps to mobiles, and the
Miermont-Ambj{\o}rn-Budd bijection~\cite{Miermont09,AmBu13} mapping
suitably labeled quadrangulations to well-labeled maps.  We start by
defining the two families related by our bijection.

\begin{figure}
  \centering
  \fig{.8}{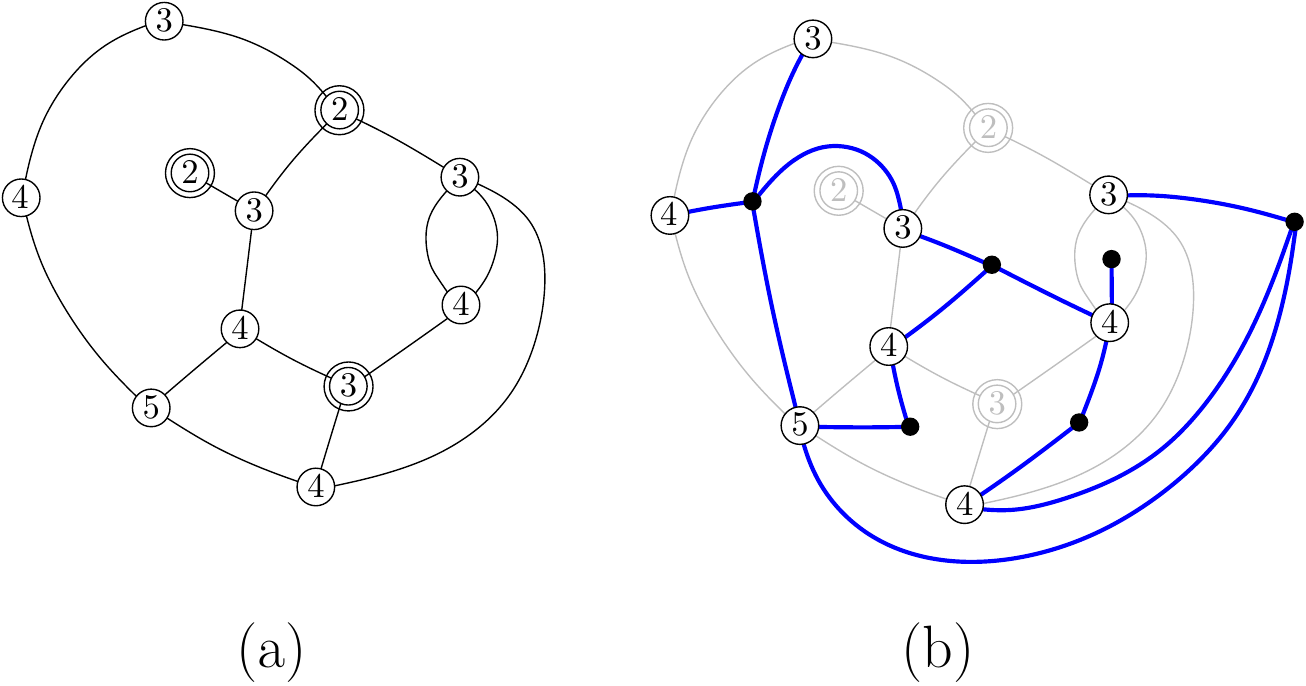}
  \caption{(a) A suitably labeled map (local min are
    surrounded). (b) Its corresponding mobile (blue edges).}
  \label{fig:bij_open_red}
\end{figure}

Recall from the \hyperref[page:defmaps]{mathematical preliminaries}
that a \emph{suitably labeled map} is a planar map whose vertices are
labeled by integers, in such a way that the labels of adjacent
vertices differ by exactly $1$. A \emph{local min}
is a vertex every neighbor of which has greater
label. See Figure~\ref{fig:bij_open_red}(a) for an example.

\begin{figure}
  \centering
  \fig{.6}{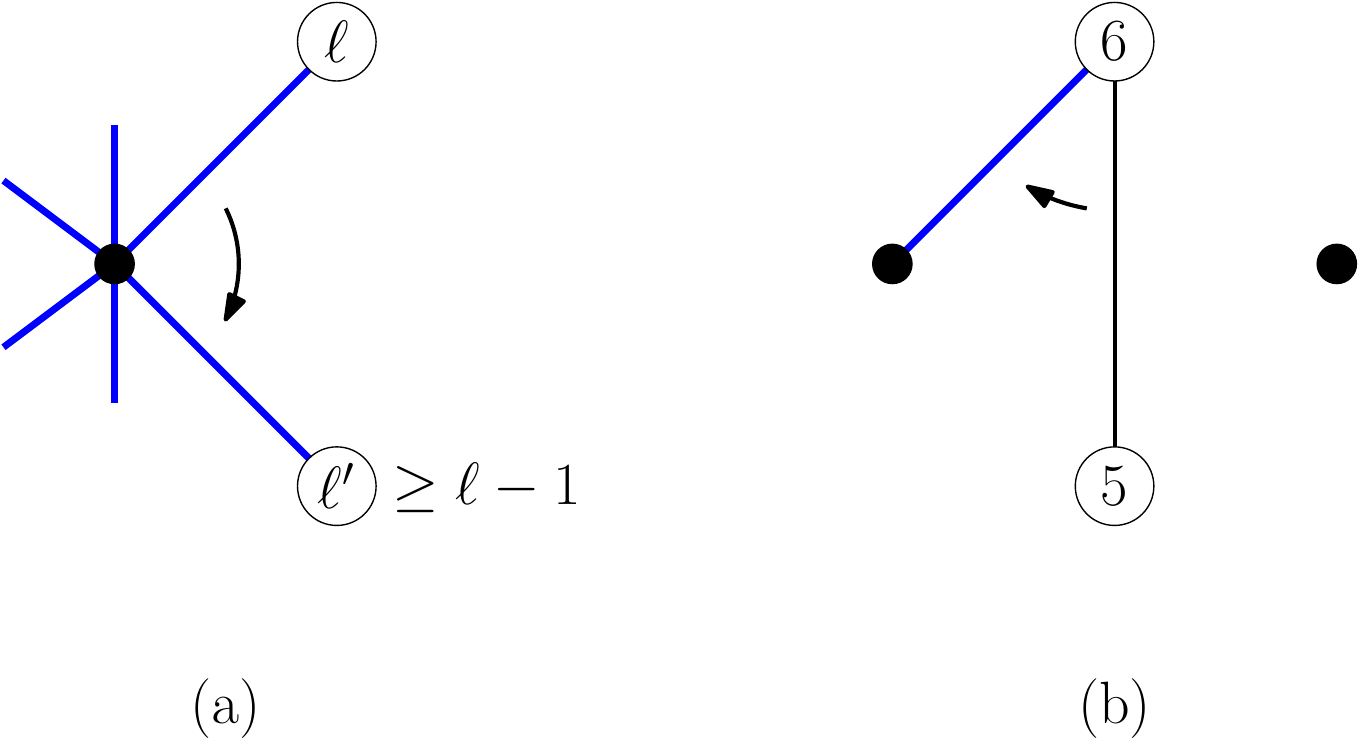}
  \caption{(a) The constraint on labels in a mobile: around an
    unlabeled (black) vertex, the labels $\ell,\ell'$ of two adjacent
    vertices that appear consecutively in clockwise order must satisfy
    $\ell'\geq \ell-1$. (b) The local rule from~\cite{mobiles}: for
    each edge of the suitably labeled map, we connect its endpoint
    with larger label (here equal to $6$) to the unlabeled vertex
    added in the face on the left (when seen from the smaller label,
    here equal to $5$).}
  \label{fig:mobile_edgerule}
\end{figure}

A \emph{mobile} is a bipartite planar map where one of the subsets of
the bipartition consists of \emph{labeled vertices} which carry an
integer label (the vertices in the other subset being
\emph{unlabeled}), subject to the constraint displayed on
Figure~\ref{fig:mobile_edgerule}(a).

To construct a mobile from a suitably labeled map, first add one
unlabeled vertex inside each face of the map, then apply the local
rule from \cite{mobiles} shown on Figure~\ref{fig:mobile_edgerule}(b),
and finally remove the local min which, by construction, are not
incident to any edge in the mobile. See
Figure~\ref{fig:bij_open_red}(b) for an example. Note that, in
\cite{mobiles}, the construction was restricted to the case where the
suitably labeled map is endowed with a ``geodesic labeling'' (to be
defined below), in which case the resulting mobile is a tree. The
extension to a general suitable labeling was done in~\citemy{gen2p},
and the mobile is not necessarily a tree anymore. In fact, its number
of faces is equal to the number of local min in the suitably labeled
map.

\begin{thm}[see {\citemy[Theorem~1]{gen2p}}]
  \label{thm:gen2p1}
  The construction described above is a bijection between suitably
  labeled maps and mobiles. Furthermore, we have the following
  correspondences between their components:
  \begin{center}
    \begin{tabular}{rcl}
      suitably labeled map & $\leftrightarrow$ & mobile, \\
      local min & $\leftrightarrow$ & face, \\
      non-local min vertex & $\leftrightarrow$ & vertex, \\
      face of degree $2k$ & $\leftrightarrow$ & unlabeled vertex of degree $k$, \\
      edge with endpoints labeled $\ell/\ell-1$ & $\leftrightarrow$ & corner at a vertex labeled $\ell$. \\
    \end{tabular}
  \end{center}
\end{thm}

We refer to \citemy{gen2p} for the proof and a description of the
reverse bijection. We now discuss two particular cases.
\begin{itemize}
\item When the suitably labeled map is a quadrangulation, then all
  unlabeled vertices in the mobile have degree two. Consequently we
  may ignore them, to obtain a well-labeled map\footnote{It is
    somewhat surprising that well-labeled maps appear both as
    generalizations of suitably labeled maps and as particular cases
    of mobiles.}. We recover a bijection which appeared in two
  different forms in \cite{Miermont09} and \cite{AmBu13}, see for
  instance \citemy[Remark~1]{gen2p} for a discussion of their
  equivalence.
\item When the suitably labeled map has a unique local min, whose
  label can be taken equal to $0$ by a global shift, then it is not
  difficult to check that the label of any vertex is necessarily equal
  to its graph distance to the local min. The labeling is then said
  \emph{geodesic}. The corresponding mobile is a tree, and we recover
  the bijection from~\cite{mobiles} between pointed bipartite maps and
  mobiles.
\end{itemize}
The intersection between these two cases corresponds to the CVS
bijection between pointed quadrangulations (endowed with their
geodesic labeling) and well-labeled trees.

\begin{rem}
  \label{rem:frust}
  In~\cite{mobiles}, we also gave more general constructions dealing
  with non necessarily bipartite maps. Lifting the bipartiteness
  assumption is important to study for instance
  triangulations~\cite[Section~8]{LeGall13}. In fact, we may adapt our
  current construction as follows: instead of a suitably labeled map,
  let us start with a well-labeled map. There may then exist some
  \emph{frustrated} edges whose two endpoints have the same label, and
  this is unavoidable if the map is not bipartite. We then modify the
  map by adding new ``square'' vertices in the middle of frustrated
  edges, as shown on Figure~\ref{fig:mobile_frustrated}: we get a
  suitably labeled map. We then construct the mobile as before: the
  square vertices (which cannot be local min) remain of degree two in
  the mobile. Furthermore, we may check that there is an additional
  constraint on the mobile: if, on
  Figure~\ref{fig:mobile_edgerule}(a), the vertex with label $\ell'$
  is square, then we must have $\ell' \geq \ell$ (instead of
  $\ell' \geq \ell -1$).  In the case where the original map had a
  unique local min, hence was endowed with its geodesic labeling, we
  recover the construction from~\cite[Section~4.2]{mobiles}, where
  square vertices were regarded as ``flagged edges'' with labels
  shifted by $-1$\footnote{As observed in
    \cite[Section~3.3]{Miermont06}, another natural convention
    consists in labeling the square vertices by half-integers, for
    instance the square vertex on Figure~\ref{fig:mobile_frustrated}
    would get label $3.5$ instead of $4$.}. It is also possible to
  recover the most general construction from~\cite[Section~3]{mobiles}
  for Eulerian maps, we leave this as an exercise to the reader.
  \begin{figure}
    \centering
    \fig{.8}{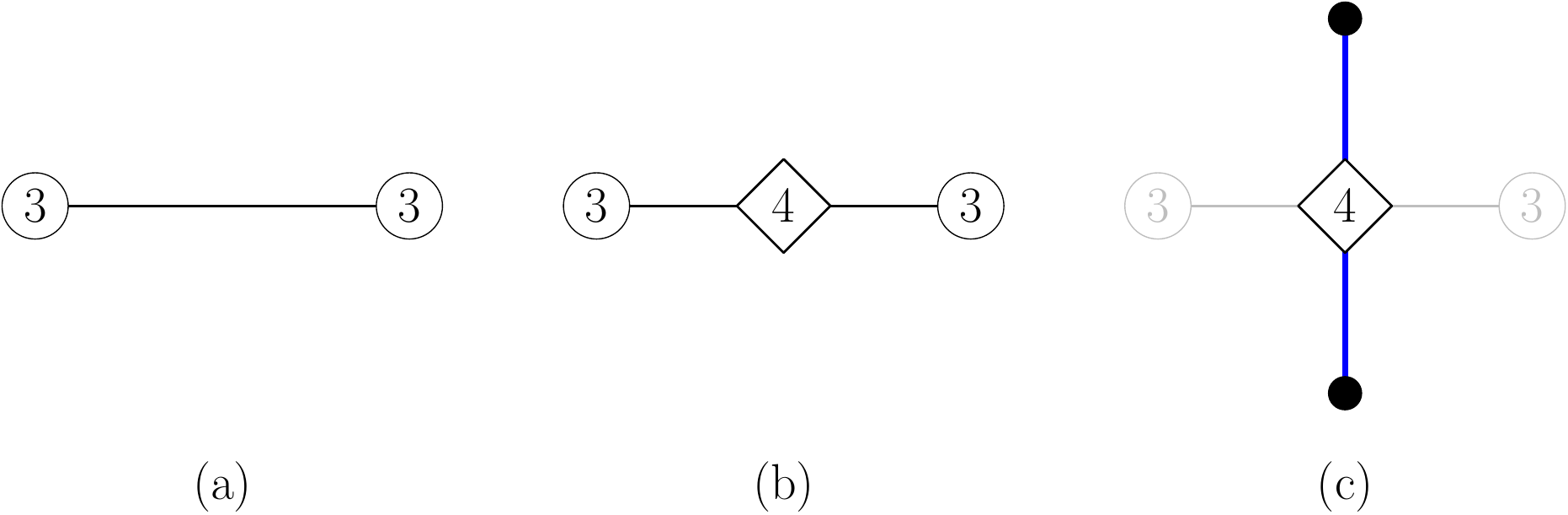}
    \caption{Dealing with frustrated edges: starting with a frustrated
      edge with endpoints labeled, say, $3$ (a), we add in its middle
      a new ``square'' vertex with label $3+1=4$ (b). This square
      vertex becomes a labeled vertex of degree two in the mobile (c).}
    \label{fig:mobile_frustrated}
  \end{figure}
\end{rem}

\begin{rem}
  The bijection extends without difficulty to maps of higher
  genus~\citemy[Section~2.7]{gen2p}. The extension of the CVS
  bijection to higher genus was done in~\cite{CMS09} and that
  of~\cite{mobiles} in~\cite{Cha09}.
\end{rem}

Theorem~\ref{thm:gen2p1} was used in~\citemy{gen2p} as the basis for
several further correspondences. In particular we adapted a neat idea
from~\cite{AmBu13} which consists in composing our bijection with its
``mirror''. The mirror bijection consists in reversing the rule of
Figure~\ref{fig:mobile_edgerule}(b): instead of connecting the vertex
with larger label to the unlabeled vertex on the left, we connect the
two other vertices. And we do so for every edge of the suitably
labeled map. In the mirror bijection, it is now the \emph{local max}
(rather than local min) which become the faces of the mobile. This has
drastic consequences in the case where the suitably labeled map is
endowed with its geodesic labeling: there is no reason that there
exists a unique local max, unless the map is ``causal'', which was one
the motivations of~\cite{AmBu13}. Therefore, the mirror mobile has in
general several faces, but it is possible to show that its labeling is
also geodesic in a suitable sense~\citemy[Proposition~2]{gen2p}. From
this observation, we have been able to compute the two-point function
of general maps and other similar families. We do not enter into the
details here, but since we have been using the term ``two-point
function'' several times without explaining it, let us try to give a
(slightly informal) definition:

\begin{Def}
  \label{Def:dist2p}
  Let $\mathcal{M}$ be a family of maps, endowed with a \emph{weight
    function} $w: \mathcal{M} \to \mathbb{R}_+$. Let us denote by
  $\ddot{\mathcal{M}}$ the set of maps in $\mathcal{M}$
  with two marked points and, for
  $m \in \ddot{\mathcal{M}}$, denote by $D(m)$ the graph
  distance between the two marked points in $m$. Then, the
  (distance-dependent) \emph{two-point function} $(T_d)_{d \geq 0}$
  associated with the weighted family $(\mathcal{M},w)$ is defined by
  \begin{equation}
    T_d := \sum_{\substack{m \in \ddot{\mathcal{M}} \\ D(m) \leq d}} w(m) \in
    \mathbb{R}_+ \cup \{+\infty\}.
  \end{equation}
  Taking $d=\infty$ means that we drop the constraint on $D(m)$ in the
  sum.
\end{Def}

When the weight function is such that $0<T_\infty<\infty$, then
$w(\cdot)/T_\infty$ defines a probability measure on
$\ddot{\mathcal{M}}$, and $T_d/T_\infty$ may be interpreted as the
distribution function of the random variable $D$.

\begin{exmp}
  \label{exmp:distgm2p}
  The two-point function of general maps considered in~\citemy{gen2p}
  corresponds to taking $\mathcal{M}$ the set of all planar maps, and
  the weight function $w(m)=t^{E(m)}$, where $E(m)$ denote the number
  of edges of $m$ and $t$ is a nonnegative real parameter. Here $T_d$
  is finite for all $d$ if and only if $t \leq 1/12$, as may be seen
  from \eqref{eq:prelmn}.
\end{exmp}

What makes Definition~\ref{Def:dist2p} slightly informal is the notion
of ``points'': ideally these would be vertices, but then we run into
the problem that a map with only two marked vertices may have
symmetries.  We usually circumvent the problem by considering pointed
rooted maps ---i.e.\ we mark one vertex and one corner--- at the price
of making the definition less symmetric. Such details do not matter
much for asymptotics.

It is immediate to extend the definition to an arbitrary number $n$ of
marked points: the (full distance-dependent) \emph{$n$-point function}
of $(\mathcal{M},w)$ depends on $\binom{n}{2}$ parameters, which
control the distances between the marked points. These distances
should obey the triangular inequalities, which immediately raises the
question of a ``natural parametrization'' of the $n$-point
function. As we shall see in the next section, a very natural
parametrization exists for $n=3$, while the situation remains widely
open for $n \geq 4$. Note that, for $n \geq 3$, the problem of
symmetries does not occur anymore: a map with $3$ marked distinct
vertices cannot have symmetries.

\section{The three-point function of quadrangulations and related results}
\label{sec:dist3P}

In this section we restrict to the case of quadrangulations, endowed
with the weight function $w(m)=g^{F(m)}$ where $F(m)$ denotes the
number of faces of $m$ and $g$ is a nonnegative real parameter smaller
than $1/12$. An explicit form for the two-point function of
quadrangulations was given in~\cite{geod}\footnote{Note that, though
  general maps and quadrangulations are in bijection, their metric
  structures hence their two- or three-point functions are \emph{a
    priori} different. But, in fact, not too much \citemy{gen2p}
  \cite{BJM14,FuGu14,Lehericy19}.}, and a few years later, together
with Emmanuel Guitter, we succeeded in computing their three-point
function in~\citemy{threepoint}. This is the story I would like to
tell here.

Our inspiration came from the paper of Grégory
Miermont~\cite{Miermont09}\footnote{The preprint was in fact posted in
  December 2007 on the arXiv. We also had the chance to learn about
  Grégory's work via the \emph{séminaire Cartes}, which had not yet
  evolved into the current \emph{journées}.} containing his
generalization of the CVS bijection, mentioned in the previous
section. It was then formulated in terms of ``sources'' and
``delays'', reflecting his motivations related to Voronoi
tessellations. Our key idea was to apply his bijection with a special
choice of sources and delays, which we now explain.

\begin{figure}
  \centering
  \fig{.5}{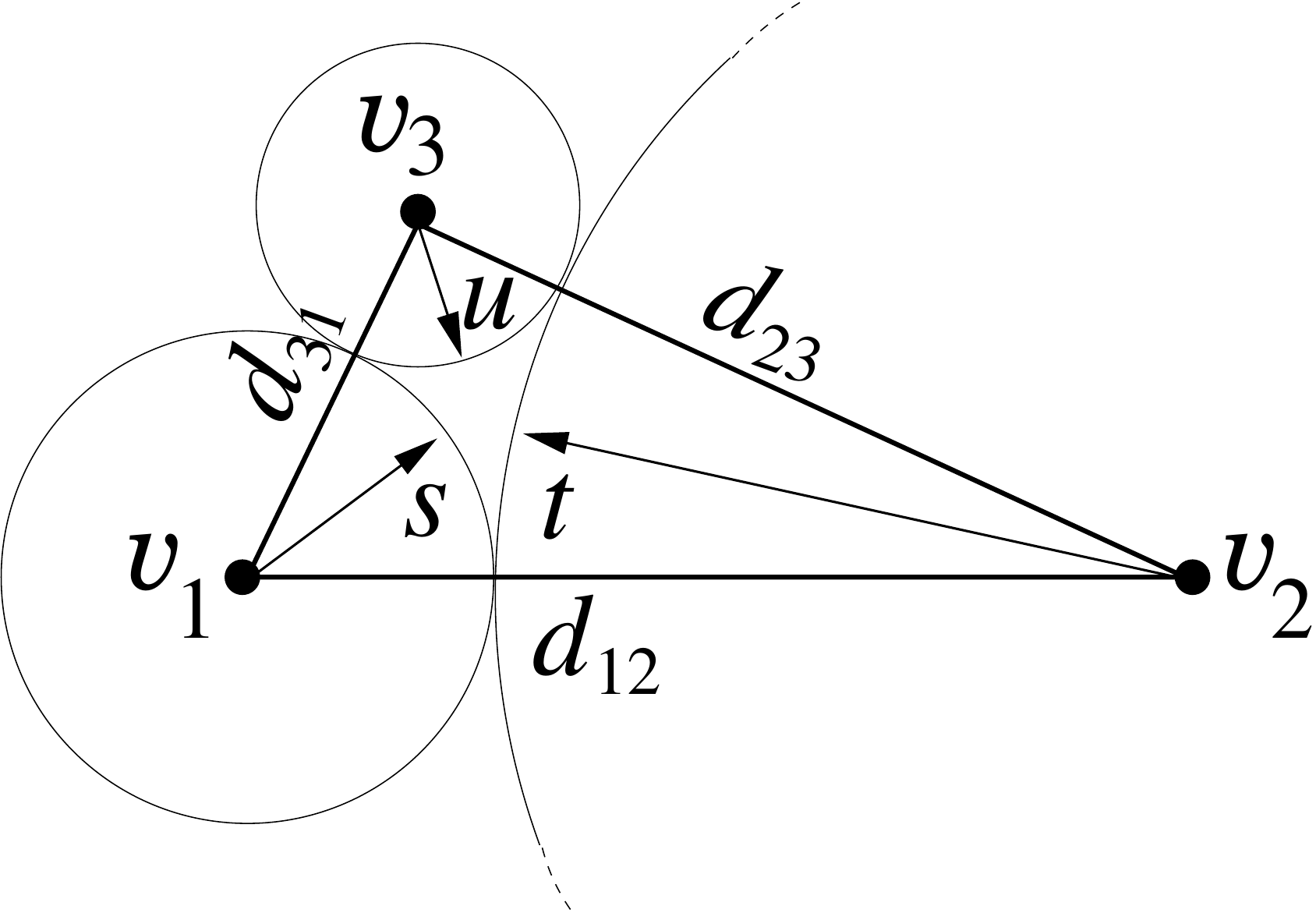}
  \caption{Geometrical interpretation of the parameters $s,t,u$ of
    \eqref{eq:distspecdel}: they correspond to the radii in the
    ``circle packing'' consisting of three circles whose centers are
    at pairwise distances $d_{12}$, $d_{23}$ and $d_{31}$.}
  \label{fig:triangle}
\end{figure}

Let us consider a quadrangulation with three marked distinct vertices
$v_1$, $v_2$ and $v_3$. As mentioned above such a quadrangulation does
not possess nontrivial symmetries. We denote by $d_{12}$ the graph
distance from $v_1$ to $v_2$ in the quadrangulation, and define
similarly $d_{23}$ and $d_{31}$. We then introduce the three
parameters $s,t,u$ such that
\begin{equation}
  \label{eq:distspecdel}
  \begin{split}
    d_{12} &= s + t \\
    d_{23} &= t + u \\
    d_{31} &= u + s \\
  \end{split}
\end{equation}
i.e.\ $s=(d_{12}-d_{23}+d_{31})/2$, etc. A geometrical interpretation
of these parameters is given on Figure~\ref{fig:triangle}. The nice
thing about $s,t,u$ is that they are ``free'', unlike
$d_{12},d_{23},d_{31}$ which are constrained by the triangular
inequalities. More precisely we simply have $s,t,u \geq 0$, and at
most one of them may vanish due to the constraint that $v_1,v_2,v_3$
are distinct. The vanishing of one parameters occurs when the points
are ``aligned'': having, say, $u=0$ simply means that $v_3$ is on a
geodesic between $v_1$ and $v_2$. Also, since we are working with
(bipartite) quadrangulations, $d_{12},d_{23},d_{31}$ are integer
numbers whose sum is even. Hence $s,t,u$ are always nonnegative
integers.

\begin{figure}
  \centering
  \fig{.8}{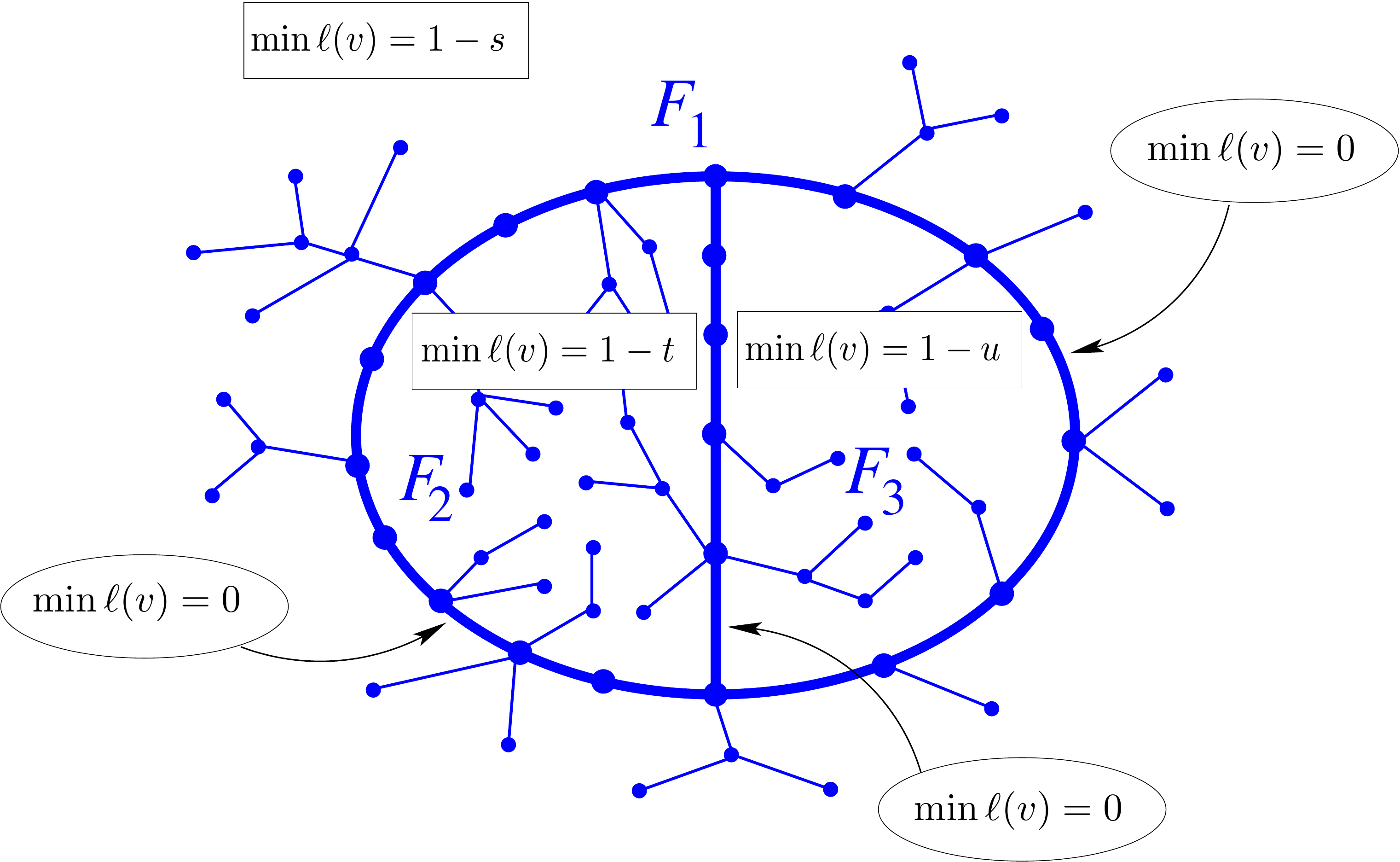}
  \caption{Structure of the well-labeled map in the case
    $s,t,u>0$: the map has three faces $F_1,F_2,F_3$, which are
    pairwise adjacent. The minimal label among vertices incident to
    $F_1$ (resp.\ $F_2,F_3$) is $1-s$ (resp.\ $1-t,1-u$). The minimal
    label among vertices incident both to $F_1$ and $F_2$ is $0$, and
    the same holds for the two other interfaces.}
  \label{fig:specialdelays}
\end{figure}

\begin{figure}
  \centering
  \fig{.8}{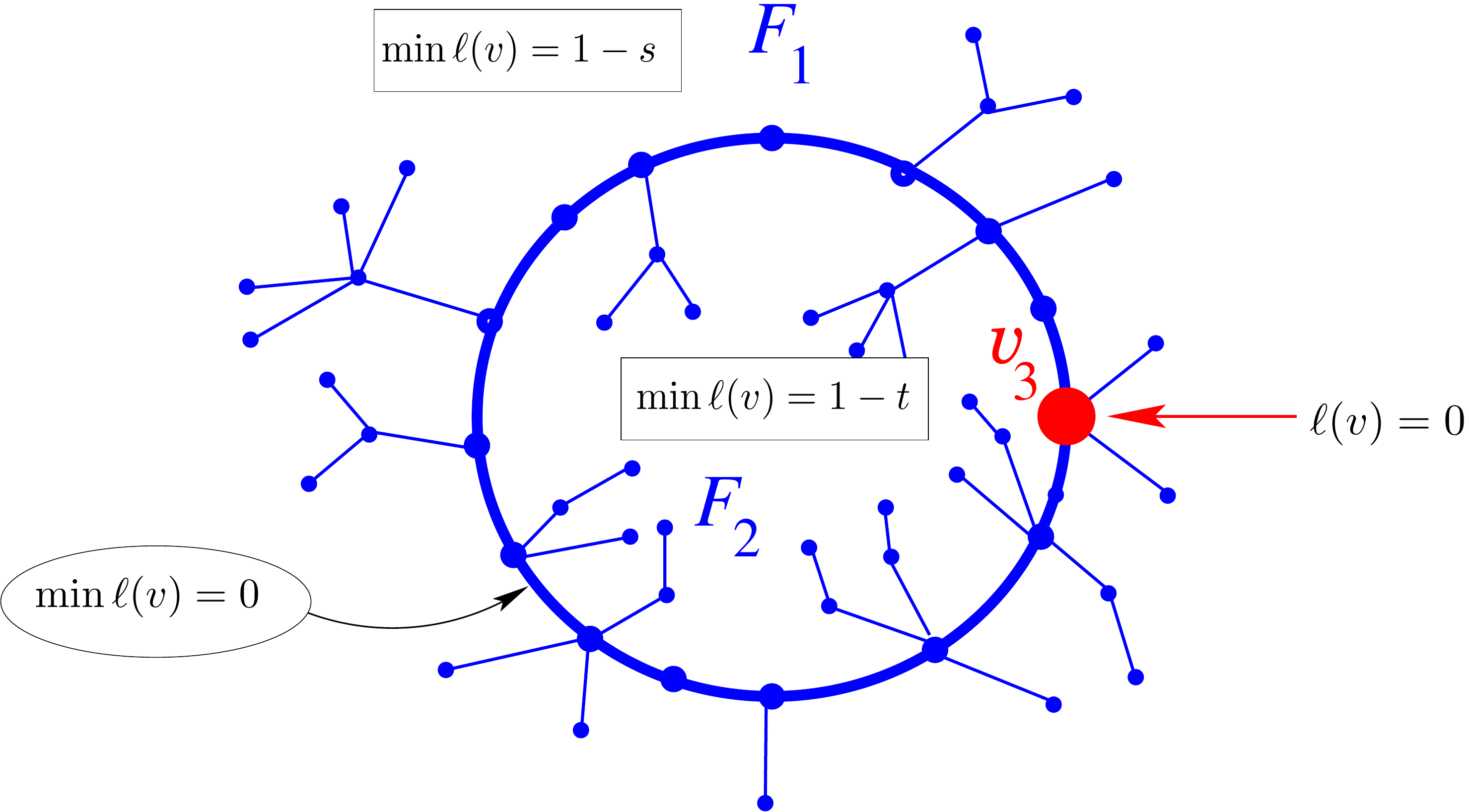}
  \caption{Structure of the well-labeled map in the case $s,t>0$ and
    $u=0$: the map has two faces $F_1,F_2$, and $v_3$ is a vertex
    incident to both. The minimal label among vertices incident to
    $F_1$ (resp.\ $F_2$) is $1-s$ (resp.\ $1-t$). The minimal label
    among vertices incident to both faces is $0$, and is attained at
    $v_3$.}
  \label{fig:spectwo}
\end{figure}

We now apply the Miermont bijection with the ``sources'' $v_1,v_2,v_3$
and the ``delays'' $-s,-t,-u$. In the language of
Section~\ref{sec:distmobiles}, this corresponds to endowing the
quadrangulation with the suitable labeling defined by
\begin{equation}
  \label{eq:dist3plabdef}
  \ell(v) = \min \{ d(v,v_1)-s, d(v,v_2)-t, d(v,v_3)-u \}
\end{equation}
where $v$ denotes an arbitrary vertex of the quadrangulation. It is
not difficult to check that $\ell$ is indeed a suitable labeling, with
generically $3$ local min at $v_1,v_2,v_3$, except in the
\emph{aligned case} where there are only $2$ (e.g., if $u=0$ then
$v_3$ is no longer a local min). By Theorem~\ref{thm:gen2p1}
(specialized to quadrangulations), we obtain a well-labeled map with
respectively $3$ or $2$ faces. Examining the labels more closely, we
find that it has the structure displayed respectively on
Figures~\ref{fig:specialdelays} or \ref{fig:spectwo}. We have
reformulated the problem of determining the three-point function of
quadrangulations into the problem of counting well-labeled maps with
these structures (with now a weight $g$ per edge). We now sketch how
to solve this counting problem.

For $d \geq 1$, let us denote by $R_d$ the generating function of
well-labeled \emph{trees} with (strictly) positive labels, rooted at a
corner labeled $d$. By shifting, trees with labels (strictly) larger
than $-s$ and rooted a corner labeled $d$ have generating function
$R_{d+s}$.  From~\cite{geod,onewall}, we know that $R_d$ is actually
the two-point function of quadrangulations, and is explicitly given by
\begin{equation}
  \label{eq:distRdquad}
  R_d = R \frac{(1-x^d)(1-x^{d+3})}{(1-x^{d+1})(1-x^{d+2})}
\end{equation}
where $R=\frac{1-\sqrt{1-12g}}{6g}$ and $x$ is root of the equation
$x+x^{-1}=\frac{1-4 gR}{4g R}$ (this equation has two reals roots for
$g \leq 1/12$, and we usually pick the root $x \leq 1$ which is
analytic at $g=0$).  

\begin{figure}
  \centering
  \fig{}{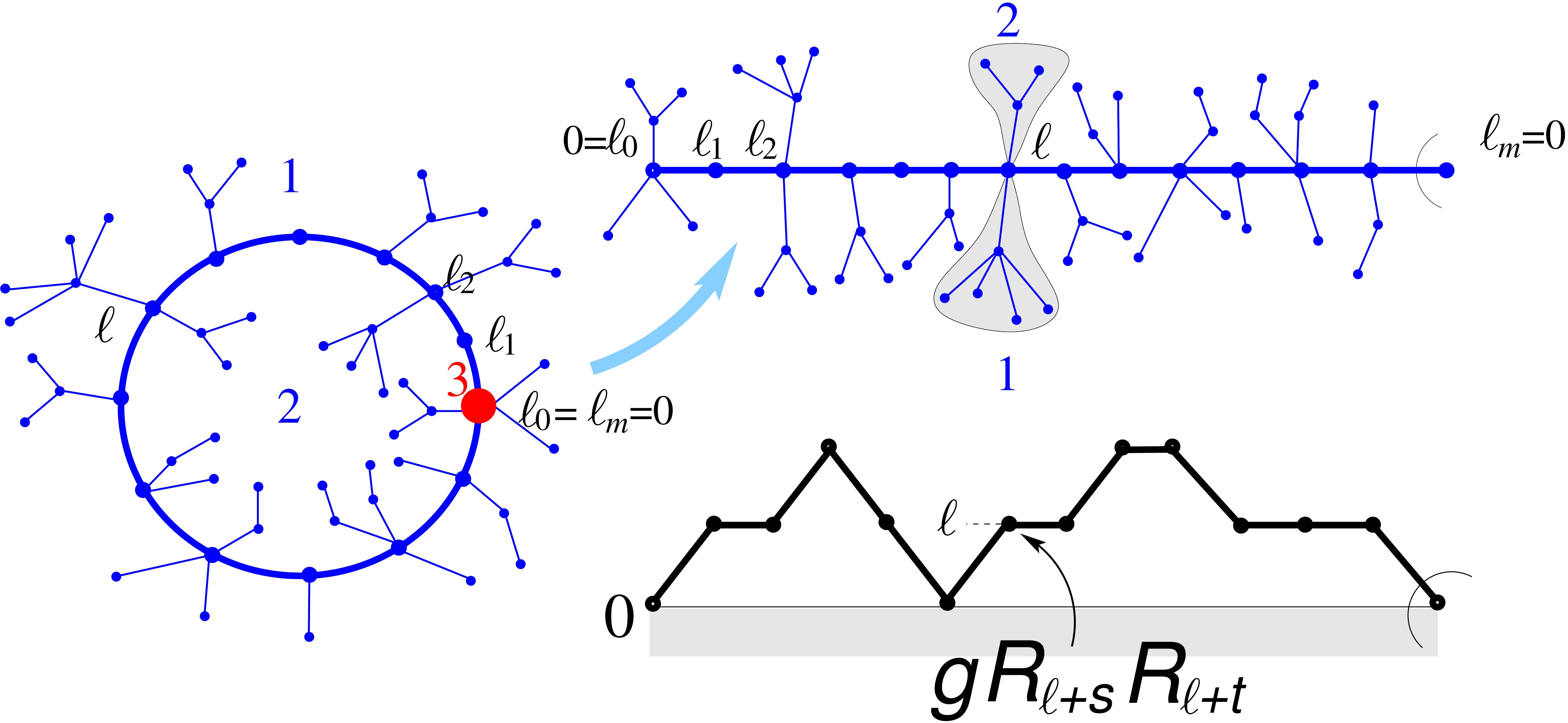}
  \caption{Cutting the well-labeled map of Figure~\ref{fig:spectwo}
    into a chain.  A chain is a well-labeled tree with two marked
    vertices with label $0$. On the branch between them, the labels
    remain nonnegative hence form a Motzkin path. Well-labeled
    subtrees are attached to both sides of the branch, and their
    labels are at least $1-s$ and $1-t$ respectively. Therefore,
    summing the weights over all possible such subtrees, the global
    contribution of a step at height $\ell$ in the Motzkin path is
    equal $g R_{\ell+s} R_{\ell+t}$. The careful reader might notice
    the issue that the minima $1-s$ and $1-t$ should be attained in
    some subtree, this is taken care of by \eqref{eq:distXstfix}.}
  \label{fig:newmotzkin}
\end{figure}

The maps of Figures~\ref{fig:specialdelays} or \ref{fig:spectwo} are
not trees, but since they have few faces it is natural to attempt to
decompose them into trees.  To remain brief, we only
detail 
the aligned case $u=0$: as displayed on Figure~\ref{fig:newmotzkin},
we cut the well-labeled map at $v_3$, to obtain a \emph{chain}, which
may be viewed as a \emph{Motzkin path} with attached subtrees. By a
standard combinatorial argument we find that the generating function
$X_{s,t}$ of such paths satisfies the recurrence equation
\begin{equation}
  \label{eq:distXsteq}
  X_{s,t} = 1 + g R_s R_t X_{s,t} + g^2 R_s R_t X_{s,t} R_{s+1} R_{t+1} X_{s+1,t+1}
\end{equation}
(which uniquely determines $X_{s,t}$ as a power series in
$g$). Remarkably, $X_{s,t}$ admits an explicit expression similar to
that of $R_d$, namely
\begin{equation}
  \label{eq:distXstexp}
  X_{s,t} =
  \frac{(1-x^3)(1-x^{s+1})(1-x^{t+1})(1-x^{s+t+3})}{
    (1-x)(1-x^{s+3})(1-x^{t+3})(1-x^{s+t+1})}.
\end{equation}
It is straightforward to check that this is indeed the solution of
\eqref{eq:distXsteq}, alternatively we give two bijective derivations
in~\citemy{threepoint} (which however take the expression
\eqref{eq:distRdquad} for $R_d$ as an input). As mentioned in the
caption of Figure~\ref{fig:newmotzkin}, $X_{s,t}$ is not quite the
generating function of the maps of Figure~\ref{fig:spectwo} because
the minimal label $1-s$ (resp.\ $1-t$) should be attained somewhere
along $F_1$ (resp.\ $F_2$). But this is easily fixed: the actual
generating function for them is
\begin{equation}
  \label{eq:distXstfix}
  \Delta_s \Delta_t X_{s,t} = X_{s,t} - X_{s-1,t} - X_{s,t-1} + X_{s-1,t-1}.
\end{equation}
Here $\Delta_s$ denotes the discrete difference operator
$\Delta_s f(s)=f(s)-f(s-1)$.

In the generic case $s,t,u>0$ the decomposition is more involved,
see~\citemy[Section~4.3]{threepoint}. Some pieces in the decomposition
are the same chains as before, but we also need to introduce so-called
\emph{$Y$-diagrams} whose generating function $Y_{s,t,u}$ depends on
the three parameters, and is explicitly given by
\begin{equation}
  \label{eq:distYstu}
  Y_{s,t,u} =
  \frac{(1-x^{s+3})(1-x^{t+3})(1-x^{u+3})(1-x^{s+t+u+3})}{
    (1-x^3)(1-x^{s+t+3})(1-x^{t+u+3})(1-x^{u+s+3})}.
\end{equation}
This expression can be proved by checking that it solves a recurrence
equation similar to \eqref{eq:distXsteq}, but unlike $X_{s,t}$ we do
not know an alternative bijective derivation. Armed with these
notations, we may now state the main result of this section:
\begin{thm}[see \citemy{threepoint}]
  The three-point function of planar quadrangulations, which counts
  such maps with three marked distinct vertices at prescribed
  distances parametrized by $s,t,u$ as in~\eqref{eq:distspecdel}, is
  equal to $\Delta_s \Delta_t \Delta_u F_{s,t,u}$ where
  \begin{equation}
    \label{eq:distFstu}
    F_{s,t,u} := X_{s,t} X_{t,u} X_{u,s} (Y_{s,t,u})^2.
  \end{equation}
\end{thm}
The role of the discrete difference operators is again to make sure
that the minimal labels $1-s,1-t,1-u$ are attained somewhere along
their respective faces in Figure~\ref{fig:specialdelays}, $F_{s,t,u}$
corresponds to a cumulative generating function.

\begin{figure}
  \centering \fig{}{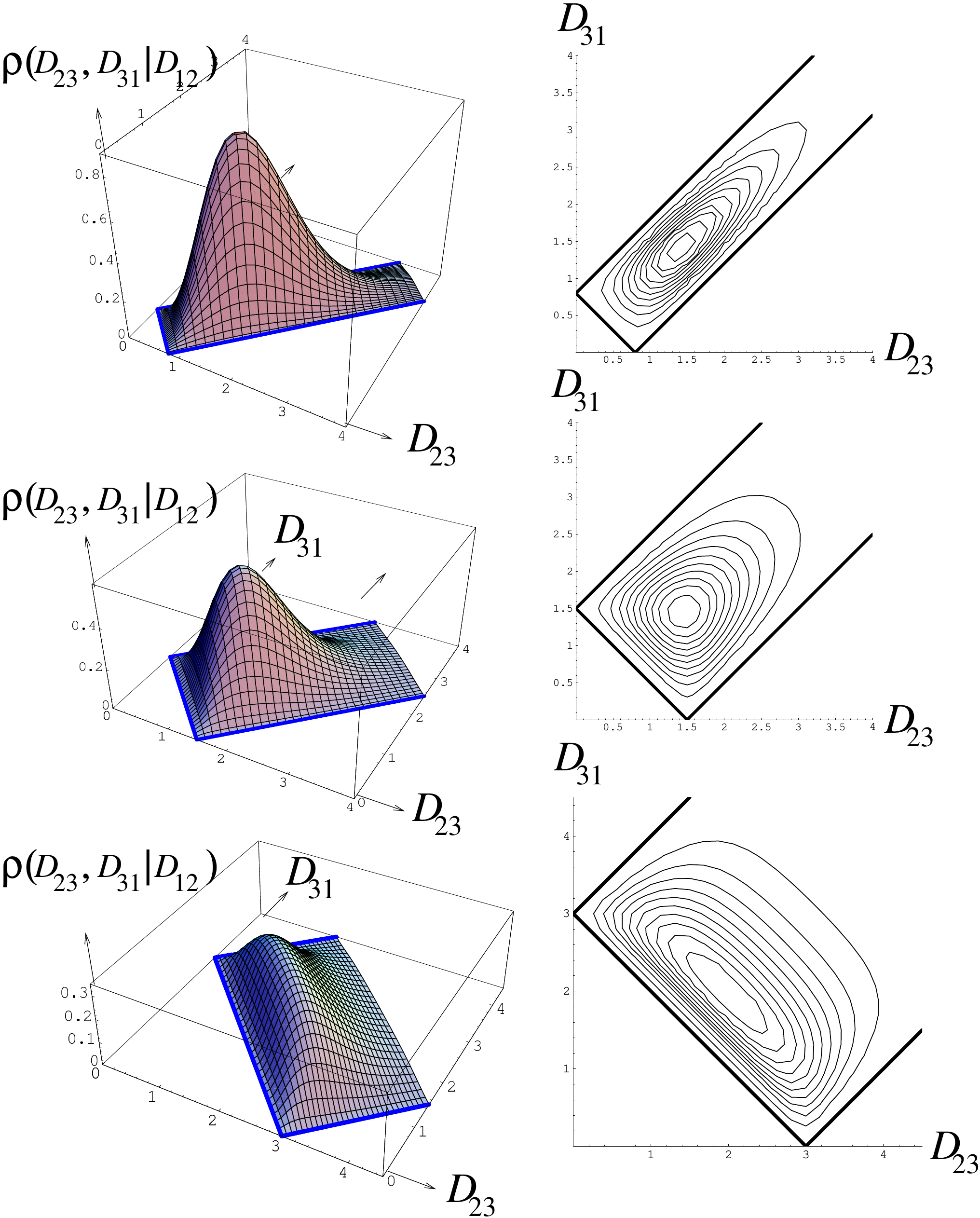}
  \caption{Plots of the probability density of $(D_{23},D_{31})$
    conditionally on $D_{12}$, for $D_{12}=0.8$, $1.5$ and $3$ (top to
    bottom). Here $D_{12}=S+T$, $D_{23}=T+U$ and $D_{31}=U+S$ denote
    the distances between three uniform points in a random
    quadrangulation with a large number $n$ of faces, rescaled by a
    factor $n^{-1/4}$.}
  \label{fig:plots}
\end{figure}

It is natural to consider the ``continuum limit'' of the three-point
function, which corresponds to a certain observable of the Brownian
map (precisely, it is the joint law of the distances between three
uniform points). The first step consists in letting $g$ tend to the
critical value $1/12$ and rescaling jointly the parameters $s,t,u$ via
\begin{equation}
  \label{eq:gstuscal}
  g = \frac{1 - \Lambda \epsilon}{12}, \qquad
  s = \lfloor S \epsilon^{-1/4} \rfloor, \quad
  t = \lfloor T \epsilon^{-1/4} \rfloor, \quad
  u = \lfloor U \epsilon^{-1/4} \rfloor
\end{equation}
where $\epsilon \to 0$ and $\Lambda,S,T,U$ are kept fixed ($\Lambda$
plays the role of a ``cosmological constant''). Then, under this
scaling, it is straightforward to check that
\begin{equation}
  F_{s,t,u} \sim \epsilon^{-1/2} \mathcal{F}(S,T,U;\sqrt{3/2} \Lambda^{1/4})
\end{equation}
where
\begin{equation}
  \mathcal{F}(S,T,U;\alpha) := \frac{3}{\alpha^2}
  \left( \frac{\sinh(\alpha S) \sinh(\alpha T) \sinh(\alpha U) \sinh(\alpha (S+T+U))}{\sinh(\alpha (S+T)) \sinh(\alpha (T+U)) \sinh(\alpha (U+S))} \right)^2.
\end{equation}
The reader might notice that the nontrivial dependency in $S,T,U$
comes from the factor $(Y_{s,t,u})^2$ in $F_{s,t,u}$, while the $X$
factors contribute a trivial factor $3$ at leading order. One may
interpret this as the fact that the $Y$-diagrams capture all the
``mass'' in the scaling limit, while the $X$-chains are negligible. It
may be seen~\citemy{threepoint,loops} that the $Y$-diagrams correspond
in the Brownian map to the two ``geodesic triangles'', namely the two
regions delimited by the three geodesics between the marked
points\footnote{These two regions are well-defined almost surely by
  the sphericity of the Brownian map~\cite{LGPa08,Miermont08} and the
  uniqueness of the geodesic between almost every pair of
  points~\cite{Miermont09}.}. $\mathcal{F}$ corresponds to the ``grand
canonical'' continuum three-point function, for probabilistic
interpretation it is preferable to return to the ``canonical''
ensemble, i.e.\ to consider the uniform distribution over
quadrangulations with a large but fixed number $n$ of faces.  This is
done by extracting the coefficient of $g^n$ in the series $F_{s,t,u}$,
which may be estimated for large $n$ by a saddle-point analysis, and
normalizing it by the total number $\frac{3^n}{2} \binom{2n}{n}$ of
quadrangulations with three marked distinct vertices. The end
result~\citemy{threepoint} is that, upon replacing $\epsilon$ by $1/n$
in the scaling \eqref{eq:gstuscal} for $s,t,u$, we have
\begin{equation}
  \label{eq:dist3pcont}
  \lim_{n \to \infty} \frac{[g^n]F_{s,t,u}}{\frac{3^n}{2} \binom{2n}{n}} = \frac{2}{i \sqrt{\pi}} \int_{-\infty}^\infty d\xi \xi e^{-\xi^2} \mathcal{F}(S,T,U;e^{-\mathrm{sgn}(\xi)i \pi/4} \sqrt{3\xi/2}).
\end{equation}
This integral is real and easy to evaluate numerically, see
Figure~\ref{fig:plots} for some plots.

To conclude this section, let us briefly my other results in the same
vein, all obtained jointly with Emmanuel Guitter. Before the
three-point function, we had considered in~\citemy{statgeod} the
statistics of the number of geodesics between two points in a
quadrangulation. These can be viewed as discrete enumerative
counterparts of the results of Miermont~\cite{Miermont09} and Le
Gall~\cite{LeGall10} about geodesics in the Brownian
map. In~\citemy{loops}, we refined the computation of the three-point
function by taking into effect the phenomenon of ``confluence'' of
geodesics, evidenced by Le Gall~\cite{LeGall10}. The full three-point
function depends on $6$ distance parameters, corresponding to the
lengths of the common and proper parts of the three geodesics between
the three marked points. We also obtain similar results for minimal
separating loops. In~\citemy{pseudoquad}, we considered
quadrangulations with a boundary, and studied variants of the
two-point function where either one or both marked points are on the
boundary. Our work was an early attempt at considering the rich
possible scaling limits of quadrangulations with a boundary, now
studied in much greater detail in~\cite{BeMi17,BMR16}. Also, we first
noticed there~\citemy[Equation~(7.1)]{pseudoquad} the connection with
continued fractions, on which we will return in the next
section. Finally, we considered in~\citemy{quadwithnoME} the two-point
function of quadrangulations without multiple edges, which we related
to the physical concept of ``minimal neck baby universes''
(minbus). On the combinatorial side, we understood how multiple edges
can be eliminated at the level of the CVS bijection, via the notion of
``well-balanced tree'', and of the two-point function, via a
substitution scheme. More involved substitution schemes will be
discussed in Chapter~\ref{chap:slices}.

\section{From the two-point function of Boltzmann maps to continued
  fractions}
\label{sec:distCF}

In this section we consider the family of (face-weighted)
\emph{Boltzmann maps}, obtained by endowing the set of all planar maps
with the weight function
\begin{equation}
  \label{eq:distboltz}
  w(m) = \prod_{f \text{ face of } m} g_{\mathrm{degree}(f)}
\end{equation}
where $(g_k)_{k \geq 1}$ is a given sequence of nonnegative real
numbers. This setting encompasses quadrangulations (take $g_k=g$ if
$k=4$ and $0$ otherwise), edge-weighted general maps (take
$g_k=t^{k/2}$ for all $k$), triangulations, etc. For simplicity, we
restrict here to the bipartite case, i.e.\ we assume that $g_k=0$ if
$k$ is odd (see Remark~\ref{rem:distCFnonbip} below for a brief
discussion of the general case). Also, some of our formulas hold only
in the case of \emph{bounded degrees}, i.e.\ when the set of $k$'s
such that $g_k \neq 0$ has a maximal element, which we denote by $M$.

The two-point function of Boltzmann bipartite maps with bounded
degrees was considered in~\cite{geod}, where we \emph{guessed} its
general formula by inferring it from the small values of $M$ (e.g.\
$4$ or $6$) which we could solve explicitly. Our original approach was
based on a certain ``integrable'' recurrence equation obeyed by the
two-point function. Here I will present a rigorous proof of our
formula, which was obtained several years later with Emmanuel Guitter
in~\citemy{hankel}. Our proof is based on the beautiful combinatorial
theory of continued fractions, developed notably by
Flajolet~\cite{Flajolet80} and Viennot~\cite{Viennot84}.

Let us first set up some conventions and notations. Here, the
two-point function $R_d$ is precisely defined as the generating
function of pointed rooted bipartite maps where both endpoints of the
root edge are at distance at most $d$ from the origin (since the map
is bipartite, one of the endpoints is strictly further from the origin
than the other, and we assume that the root edge is oriented towards
it). We conventionally set the constant term of $R_d$ to $1$, which
makes the forthcoming expressions nicer. By Theorem~\ref{thm:gen2p1}
specialized to the geodesic labeling, we find that $R_d$ is equal to
the generating function of (one-face) mobiles whose labels are all
positive, and which are rooted on a corner at a vertex labeled
$d$. The Boltzmann weight \eqref{eq:distboltz} simply amounts to
weighting unlabeled vertices in the mobile (an unlabeled vertex of
degree $k$ has weight $g_{2k}$). For $d=\infty$, the two-point
function reduces to the generating function $R$ of pointed rooted maps
(or mobiles without the positivity constraint), which is the smallest
positive, possibly infinite, root of the equation
\begin{equation}
  \label{eq:distCFReq}
  R = 1 + \sum_{k=1}^\infty \binom{2k-1}{k} g_{2k} R^k.
\end{equation}
From now on we assume that the weights $(g_{2k})_{k \geq 0}$ are such
that $R<\infty$.

\begin{thm}[see \cite{geod} and \citemy{hankel}]
  \label{thm:distCF}
  The two-point function of Boltzmann bipartite maps with degrees
  bounded by $M=2p+2$ is given by
  \begin{equation}
    \label{eq:distRdform}
    R_d = R \frac{u_d u_{d+3}}{u_{d+1} u_{d+2}}
  \end{equation}
  where
  \begin{equation}
    \label{eq:distudet}
    u_d = \det_{1 \leq i,j \leq p} \left(y_i^{d/2+j-1} - y_i^{-d/2-j+1}\right),
  \end{equation}
  and $y_1,y_1^{-1},\ldots,y_p,y_p^{-1}$ are the roots of the
  \emph{characteristic equation}
  \begin{equation}
    \label{eq:distchareq}
    1 = \sum_{k=1}^{p+1} g_{2k} R^{k-1} \sum_{m=0}^{k-1} \binom{2k-1}{k-m-1} \sum_{j=-m}^m y^j.
  \end{equation}
\end{thm}

In the case of quadrangulations ($p=1$), we recover the
expression~\eqref{eq:distRdquad}. An interesting feature of our
general expression is that the size of the determinant
in~\eqref{eq:distudet} does not depend on the distance parameter $d$,
which is useful for studying its asymptotic behavior.

In a nutshell, the key observation for proving
Theorem~\ref{thm:distCF} is that the sequence $(R_d)_{d \geq 1}$
arises as the continued fraction expansion of the (well-known)
generating function $W_0$ of maps with a boundary. From the general
theory of continued fractions, it follows that $R_d$ admits an
expression of the form~\eqref{eq:distRdform} with $u_d$ a Hankel
determinant whose entries are given by the series expansion of $W_0$.
We finally use the specific form of these entries to identify the
Hankel determinant as a symplectic or odd orthogonal Schur function,
which yields~\eqref{eq:distudet} using the Weyl character formula.

\begin{figure}
  \centering
  \fig{}{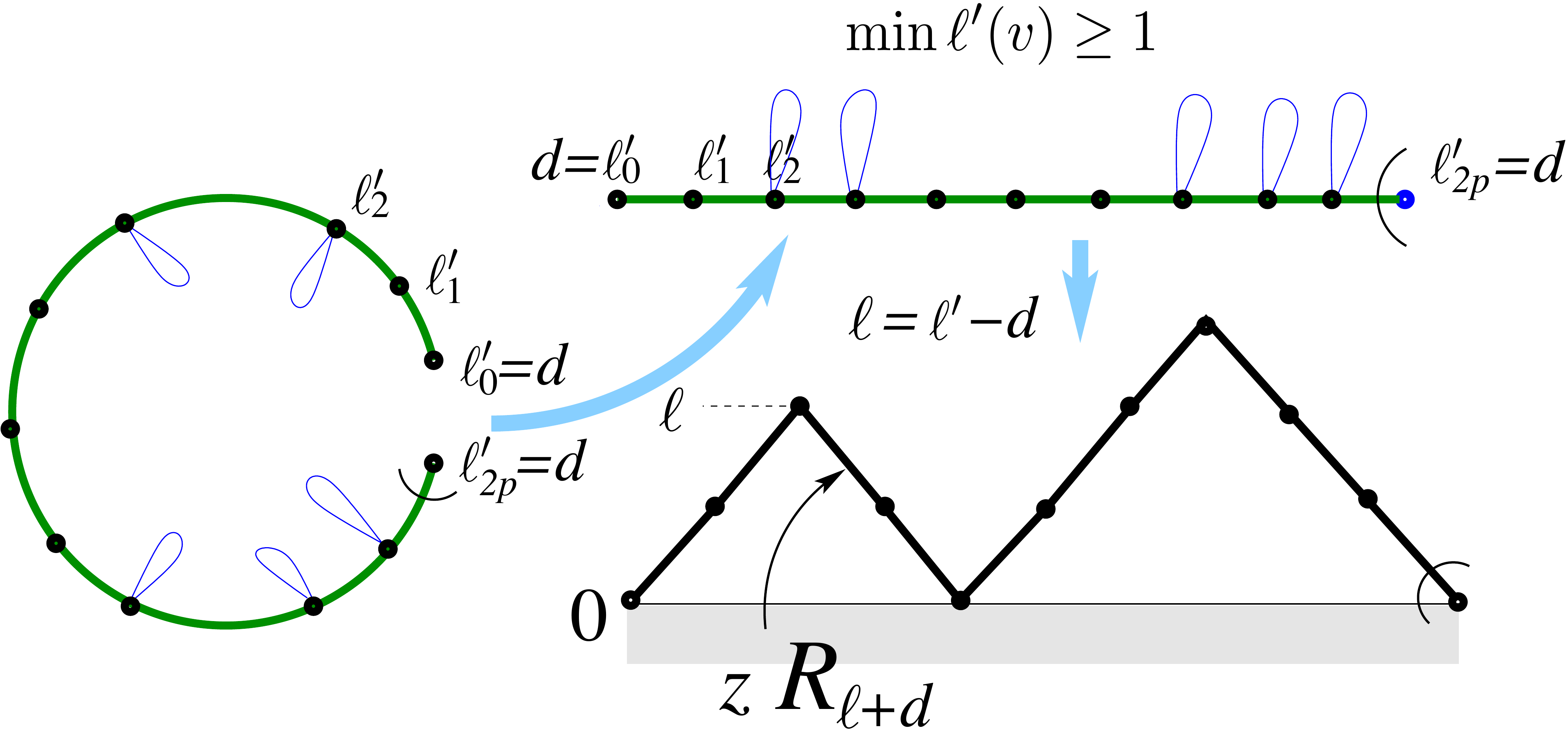}
  \caption{The decomposition of the mobile associated with a map with
    a boundary counted by $W_d$. The sequence of labels around the
    special unlabeled vertex (coming from the root face) is coded by a
    Dyck path, whose down-steps correspond to the subtrees of the
    mobile. The global contribution of all possible subtrees for a
    down-step from height $\ell$ is $z R_{\ell+d}$. Notice the
    similarity with Figure~\ref{fig:newmotzkin}.}
  \label{fig:dyck}
\end{figure}

Let us now provide a few more details. As mentioned above, the
connection with continued fractions was first observed
in~\citemy{pseudoquad}, as a byproduct of the study of distances in
maps with a boundary\footnote{The observation was initially made in
  the case of quadrangulations, but it immediately extends to
  Boltzmann maps.}.  More precisely, let us consider pointed rooted
maps and regard the root face as a boundary. Since we wish to control
the boundary length (degree) $2n$ separately, we exclude by convention
the root face from the product \eqref{eq:distboltz}, and rather assign
it a weight $z^n$ with $z$ an extra parameter\footnote{This is
  common in the context of Tutte equations, where $z$ appears as the
  ``catalytic variable''~\cite{BoJe06}.}. We denote by $W_d$ the
generating function of such weighted maps in which the distance from
the origin to the root vertex is at most $d$, and no vertex incident
to the boundary lies closer to the origin than the root vertex. In
particular, for $d=0$, the origin coincides with the root vertex,
hence $W_0$ can be regarded as the generating function of maps with
a boundary. By Theorem~\ref{thm:gen2p1}, $W_d$ is also the generating
function of certain (one-face) mobiles with positive labels.  Using a
decomposition of these mobiles illustrated on Figure~\ref{fig:dyck}
---which is in the same spirit as the decomposition used to arrive at
\eqref{eq:distXsteq}--- we find that $W_d$ counts certain weighted
Dyck paths and, as such, satisfies the recurrence relation
\begin{equation}
  W_d = 1 + z R_{d+1} W_d W_{d+1}.
\end{equation}
By iterating this recurrence, we find that $W_d$ admits the
Stieljes-type continued fraction expansion
\begin{equation}
  \label{eq:distWdfrac}
  W_d = \cfrac{1}{1-\cfrac{z R_{d+1}}{1-\cfrac{z R_{d+2}}{1-\ddots}}}
\end{equation}
and in particular the \emph{whole} sequence $(R_d)_{d \geq 1}$ arises
as the \emph{continued fraction expansion} of $W_0$ with respect to
the variable $z$.

It now follows from the general theory of continued fractions that
$R_d$ can be expressed in terms of the Hankel determinants formed with
the coefficients of the \emph{power series expansion} of $W_0$ with
respect to the variable $z$. More precisely, let us denote by $w_n$
the coefficient of $z^n$ in $W_0$, which counts rooted maps with a boundary
of length $2n$ (we have $w_0=1$, corresponding to the map reduced to a
single vertex). Then, for $n \geq 0$, we form the \emph{Hankel
  determinants}
\begin{equation}
  \label{eq:distCFhn}
  h_n^{(0)}=\det_{0 \leq i,j \leq n} w_{i+j}, \qquad
  h_n^{(1)}=\det_{0 \leq i,j \leq n} w_{i+j+1}
\end{equation}
and we have the relations
\begin{equation}
  \label{eq:distCFRdhn}
  R_{2n+1} = \frac{h_n^{(1)}/h_{n-1}^{(1)}}{h_n^{(0)}/h_{n-1}^{(0)}}, \qquad R_{2n+2} = \frac{h_{n+1}^{(0)}/h_n^{(0)}}{h_n^{(1)}/h_{n-1}^{(1)}}
\end{equation}
where by convention $h_{-1}^{(0)}=h_{-1}^{(1)}=1$.  See for
instance~\cite{Viennot84} for a nice combinatorial proof of these
formulas using nonintersecting lattice paths.

To proceed further we have to use the specific form of $w_n$, which is
essentially known in the bipartite case since
Tutte~\cite{Tutte62}\footnote{We pointed out in~\citemy{hankel} that
  the generating function $W_0 = \sum_{n \geq 0} w_n z^n$ counting
  maps with a boundary is arguably the most important object in the
  enumerative theory of planar maps: it plays a fundamental role in
  Tutte equations, matrix integrals, etc. In some talks,
  Emmanuel had the idea of
  picturing it as an iceberg, whose hidden part was the continued
  fraction expansion.}. It reads explicitly
\begin{equation}
  \label{eq:distCFwnexp}
  w_n = R^n \sum_{q \geq 0} a_q \Cat(n+q)
\end{equation}
where $R$ is the generating function of pointed rooted maps considered
before, $\Cat(n):=\frac{(2n)!}{n!(n+1)!}$ denotes the $n$-th Catalan
number, and
\begin{equation}
  \label{eq:distCFaq}
  a_q := \delta_{q,0} - \sum_{k \geq q+1} \binom{2k-2q-2}{k-q-1} g_{2k} R^k.
\end{equation}
See~\citemy[Section~3]{hankel} for three different proofs
of~\eqref{eq:distCFwnexp} in the more general nonbipartite
setting\footnote{A relatively short derivation
  of~\eqref{eq:distCFwnexp} is by
  ``depointing''~\citemy[Section~3.1]{hankel}: let $R(v)$ denote the
  generating function of pointed rooted maps with an extra weight $v$
  per vertex. Then using mobiles one may see that (i) $R(v)$ satisfies
  the equation obtained from~\eqref{eq:distCFReq} by replacing the
  term $1$ in the right-hand side by $v$ and (ii) we have
  $w_n = \binom{2n}{n} \int_0^1 R(v)^n dv$. From this we
  get~\eqref{eq:distCFwnexp} through the change of variable
  $v \to R$.}. We now form the Hankel
determinants~\eqref{eq:distCFhn}: first we dispose of the prefactor
$R^n$ in $w_n$, by noting that it just yields a prefactor $R$ in the
ratios~\eqref{eq:distCFRdhn}. Therefore we find that $R_d$ has the
form \eqref{eq:distRdform}, with $u_d$ given by the reduced Hankel
determinants
\begin{equation}
  \label{eq:distCFhankred}
  u_{2d+1} = \det_{0 \leq i,j \leq d-1} \tilde{w}_{i+j}, \qquad
  u_{2d+2} = \det_{0 \leq i,j \leq d-1} \tilde{w}_{i+j+1}
\end{equation}
with $\tilde{w}_n=\sum_{q=0}^\infty a_q \Cat(n+q)$. This very specific
form can be exploited to rewrite the determinants in
``Toeplitz-Hankel'' form: for $\ell,i,j$ nonnegative integers, let
$\paths{\ell}{i}{j}:=\binom{\ell}{\frac{\ell+i-j}{2}}-\binom{\ell}{\frac{\ell+i+j+2}{2}}$
denote the number of sequences $(i=x_0,x_1,\ldots,x_{\ell-1},x_\ell=j)$ such that
$|x_k-x_{k-1}|=1$ and $x_k \geq 0$ for all $k$ (we have
$\Cat(n)=p_{2n,0,0}$), then it is clear that
\begin{equation}
  \begin{split}
    \Cat(i+j+q)&=\sum_{h,h'\geq 0} \paths{2i}{0}{2h} \paths{2q}{2h}{2h'} \paths{2j}{2h'}{0}, \\
    \Cat(i+j+q+1)&=\sum_{h,h'\geq 0} \paths{2i+1}{0}{2h+1} \paths{2q}{2h+1}{2h'+1} \paths{2j+1}{2h'+1}{0}.
  \end{split}
\end{equation}
This allows to factor out unitriangular matrices from the Hankel
matrices, and we obtain the expressions
\begin{equation}
  \label{eq:distCFTH}
  u_{2d+1} = \det_{0 \leq i,j \leq d-1} (b_{i-j}-b_{i+j+1}), \qquad
  u_{2d+2} = \det_{0 \leq i,j \leq d-1} (b_{i-j}-b_{i+j+2})
\end{equation}
where $b_n:=\sum_{q \geq 0} a_q \binom{2q}{q+n}$. Note that $b_{-n}=b_n$.

So far we have not used the assumptions of bounded degrees. When this
is the case, taking the maximal degree $M=2p+2$ as in
Theorem~\ref{thm:distCF}, then it is clear from \eqref{eq:distCFaq}
that $a_q=0$ for $q>p$, hence $b_n=0$ for $|n|>p$. Hence, the matrices
in \eqref{eq:distCFTH} are banded. Let us consider the characteristic
equation
\begin{equation}
  \label{eq:distchareqb}
  \sum_{n=-p}^p b_n y^n = 0
\end{equation}
which may be rewritten in the form~\eqref{eq:distchareq} by simple
manipulations: it has $2p$ roots $y_1,y_1^{-1},\ldots,y_p,y_p^{-1}$,
and the $b_n$'s, hence the $u_d$'s, are symmetric functions of
them. It is then possible to deduce the form of $u_d$ announced in
Theorem~\ref{thm:distCF} by some algebraic considerations: essentially
this boils down to showing that the expressions \eqref{eq:distCFTH}
(viewed as polynomials in the $y_i$'s) and \eqref{eq:distudet} have
the same zeros.
But, as pointed out to us by Christian Krattenthaler, our determinants
may actually be identified with \emph{odd-orthogonal} and
\emph{symplectic Schur functions} of rectangular partitions, namely
\begin{equation}
  \label{eq:distCFuSchur}
  u_{2d+1} \propto \mathrm{o}_{2p+1}((n+1)^d;y), \qquad
  u_{2d+2} \propto \mathrm{sp}_{2p}((n+1)^d;y),
\end{equation}
see~\citemy[Section~5]{hankel} and references therein for more
details. The expressions~\eqref{eq:distCFTH} correspond to
Jacobi-Trudi-type formulas, which express the Schur functions in terms
of the elementary symmetric polynomials, while~\eqref{eq:distudet}
corresponds to the Weyl character formula (where the two parities can
be put in a common form). Note that there are normalizing constants
in~\eqref{eq:distCFuSchur}, but they cancel in the
ratios~\eqref{eq:distCFRdhn}. This completes the proof of
Theorem~\ref{thm:distCF}.

\begin{rem}
  \label{rem:distCFnonbip}
  In~\citemy{hankel} we actually treated the more general case of
  Boltzmann maps which are not necessarily bipartite. The main
  complication is that there are now \emph{two} two-point functions to
  consider: $R_d$ defined as before, and another $S_d$ which
  essentially counts pointed rooted maps where the root edge is
  frustrated in the sense of Remark~\ref{rem:frust}. The continued
  fractions that we encounter are of \emph{Jacobi-type} instead of
  Stieljes-type, the combinatorial reason~\cite{Flajolet80} being that
  the decomposition illustrated on Figure~\ref{fig:dyck} now involves
  Motzkin paths instead of Dyck paths. We end up with an expression of
  the form $R_d=R \frac{u_d u_{d+2}}{u_{d+1}^2}$ where $u_d$ is still
  a rectangular symplectic Schur function, but with a larger number of
  variables than in the bipartite case ($M-2$ rather than
  $(M-2)/2$). $S_d$ may be expressed in terms of a
  ``nearly-rectangular'' symplectic Schur in the same variables. See
  \citemy[Section~2]{hankel} for more details. In the bipartite case
  there are ``miraculous'' factorizations\citemy[Section~5]{hankel}
  which lead to the expressions presented here.
\end{rem}

\begin{rem}
  In the case of quadrangulations, we may provide a combinatorial
  interpretation of the simplest formula~\eqref{eq:distRdquad} via 1D
  dimers. There is an analogous formula for triangulations.
  See~\citemy[Section~6]{hankel} for more details.
\end{rem}

\section{Conclusion and perspectives}
\label{sec:distconc}

To conclude this chapter, let me first make a personal comment: I feel
that the results presented or evoked here form the actual completion
of the line of research initiated in~\cite{geod} during my doctoral
thesis. With Emmanuel we have obtained exact discrete expressions for
several quantities related to the distance, studied their continuum
limit, and finally obtained an explanation for the existence of such
formulas through the combinatorial theory of continued
fractions. Still, our approach left several questions which, to the
best of my knowledge, remain open as of today.

First, it is natural to ask for an expression of the three-point
function of Boltzmann maps, beyond the case of quadrangulations
treated in this chapter and the case of general maps treated
in~\cite{FuGu14}. It seems that, at least in the bipartite case, we
may still employ the suitable labeling \eqref{eq:dist3plabdef} and
reduce the problem to the enumeration of certain mobiles with three
faces. But we do not know how manageable their general structure will
be. It seems that the notion of ``chains'', and the expression for
their generating function $X_{s,t}$, extend naturally to the Boltzmann
setting. But this is less clear for $Y$-diagrams. A different but
related question is whether we may directly obtain the scaling
limit~\eqref{eq:dist3pcont} of the three-point function by a
continuous approach, directly at the level of the Brownian map or why
not Liouville quantum gravity/Quantum Loewner
Evolution~\cite{MiSh16QLE}. Let me mention that, for the two-point
function, such calculation was essentially done by
Delmas~\cite{Delmas03}.

Then, we may of course consider the $n$-point function for $n \geq
4$. As mentioned above, it is already unclear what a natural
parametrization of the $\binom{n}{2}$ pairwise distances,
generalizing~\eqref{eq:distspecdel}, would be. We might still renounce
to keep track of all distances, but just only consider a subset of
them, for instance between the nearest neighbours. This seems similar
in spirit to the original idea of Miermont~\cite{Miermont09} of
studying Voronoi-like tessellations on quadrangulations. Let me
mention the fascinating recent conjecture of Chapuy~\cite{Cha17}
concerning the masses of the cells in Voronoi tessellations of
Brownian surfaces. The simplest case of this conjecture was proved by
Guitter~\cite{Guitter17} using techniques similar to those presented
in this chapter. I cannot help noting that, as apparent on
Figure~\ref{fig:triangle}, our trick for the three-point function
involves a circle packing, and we might wonder whether more involved
circle packings might play a role for more points.

Finally, several points remain open in the continued fraction approach
of Section~\ref{sec:distCF}. While we have seen that the two-point
function appears naturally in this approach, we might ask for a
continued fraction interpretation of the three-point function, or of
the intermediate quantities $X_{s,t}$ and $Y_{s,t,u}$ encountered in
its computation. On the other hand, the generating function $W_d$
of~\eqref{eq:distWdfrac} corresponds to ``truncations'' of the
continued fractions, but we do not know its general expression beyond
the case of quadrangulations~\citemy{pseudoquad}. More generally,
there is a deep connection between continued fractions and orthogonal
polynomials, but the implications of this connection for maps have not
been much investigated. Let me mention that the orthogonal polynomials
arising from our approach are \emph{different} from those usually
considered in matrix models~\citemy[Section~7.2]{hankel}. Another
question concerns the extension of our approach to Eulerian maps, for
which it is natural to generalize the Boltzmann
weight~\eqref{eq:distboltz} by weighting differently the two types of
faces. Rather than continued fractions one should then consider
``multicontinued fractions''. Together with Marie Albenque, we made a
preliminary study of such fractions~\citemy{constfpsac}, but much
remains to be understood. Let me also mention a work in progress by
Bertrand Eynard and Emmanuel Guitter, which should answer some of
these questions via the framework of classical
integrability\cite{EynardTalk}. Finally, it would be interesting to
analyze the two-point function of Boltzmann maps without the
assumption of bounded degrees, in order to deduce the continuum
two-point function of stable maps~\cite{LGMi11}. With Emmanuel Guitter
and Grégory Miermont we made a first, but so far unsuccessful, attempt
by trying to guess a (pseudo-)differential equation for the continuum
one.  Another possible way would be to analyze the asymptotics of the
Hankel determinants~\eqref{eq:distCFhn}, which should present all sorts
of interesting technical challenges.

This concludes the list of questions that arise in line with the point
of view developed in this chapter. The reader is invited to consult
the references given in Section~\ref{sec:distcontext} for the many
other approaches to the study of distances in random planar maps that
have been developed since, and their current challenges.

\chapter{The slice decomposition of planar maps}
\label{chap:slices}

In this chapter we return to the enumerative and bijective theory of
planar maps. The notions of distances and geodesics will still play an
important role, but as technical tools rather than objects of study.
The publications closest to this topic
are~\citemy{HObipar,fomap,OHRMT,irredmaps,irredsuite}, in addition to
the papers discussed in Chapter~\ref{chap:dist} that already contain a
fair amount of bijections. Here I will focus on one specific topic,
the slice decomposition, which first made brief appearances in
\citemy{pseudoquad,hankel,constfpsac} as a reformulation of the
\cite{mobiles} bijection, and whose real significance was understood
in~\citemy{irredmaps}.

My purpose is here to give a ``modern'' presentation of the subject,
with an emphasis on bijections. It is as self-contained as possible,
except for some detailed proofs which should be looked for
in~\citemy{irredmaps}. After recalling the context in
Section~\ref{sec:slicecontext}, I will introduce the theory of general
slices in Section~\ref{sec:slicegen}. I will then discuss irreducible
slices in Section~\ref{sec:sliceirr}, before concluding with some
perspectives in Section~\ref{sec:sliceconc}.

\section{Context}
\label{sec:slicecontext}

The enumerative and bijective theory of planar maps has recently been
the subject of a very nice review by Gilles
Schaeffer~\cite{Schaeffer15}, containing an extensive bibliography. I
feel therefore exempted from having to recall the whole context, and
may focus on the more specific line of research motivating this
chapter.

While fairly general, the ``map-mobile'' correspondence presented in
Section~\ref{sec:distmobiles} is only one among the many bijections
that are known in literature. In particular, it is not well-suited for
the study of maps with connectivity or girth constraints, for which
specific bijections have been
found~\cite{JaSc98,theseSchaeffer,DDP00,PoSc03,PoSc06,Fusy07,FPS08,Fusy09}.
It does not apply either easily to tree-rooted
maps~\cite{Bernardi07}\footnote{In~\citemy{fomap} we proposed an
  extension of the bijection of~\cite{mobiles} for maps decorated with
  spanning trees, Ising spins or hard particles (on bipartite maps),
  but our construction is arguably complicated. Note that Ising spins
  and hard particles on nonbipartite maps can be treated through the
  Eulerian version of the bijection of~\cite{mobiles}.}. Recently,
several authors developed a general ``bijective canvas'' ---this name
being borrowed from~\cite{Schaeffer15}---
which, given a family
of maps, splits the task of designing a bijection for it into two,
more systematic, subtasks:
\begin{enumerate}
\item characterize the family of maps at hand in terms of the
  existence of certain orientations, following the theory of
  $\alpha$-orientations developed by
  Felsner~\cite{Felsner04}\footnote{See also the earlier work of
    Propp~\cite{Propp93}. Bernardi~\cite{Bernardi07} was one of first
    to realize the relevance of orientations for the bijective
    approach, see also~\cite{BeCh11} for the higher genus extension of
    his bijection.},
\item apply to these orientations one of the two known unified
  constructions: either the Bernardi-Fusy construction which is based
  on a suitably generalized notion of
  mobiles~\cite{BeFu12a,BeFu12b,BeFu14,BeFu18}, or the
  Albenque-Poulalhon construction based on blossoming
  trees~\cite{AlPo15}\footnote{In this document I do not consider at
    all blossoming trees, though I also investigated them during my
    doctoral thesis and shortly after in~\citemy{HObipar}. I remain
    puzzled by the fact that the blossoming bijection of~\cite{census}
    appears in the intersection of both unified
    constructions~\cite[Section~7.3]{BeFu12b} \cite[p.~14]{AlPo15}.}
  \footnote{See also the recent paper~\cite{Lepoutre19} which gives a
    blossoming bijection in any genus, building on Propp's theory of
    orientations~\cite{Propp93}.}.
\end{enumerate}
Many, if not all, previously known bijections for planar maps may be
recovered through the bijective canvas.

In this chapter, we propose another general bijective approach which
is based on geodesics rather than orientations. The fundamental idea
is that maps can be canonically decomposed along leftmost geodesics,
and we call slices the pieces appearing in this
decomposition. Initially slices were conceived as a by-product of the
CVS/BDG bijection: in the algorithm which produces a map out of a
mobile, a slice is the portion of the map which arises from a specific
subtree. This follows easily from the well-known observation that the
sequence of ``successors'' of a corner of the mobile forms a leftmost
geodesic in the map. We used slices in~\citemy[Appendix~A]{hankel}
and~\citemy[Section~4]{constfpsac} as a shortcut to reformulate
certain combinatorial arguments without resorting to mobiles. But we
also realized that slices remain convenient in some situations where
mobiles might be impractical. A first example
is~\citemy[p.~33]{pseudoquad}, where slices are used to decompose
quadrangulations with self-avoiding (simple) boundaries, whose
corresponding mobiles are not so easy to characterize. But the virtue
of slice decomposition was only fully realized in~\citemy{irredmaps}
where we studied ``irreducible'' maps, which are even less easy to
study using mobiles\footnote{In~\cite[Section~1.1]{BeFu12b}, Bernardi
  and Fusy mention that their approach based on generalized mobiles
  may be extended to irreducible maps, but cite a paper which has not
  appeared yet.}.  This shows that slice decomposition
is~\emph{robust}. It also seems to behave well when passing to the
scaling limit: in~\cite{LeGall13}, Le Gall used slices (which he calls
discrete maps with geodesic boundaries or DMGB) as an important
ingredient in his proof. Bettinelli and Miermont use them again to
study Brownian disks~\cite{BeMi17}.

While it does not seem directly related with the topic of this
chapter, I will conclude this overview of the context by mentioning
the current common belief that planar maps decorated with models of
statistical physics ---such as the $O(n)$ loop model we will discuss
in Chapter~\ref{chap:on}--- are deeply related with bidimensional
walks confined in a quadrant. One of the first to realize the
importance of this idea is Sheffield in his ``hamburger-cheeseburger
bijection''~\cite{Sheffield16} which motivated the ``mating-of-trees''
approach to Liouville quantum gravity~\cite{DMS14}. There are several
known bijections of this style ---see the references given in the last
section of~\cite{Schaeffer15}--- and the list keeps
growing~\cite{KMSW15,Budd17talk,GKMW18,BHS18}. I mention a possible
connection with slice decomposition in Section~\ref{sec:sliceconc}.

\section{General slices}
\label{sec:slicegen}

\begin{figure}
  \centering
  \includegraphics{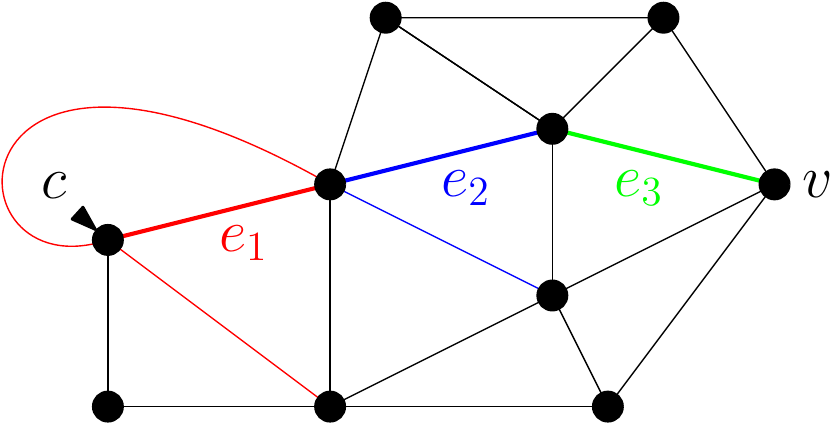}
  \caption{Construction of the leftmost geodesic from the corner $c$
    to the vertex $v$. We start at the vertex incident to $c$, and
    consider all the incident edges leading to a vertex strictly
    closer to $v$ (these edges are here shown in red). We pick $e_1$
    as the leftmost one (i.e.\ $e_1$ is the first red edge encountered
    when turning clockwise around the vertex incident to $c$). We then
    move to the other endpoint of $e_1$, and consider again all the
    incident edges leading to a vertex strictly closer to $v$ (these
    edges are here shown in blue). We pick $e_2$ as the leftmost one,
    using now $e_1$ as the reference, and so on until we reach
    $v$. The edges $e_1$, $e_2$, $\ldots$, form the wanted leftmost
    geodesic. Note that this construction only relies on the existence
    of a local orientation at each vertex.}
  \label{fig:leftgeod}
\end{figure}

The most important part of this section consists of definitions: what
slices are and which operations we may perform on them. Let us first
observe that, in a planar map, or more generally an orientable map,
there is a well-defined notion of \emph{leftmost geodesic} from a
corner $c$ to a vertex $v$. As illustrated on
Figure~\ref{fig:leftgeod}, we may construct it step-by-step by
starting from the vertex incident to $c$ and by moving, at each step,
to the leftmost adjacent vertex lying strictly closer to $v$, where we
use the direction we are coming from as a reference. We now introduce
slices and some related terminology.

\begin{figure}
  \centering
  \fig{}{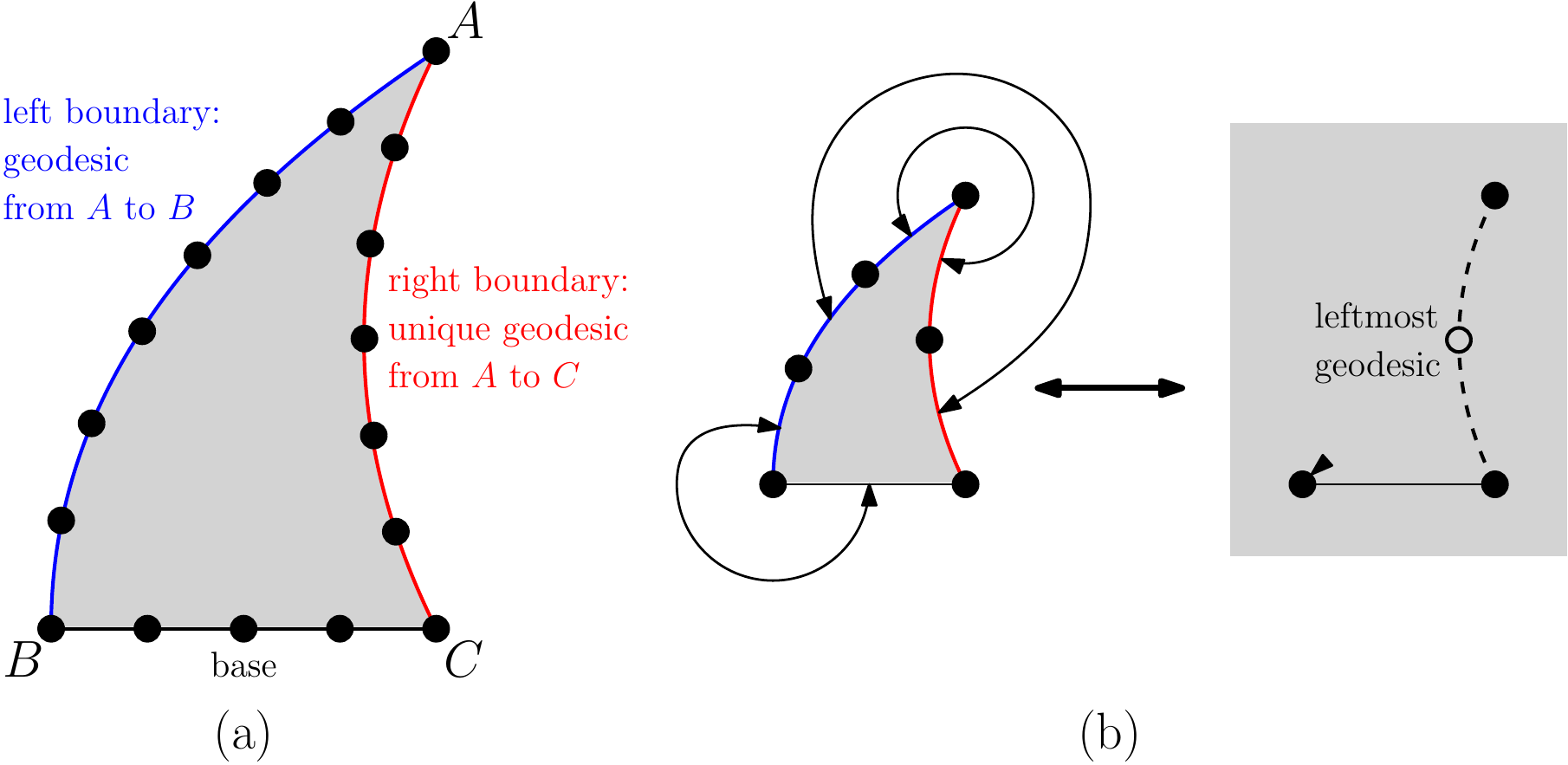}
  \caption{(a) General structure of a slice. (b) The one-to-one
    correspondence between elementary slices and pointed rooted maps.}
  \label{fig:slices}
\end{figure}

\begin{Def}[see Figure~\ref{fig:slices}(a) for an illustration]
  \label{Def:slice}
  A \emph{slice} is a planar map with a boundary, with three
  distinguished outer corners ---denoted here $A$, $B$, and $C$---
  which appear in counterclockwise around the outer face and split its
  contour in three portions:
  \begin{itemize}
  \item the \emph{left boundary} $AB$, which is a geodesic between $A$
    and $B$,
  \item the \emph{right boundary} $AC$, which is the unique geodesic
    between $A$ and $C$,
  \item and the \emph{base} $BC$.
  \end{itemize}
  The vertex incident to $A$ is called the \emph{apex}, the
  \emph{width} is the length of the base, the \emph{depth} is the
  length of the left boundary, and the \emph{tilt} is the depth minus
  the length of the right boundary. A slice of width $1$ is said
  \emph{elementary}, a slice of arbitrary width is said
  \emph{composite}.  The unique elementary slice of tilt $-1$ is the
  \emph{trivial slice} reduced to a single edge whose two end corners
  are $A=B$ and $C$. It is different from the \emph{empty slice},
  which is the elementary slice of tilt $+1$ reduced to a single edge
  whose two end corners are $B$ and $A=C$. Note that, by the
  triangular inequality, the tilt is smaller than or equal to the width,
  hence an nontrivial elementary slice may have tilt $0$ or $1$.
\end{Def}

In~\citemy{hankel,irredmaps}, it is assumed that all slices are
elementary, but here we find convenient to also consider composite
slices. Note that we impose no constraint on the base, it is not even
assumed that it forms a simple path, and the graph distance from $B$
to $C$ may be smaller than the width.

For simplicity, we restrict from now on on to bipartite maps, and only
briefly discuss the general case in Remark~\ref{rem:slicenonbip} at
the end of this section. Bipartiteness implies that the width and the
tilt have the same parity, in particular a nontrivial elementary slice
has tilt $1$.

It was noted in~\citemy{hankel} that there is a one-to-one
correspondence between elementary slices and pointed rooted maps: we
go from an elementary slice to a pointed rooted map by identifying the
boundary edges together as illustrated on
Figure~\ref{fig:slices}(b). We must exclude the trivial and empty
slices which are pathological. After the identification, the base
becomes the root edge and the apex becomes the origin. The
correspondence is one-to-one because the merged boundaries form the
leftmost geodesic going from the root to the origin, which can be
recovered canonically from the pointed rooted map
structure\footnote{In all rigor, we only consider here pointed rooted
  maps such that the leftmost geodesic from the root to the origin
  starts with the edge immediately left of the root corner. In the
  bipartite case, these form precisely half of all pointed rooted maps
  (just reverse the direction of the root edge to obtain the
  others). Note that this remark was already made in
  Section~\ref{sec:distCF} when defining the two-point
  function.}. From this correspondence, we deduce that the generating
functions of elementary slices of tilt $1$, weighted by the Boltzmann
weight~\eqref{eq:distboltz} ---where we do not include the outer face
in the product--- is equal to $R$ as defined in
Section~\ref{sec:distCF}. Note that $R$ has constant term $1$,
corresponding to the empty slice.  The generating function of
elementary slices of depth at most $d$ is nothing but the two-point
function $R_d$.

\begin{figure}
  \centering
  \fig{}{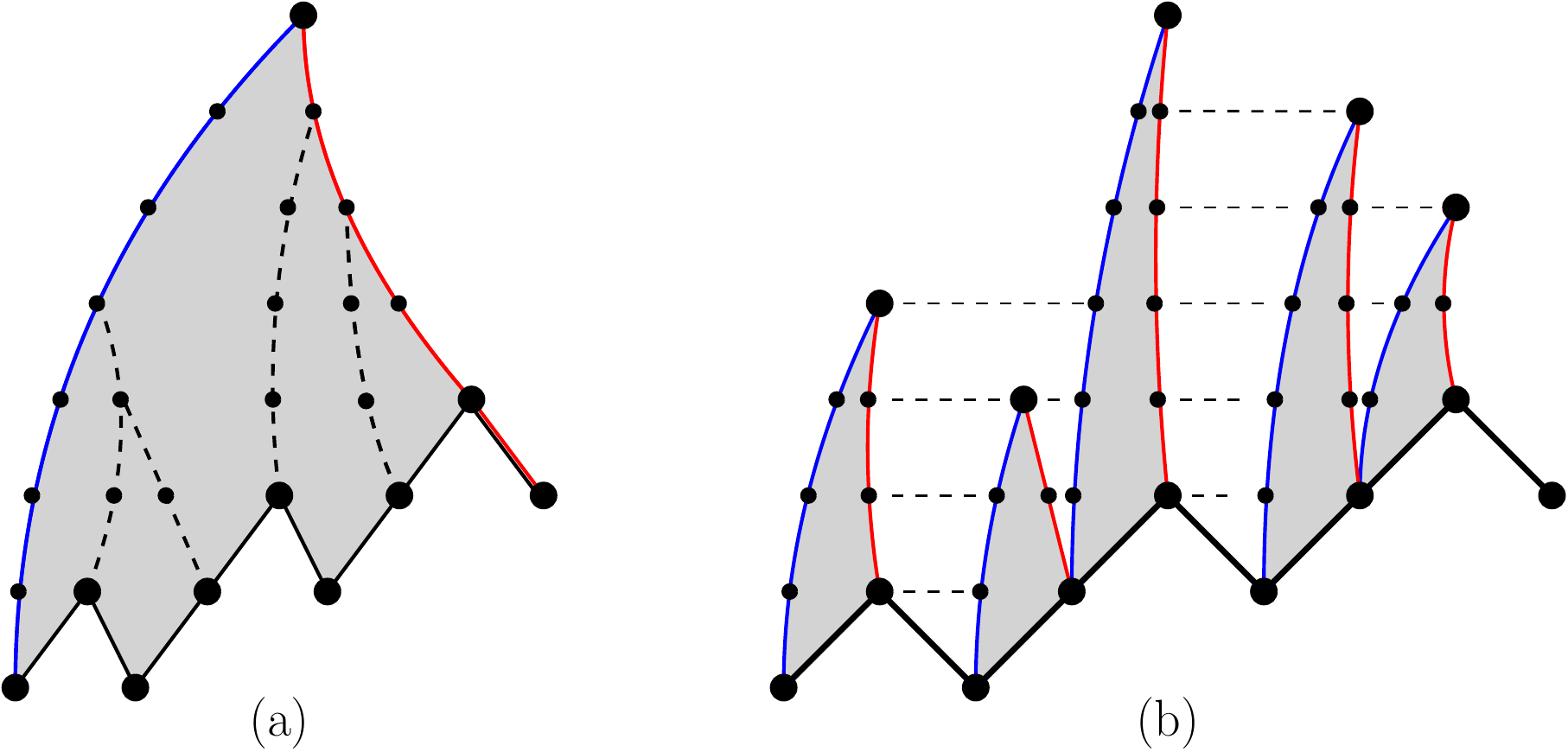}
  \caption{(a) The decomposition of a composite slice into elementary
    slices, by cutting along the leftmost geodesics from the base
    vertices. (b) The corresponding ``path of elementary slices''.}
  \label{fig:slicedecomp}
\end{figure}

We now consider composite slices and describe the important operation
of decomposing them into elementary slices. See the illustration on
Figure~\ref{fig:slicedecomp}(a). We start from a composite slice with
apex $A$ and base $BC$, and denote by $v_0,v_1,\ldots,v_n$ the base
vertices read from $B$ to $C$. To the base vertex $v_i$ we associate a
label $\ell_i:=d(v_0,A)-d(v_i,A)$, where $d$ denotes the graph
distance in the slice. Note that $n$ is the width, $d(v_0,A)$ is the
depth, $\ell_n$ is the tilt, and we have $|\ell_i-\ell_{i-1}|=1$ for
any $i=1,\ldots,n$. We now cut the slice along each leftmost geodesic
from a base vertex to the apex $A$. This has the effect of cutting the
composite slice into elementary slices. More precisely, if we denote
by $v'_i$ the first vertex common to the leftmost geodesics started at
$v_{i-1}$ and $v_i$, then these two geodesics delimit a slice of width
$1$, base $v_{i-1} v_i$, apex $v'_i$ and tilt $\ell_i-\ell_{i-1}$. If
$\ell_i-\ell_{i-1}=-1$, the slice is trivial. If
$\ell_i-\ell_{i-1}=1$, we obtain a nontrivial slice, which is possibly
empty if the geodesic from $v_{i-1}$ starts with the base edge
$v_{i-1}v_i$. To summarize, we end up with a label sequence
$0=\ell_0,\ell_1,\ldots,\ell_n$ with increments $\pm 1$, and a
nontrivial elementary slice for each up-step of the label sequence
(and the trivial slice for each down-step). We may call such object a
``path of elementary slices'', and it is not difficult to check that
the correspondence is one-to-one as we may reconstruct the composite
slice by gluing back the elementary slices together as shown on
Figure~\ref{fig:slicedecomp}(b). An immediate consequence of the
decomposition is that the generating function of composite slices of
width $n$ and tilt $\ell$ is equal to
$\binom{n}{\frac{n+\ell}{2}} R^{\frac{n+\ell}{2}}$. It is possible to
keep track of the depth of the composite slice, by expressing it in
terms of the labels and the depths of the elementary slices, but we
will not enter into such considerations here.

As an application of this decomposition, we easily recover the
equation~\eqref{eq:distCFReq} satisfied by $R$: the term $1$ on the
right-hand side corresponds to the empty slice, and the term of index
$k$ in the sum corresponds to elementary slices where the base edge is
incident to an inner face of degree $2k$. Indeed, for such a slice,
removing the base edge yields a composite slice of width $2k-1$ and
tilt $1$.

\begin{figure}
  \centering
  \fig{.9}{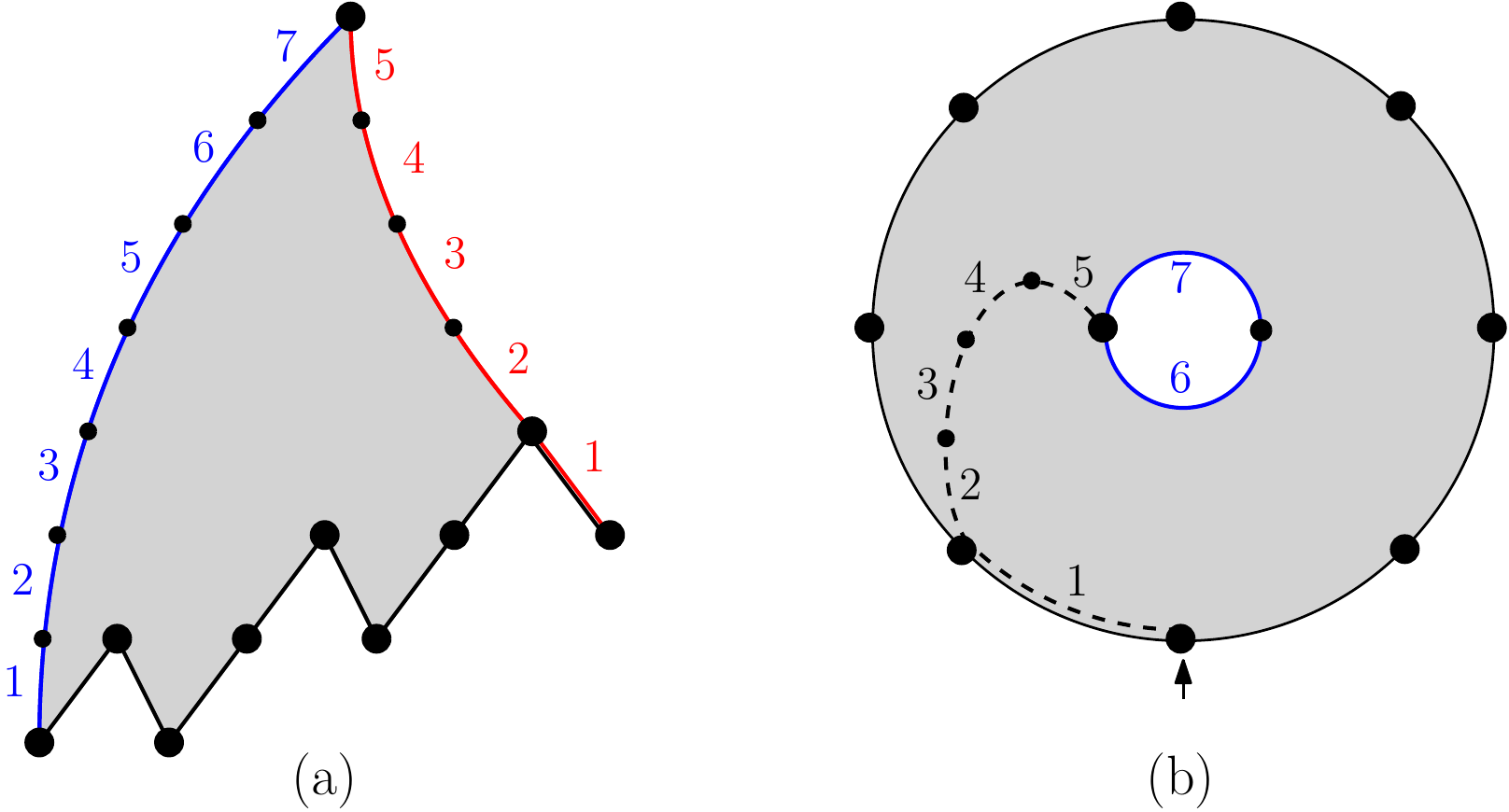}
  \caption{(a) Wrapping a composite slice of tilt $+2$: pairs of edges
    with the same label are identified. They become a path joining the
    outer face to the central face (formed by the $2$ unmatched left
    boundary edges) in the resulting $2$-annular map (b).}
  \label{fig:slicewrap}
\end{figure}

We now describe another important operation on composite slices called
\emph{wrapping}, which we illustrate on Figure~\ref{fig:slicewrap}. It
consists in identifying boundary edges as in the procedure of
Figure~\ref{fig:slices}(b), but the identifications are
different. Namely, we only identify together the left and right
boundaries, starting from the base. When the tilt is nonzero, some
edges of the longer boundary remain unmatched, and they delimit a new
face which we call the \emph{central} face. The unmatched base edges
delimit another new face which we consider as the outer face, and
which we root at the former position of the base endpoints. When the
tilt is zero, there is no central face but an origin which is the
former apex. Before stating the conclusion as a theorem, we first need
to characterize the resulting maps.

\begin{Def}
  \label{Def:sliceannular}
  Given a positive integer $\ell$, a \emph{$\ell$-annular map} is a
  rooted map with an extra distinguished face of degree $\ell$, called
  the \emph{central} face, such that any cycle winding around the
  central face has length at least $\ell$. A
  \emph{$\ell$-strict-annular map} is a $\ell$-annular map such that
  the contour of the central face is the unique winding cycle of
  length $\ell$.
\end{Def}

Note that we allow for odd values of $\ell$, which corresponds to the
``quasi-bipartite'' case where both the central and the outer face
have odd degrees (but all other faces have even degrees as
usual). Pointed rooted maps may be thought as the case $\ell=0$.

\begin{thm}[see~{\citemy{hankel} and \citemy{irredmaps}}]
  \label{thm:sliceannular}
  The wrapping operation is a one-to-one correspondence between:
  \begin{itemize}
  \item composite slices of zero tilt and pointed rooted maps,
  \item composite slices of positive tilt $\ell$ and $\ell$-annular maps,
  \item composite slices of negative tilt $-\ell$ and
    $\ell$-strict-annular maps.
  \end{itemize}
  In all cases, the width of the composite slice is equal to the
  degree of the outer face of the corresponding map.  The
  corresponding generating functions for a prescribed width/outer
  degree $n$ are respectively $\binom{n}{n/2} R^{n/2}$,
  $\binom{n}{(n+\ell)/2} R^{(n+\ell)/2}$ and
  $\binom{n}{(n-\ell)/2} R^{(n-\ell)/2}$, where $R$ is determined
  by~\eqref{eq:distCFReq}.
\end{thm}

The case of zero tilt is rather straightforward and was given
in~\citemy{hankel}. The case of nonzero tilt is more involved and
requires an argument which was first given
in~\citemy[Section~7.2]{irredmaps} for the case of irreducible maps
considered in the next section. As observed by the anonymous referee
---which I would like to thank again for his/her most insightful
report--- the argument is actually more general, and I will now sketch
it in the simpler current setting.
\begin{figure}
  \centering
  \fig{.9}{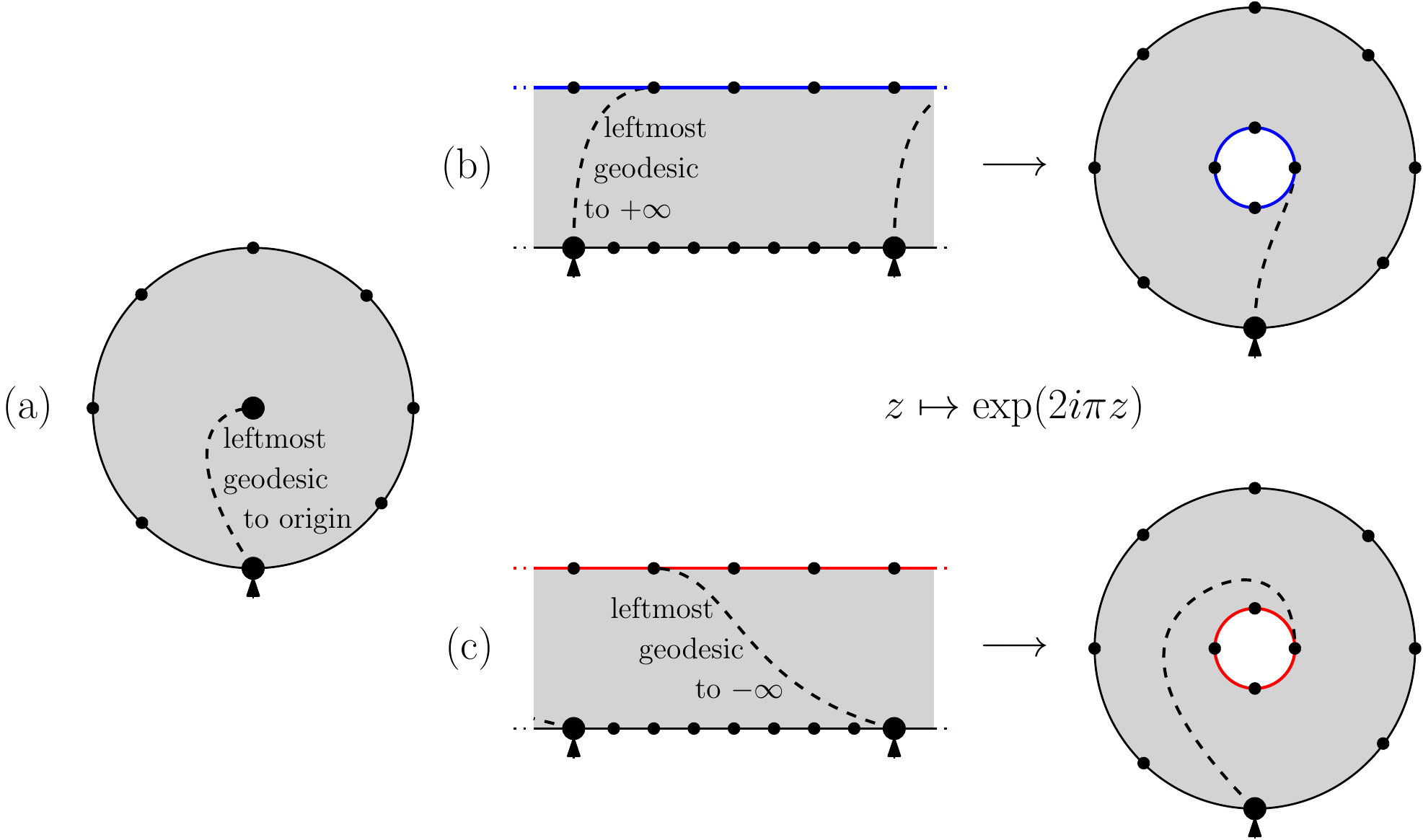}
  \caption{Selecting a suitable leftmost geodesic in (a) a pointed
    rooted map, (b) a $\ell$-annular map and (c) a
    $\ell$-strict-annular map. In the annular cases, the leftmost
    geodesics are actually constructed on the ``lift''.}
  \label{fig:sliceannular}
\end{figure}

The main difficulty consists in defining the inverse operation of wrapping,
i.e.\ reconstructing a composite slice from a map. The basic idea is
that, as before, the left and right boundaries of the slice should
somehow correspond to a ``leftmost geodesic'' in the map. This works
straightforwardly for a pointed rooted map, where we have to consider
the leftmost geodesic from the root to the origin. For a
$\ell$-(strict-)annular map, we actually have to introduce the
\emph{lift} of the map, which may be seen as its preimage through the
exponential mapping $z \mapsto \exp(2i\pi z)$, assuming that the map
is drawn on the complex plane with the point $z=0$ within the central
face. The lift is an infinite but periodic map, so its properties
remain manageable. We then consider the leftmost geodesic going from
(a preimage of) the root to $+\infty$ (for a $\ell$-annular map) or to
$-\infty$ (for a $\ell$-strict-annular map). It follows from
Definition~\ref{Def:sliceannular} that such geodesic is indeed
well-defined, and that it eventually reaches the central face and
continues following its contour forever. See
Figure~\ref{fig:sliceannular} for an illustration,
and~\citemy{irredmaps} for more details. We now simply have to cut
along the finite portion of the geodesic between the root and the
coalescence point to obtain a composite slice, and it is now
elementary to check that this is indeed the inverse operation of
wrapping.

Theorem~\ref{thm:sliceannular} may be seen as a first illustration of
the relevance of slice decomposition. Its enumerative consequences can
also be obtain as a particular case of~\cite[Theorem~34]{BeFu12b}
which relies on so-called ``$b$-dibranching mobiles'', but I consider
the present approach to be more transparent. Slice decomposition is
actually able to treat the general setting of~\cite{BeFu12b} as we
will see in the next section. But let us now finish this section with
some remarks.

First, notice that the generating function of $\ell$-annular slices is
equal to $R^\ell$ times that of $\ell$-strict-annular slices, for any
outer degree $n$. This calls for a combinatorial explanation, which is
as follows: given a $\ell$-annular slice, we may decompose it
bijectively along its \emph{outermost} winding cycle of length $\ell$,
to obtain a $\ell$-strict-annular slice (the exterior of the cycle)
and a $\ell$-annular slice of outer degree $\ell$ (the interior of the
cycle). The latter slices have generating function $R^{\ell}$, as seen
by taking $n=\ell$ in Theorem~\ref{thm:sliceannular}.

Elaborating on this idea, we may enumerate general annular maps with
prescribed boundary lengths ---i.e.\ quasi-bipartite maps with two
marked corners incident to distinct faces of respective prescribed
degrees $m$ and $n$, with $m+n$ even and no constraint on the winding
cycles--- through the generating function
\begin{equation}
  \label{eq:sliceannulgen}
  \begin{split}
    A_{m,n} :&= \sum_{\substack{0 \leq \ell \leq \min(m,n) \\ \ell + m
        \text{ even}}}
    \ell \times \binom{m}{(m-\ell)/2} R^{(m-\ell)/2} \times \binom{n}{(n+\ell)/2} R^{(n+\ell)/2} \\
    &= \frac{2}{m+n} \cdot \frac{m!}{\lfloor \frac{m}{2} \rfloor!
      \lfloor \frac{m-1}{2} \rfloor!} \cdot \frac{n!}{\lfloor
      \frac{n}{2} \rfloor! \lfloor \frac{n-1}{2} \rfloor!} \cdot
    R^{(m+n)/2}.
  \end{split}
\end{equation}
Indeed, considering such an annular map and treating the face of
degree $m$ as the outer face, we may decompose it along the outermost
cycle of minimal length, which we denote by $\ell$, and obtain a
$\ell$-strict-annular map of outer degree $m$ (the exterior of the
cycle) and a $\ell$-annular map of outer degree $n$ (the interior of
the cycle, after inversion). This decomposition is $\ell$-to-one
because there are $\ell$ ways to glue back the two pieces
together. This explains the first line of \eqref{eq:sliceannulgen} and
we pass to the second line by a simple identity on binomial
numbers. Note that the formula remains meaningful in the case $m=0$,
where we recover the generating function of pointed rooted maps with
outer degree $n$.

The formula~\eqref{eq:sliceannulgen} appears
in~\cite[Corollary~3.1.2]{Eynard16} under different notations. A more
general formula of Collet and Fusy~\cite{CoFu12} counts
quasi-bipartite maps with an arbitrary number of boundaries of
prescribed lengths: it may be obtained from~\eqref{eq:sliceannulgen}
by taking derivatives with respect to the face weights
$(g_{2k})_{k \geq 0}$ and by using \eqref{eq:distCFReq}, or
by integrating in the case of one boundary to recover the
formula~\eqref{eq:distCFwnexp}.

\begin{rem}
  \label{rem:slicenonbip}
  The discussion of this section extends naturally to the nonbipartite
  case, up to some complications which we list here. First, the tilt
  of a slice is no longer necessarily of the same parity as the
  width. Therefore, there exists another type of elementary slice
  which has zero tilt. The generating function of such slices with
  depth at most $d$ is nothing but the quantity $S_d$ mentioned in
  Remark~\ref{rem:distCFnonbip}. Pairs of such slices may be glued to
  obtain a pointed rooted map where the root edge is frustrated,
  see~\citemy[Figure~17]{hankel}. The decomposition of a composite
  slice now involves a label sequence with increments $-1,0,+1$, and
  elementary slices of zero tilt are attached to the
  $0$-increments. The wrapping operation works unchanged. Generating
  functions of composite slices and annular maps are now polynomials
  in $R$ and $S:=S_\infty$, whose expressions are elementary but
  longer than those of the bipartite case $S=0$. We therefore leave
  them as exercises to the reader.
\end{rem}

\section{Irreducible slices}
\label{sec:sliceirr}

In this section we explain how to adapt the slice decomposition in the
presence of ``irreducibility'' constraints. Our discussion follows for
a good part that of the previous section, and emphasizes what the
new phenomena related to irreducibility are. The presentation will
therefore be different from~\citemy{irredmaps}. But let us first
define what an irreducible map is.

\begin{Def}
  \label{Def:sliceirred}
  Given a nonnegative integer $d$, a map (resp.\ a map with a
  boundary) is said \emph{$d$-irreducible} if it does not contain any
  cycle of length smaller than $d$, and any cycle of length $d$ is the
  contour of a face (resp.\ is the contour of an inner face).  An
  \emph{irreducible $d$-angulation} is a $d$-irreducible map where
  every face has degree $d$.
\end{Def}

Note that $d$ is unrelated to the depth considered previously. The
notion of irreducibility is closely related to the girth: a
$d$-irreducible map has girth at least $d$, and a map of girth at
least $d$ is a $(d-1)$-irreducible map containing no face of degree
$d-1$.

Irreducible triangulations and quadrangulations were first enumerated
by respectively Tutte~\cite{Tutte62tri} and Mullin and
Schellenberg~\cite{MuSc68}. Note that these authors do not use the
word ``irreducible'' but ``simple'', which probably did not have its
current acceptation in the sixties. Their approach is based on a
substitution procedure ---see for instance~\cite[I.6.2]{FlSc09} for
general background--- by deducing the generating function of
irreducible tri- or quadrangulations from that of simple ones.  In our
initial investigation of irreducible maps, we actually followed their
approach by devising an iterated substitution procedure for
enumerating $d$-irreducible maps, with arbitrary $d$ and prescribed
face degrees. We do not enter into the details and refer the
interested reader to~\citemy[Sections~2 and 3]{irredmaps}. But we then
realized that some of the auxiliary quantities that we needed for our
computations could be interpreted combinatorially as slice generating
functions. This allowed us to rederive our enumerative results
bijectively, and recover in particular the bijections
of~\cite{Fusy09,FPS08} for irreducible triangulations and
quadrangulations. We note in passing that irreducible quadrangulations
are the ``medials'' ---recall Figure~\ref{fig:quadmap}--- of
$3$-connected maps~\cite[Section~5]{MuSc68}, which themselves
correspond to convex three-dimensional polyhedra by Steinitz's
theorem.

We now consider $d$-irreducible slices, defined by combining
Definitions~\ref{Def:slice} and \ref{Def:sliceirred} (note that a
slice has a boundary). The Boltzmann weight of a $d$-irreducible slice
is defined as in~\eqref{eq:distboltz}, where as before we do not
include the outer face in the product. Here the weight only involves
the $g_k$'s for $k \geq d$. We again restrict here to the bipartite
case, so we take $d=2b$ and $g_k=0$ for $k$ odd.  We denote by
$R^{(d)}$ the generating function of bipartite $d$-irreducible
nontrivial elementary slices.

The first new phenomenon is that it is no longer possible to identify
$R^{(d)}$ with the generating function of $d$-irreducible pointed
rooted maps. Indeed, gluing the boundaries as in
Figure~\ref{fig:slices}(b) might create ``short cycles'' that wind
around the origin.

Fortunately, $d$-irreducibility poses no problem for the decomposition
of Figure~\ref{fig:slicedecomp}: a composite slice is $d$-irreducible
if and only if all the elementary slices appearing in its
decomposition are $d$-irreducible. One direction is obvious, the other
relies on the observation that, when gluing $d$-irreducible elementary
slices along their left/right boundaries as on
Figure~\ref{fig:slicedecomp}(b), then we cannot create ``short
cycles''.  Indeed, since the glued boundaries are geodesics, and since
every cycle in the composite map that crosses one of them must cross
it another time, we see that for any cycle in the composite map there
exists a shorter cycle in each elementary slice that it visits.  We
deduce that the generating function of $d$-irreducible composite
slices of width $n$ and tilt $\ell$ is
$\binom{n}{\frac{n+\ell}{2}} (R^{(d)})^{\frac{n+\ell}{2}}$. That is,
we only change $R$ into $R^{(d)}$ with respect to the expression of
Section~\ref{sec:slicegen}.

The main complication arises when we attempt to find an analogue of
equation~\eqref{eq:distCFReq} for $R^{(d)}$ by removing the base edge
in a $d$-irreducible elementary slice. The composite slice we obtain
by removing the base edge is still $d$-irreducible, but it satisfies
extra constraints due to cycles going through the base edge in the
elementary slice. We therefore have to find another decomposition,
which may be done by introducing generalized elementary slices, which
we called $k$-slices in~\citemy{irredmaps} but which we call here
quasi-slices.

\begin{figure}
  \centering
  \fig{}{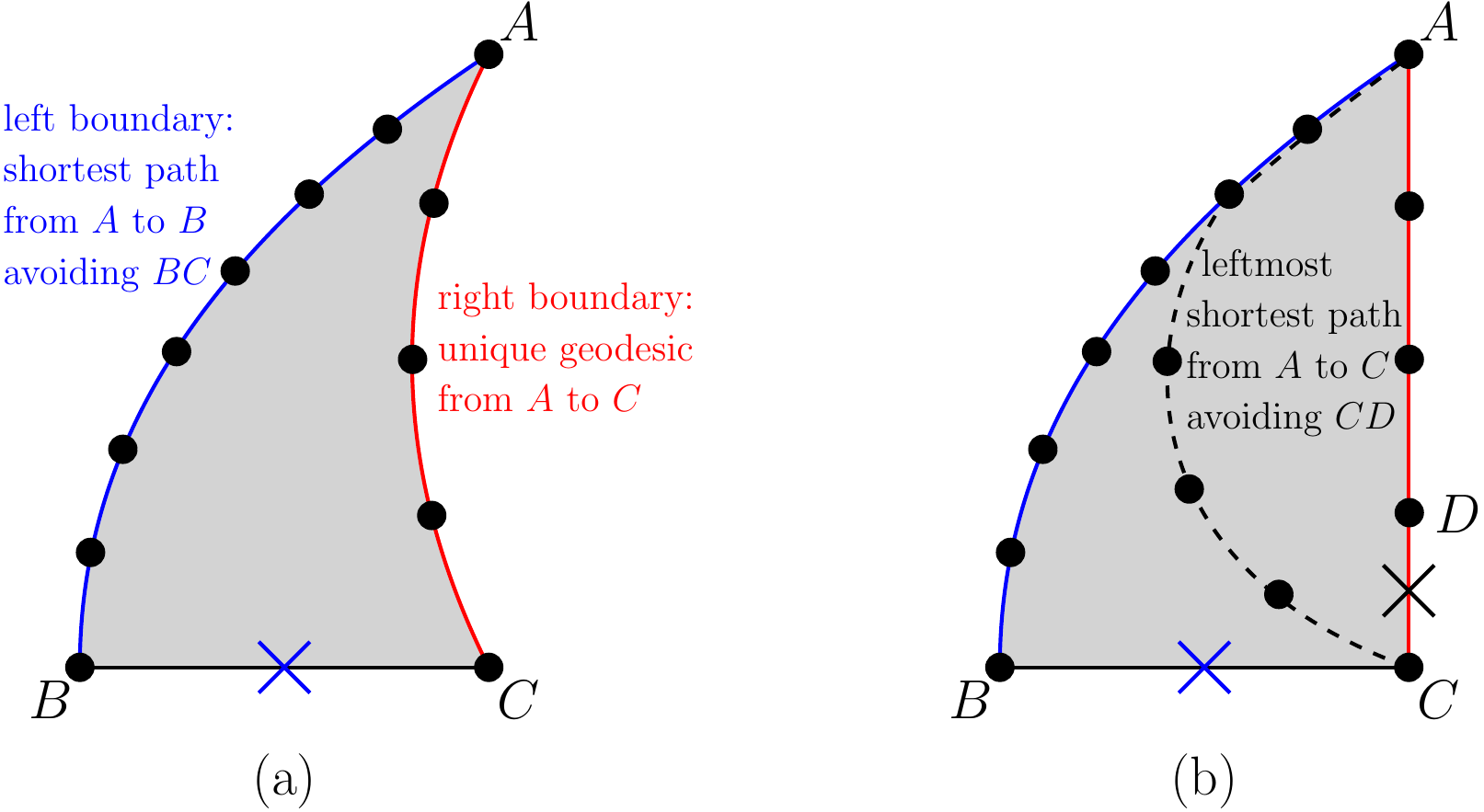}
  \caption{(a) General structure of a quasi-slice. (b) The
    decomposition of a quasi-slice with ``small'' tilt: we cut along
    the leftmost shortest path from $A$ to $C$ avoiding the first
    vertex $D$ of the right boundary.}
  \label{fig:slicequasi}
\end{figure}

A \emph{quasi-slice} is defined as an elementary slice in
Definition~\ref{Def:slice}, with the sole difference that the left
boundary is no longer assumed to be a true geodesic between $A$ and
$B$, but only a shortest path among paths \emph{avoiding the base
  edge} $BC$. See Figure~\ref{fig:slicequasi}(a) for an
illustration. The tilt may now take values other than $\pm 1$.  Since
we have not changed the assumption on the right boundary, the tilt is
still at least $-1$ with equality if and only if the quasi-slice is
reduced to the trivial slice, but we may now have a tilt larger than
$1$. By bipartiteness, it must be an odd number, and we denote by
$U_k^{(d)}$ the generating function of $d$-irreducible quasi-slices of
tilt $2k+1$. We have $U_{-1}^{(d)}=1$ and $U_0^{(d)}=R^{(d)}-1$, as
the empty slice should not be regarded as a quasi-slice. We now
explain how to decompose a quasi-slice, which is done in two ways
depending on $k$ (recall that $d=2b$).
\begin{itemize}
\item For $k \geq b$, the decomposition consists as before in removing
  the base edge. We obtain a $d$-irreducible composite slice of tilt
  $2k+1 \geq d+1$, and we easily check that the decomposition is now
  bijective because of the large tilt (in the reverse bijection we do
  not risk creating short cycles). We deduce that
  \begin{equation}
    \label{eq:sliceqlargetilt}
    U_k^{(d)} = \sum_{\ell \geq k+1} g_{2\ell} \binom{2\ell-1}{\ell+k} (R^{(d)})^{\ell+k}, \qquad k \geq b.
  \end{equation}
  In particular this quantity vanishes in the case of irreducible
  $d$-angulations, since we shall take $g_{2\ell}=0$ for all
  $\ell \geq b$.
\item For $0 \leq k \leq b-1$, we have to consider a new decomposition
  illustrated on Figure~\ref{fig:slicequasi}(b). First,
  $d$-irreducibility implies that the length of the right boundary is
  positive (except in the case $k=b-1$ where the slice might also be
  reduced to a single $d$-gon). Hence, there is a first vertex $D$ on
  the right boundary, and we now cut the quasi-slice along the
  leftmost shortest path from $A$ to $C$ avoiding $CD$. We obtain
  generically two quasi-slices, with respective bases $BC$ and $CD$
  and tilts $2i+1$ and $2j+1$, where
  $-1 \leq i < k$ and $j=k-i$. We deduce that
  \begin{equation}
    \label{eq:sliceqsmalltilt}
    U_k^{(d)} = g_{2b} \delta_{k,b-1} + U_{k+1}^{(d)} + \sum_{j=1}^{k} U_{k-j}^{(d)} U_{j}^{(d)}, \qquad 0 \leq k \leq b-1.
  \end{equation}
\end{itemize}
Note that, combining \eqref{eq:sliceqlargetilt} and
\eqref{eq:sliceqsmalltilt}, we get a closed system of equations for
$U_0^{(d)}=R^{(d)}-1,U_1^{(d)},\ldots,U_b^{(d)}$. It is actually
possible to eliminate the last $b$ quantities,
see~\citemy[Section~5.4]{irredmaps} for details, and we end up with the
following explicit equation for $R^{(d)}$:
\begin{equation}
  \label{eq:sliceRdeq}
  -\frac{1}{b} \sum_{p=1}^b \binom{b}{p} \binom{b}{p-1} \left( 1 - R^{(d)} \right)^p =
  g_{2b} + \sum_{\ell \geq k+1} g_{2\ell} \binom{2\ell-1}{\ell+k} (R^{(d)})^{\ell+k}.
\end{equation}
It should be seen as the $d$-irreducible counterpart
of~\eqref{eq:distCFReq}. Let us mention that, in the decomposition of
quasi-slices, it is possible to keep track of the depth and obtain
refined generation functions that are essentially two-point functions
of $d$-irreducible maps, see~\citemy[Section~8]{irredmaps}.

We turn to the wrapping operation of Figure~\ref{fig:slicewrap}:
it actually works without any modification once we realize that we
should treat differently the winding and nonwinding cycles. This is
the clever observation of our anonymous referee, which we mentioned
above. Let us define the resulting objects.

\begin{Def}
  \label{Def:slicedlirrann}
  Given two nonnegative integers $d,\ell$, a
  \emph{$(d,\ell)$-quasi-irreducible annular map} is a rooted map, where the
  outer face is regarded as a boundary, and an extra distinguished
  \emph{central} face of degree $\ell$ ---or a marked central vertex
  if $\ell=0$--- such that:
  \begin{itemize}
  \item the length of any cycle not winding around the central
    face/vertex is at least $d$, with equality if and only if the
    cycle is the contour of a face,
  \item for $\ell>0$, the length of any cycle winding around the
    central face is at least $\ell$.
  \end{itemize}
  A \emph{$(d,\ell)$-irreducible annular map} is defined in the same
  way, with the extra condition that the only winding cycle of length
  $\ell$ is the contour of the central face.
\end{Def}

For $d=0$, we recover the notion of $\ell$-(strict)-annular map of the
previous section. For $d=\ell$, a $(d,d)$-irreducible annular map is
nothing but a $d$-irreducible map with a boundary and an extra marked
face of degree $d$. We again allow $\ell$ to be odd, which corresponds
to the quasi-bipartite case where the central and outer faces are the
only two faces of odd degree.  We may now state the $d$-irreducible
analogue of Theorem~\ref{thm:sliceannular}.

\begin{thm}[see~{\citemy[Section~9.3]{irredmaps}}]
  \label{thm:sliceannularirr}
  The wrapping operation is a one-to-one correspondence between:
  \begin{itemize}
  \item $d$-irreducible composite slices of nonpositive tilt $-\ell$ and
    $(d,\ell)$-irreducible annular maps,
  \item $d$-irreducible composite slices of nonnegative tilt $\ell$ and
    $(d,\ell)$-quasi-irreducible annular maps.
  \end{itemize}
  In all cases, the width of the composite slice is equal to the
  degree of the outer face of the corresponding map.  The
  corresponding generating functions for a prescribed width/outer
  degree $n$ are respectively
  $\binom{n}{(n-\ell)/2} (R^{(d)})^{(n-\ell)/2}$ and
  $\binom{n}{(n+\ell)/2} (R^{(d)})^{(n+\ell)/2}$, where $R^{(d)}$ is
  determined by~\eqref{eq:sliceRdeq}.

  As a consequence, the generating function $F_{2m}^{(d)}$ of
  $d$-irreducible maps with a boundary of degree $2m$ satisfies the
  ``pointing'' formula
  \begin{equation}
    \label{eq:sliceirrpoint}
    \frac{\partial F_{2m}^{(d)}}{\partial g_d} = \binom{2m}{m-b} (R^{(d)})^{m-b}, \qquad d=2b.
  \end{equation}
\end{thm}

The proof is essentially the same as that of
Theorem~\ref{thm:sliceannular}. The only new thing to check is that we
cannot create ``short'' nonwinding cycles when identifying the left
and right boundary as in Figure~\ref{fig:slicewrap}, but the reasoning
is the same as for the decomposition of $d$-irreducible composite
slices into elementary slices.

We now conclude this section by some remarks, several of which are
direct adaptations of the remarks made after
Theorem~\ref{thm:sliceannular}. The generating function of
$(d,\ell)$-quasi-irreducible annular maps is equal to
$(R^{(d)})^{\ell}$ times that of $(d,\ell)$-irreducible annular maps,
for the same combinatorial reason as before. We may also enumerate
$d$-irreducible quasi-bipartite annular maps with prescribed boundary lengths
$m,n > d$ by decomposing along the outermost shorter cycle which must have a
length $\ell > d$, to get the generating function
\begin{equation}
  \label{eq:sliceannulgenirr}
  \begin{split}
    A_{m,n}^{(d)} :&= \sum_{\substack{2b < \ell \leq \min(m,n) \\ \ell + m
        \text{ even}}}
    \ell \times \binom{m}{(m-\ell)/2} (R^{(d)})^{(m-\ell)/2} \times \binom{n}{(n+\ell)/2} (R^{(d)})^{(n+\ell)/2} \\
    &= \frac{2}{m+n} \cdot \frac{m!}{\lfloor \frac{m+d}{2}  \rfloor!
      \lfloor \frac{m-1-d}{2} \rfloor!} \cdot \frac{n!}{\lfloor
      \frac{n+d}{2} \rfloor! \lfloor \frac{n-1-d}{2} \rfloor!} \cdot
    (R^{(d)})^{(m+n)/2}.
  \end{split}
\end{equation}
This is the $d$-irreducible analogue of the
formula~\eqref{eq:sliceannulgen}, which we recover for $d=0$. By
differentiating with respect to the face weights $(g_{2k})_{k > d}$,
we obtain a $d$-irreducible analogue of the Collet-Fusy
formula~\cite{CoFu12}. We actually adapted their bijective proof to
the irreducible setting in~\citemy{irredsuite}, but we will not enter
into the details here. Note that, to obtain the generating function
$F_n^{(d)}$ of $d$-irreducible maps with a single boundary, we may
either integrate~\eqref{eq:sliceannulgenirr} with respect to $g_m$ for
some $m>d$, or integrate~\eqref{eq:sliceirrpoint} at $n=2m$ with
respect to $g_d$. The lengthy expression of $F_n^{(d)}$ is given
at~\citemy[Equation~(3.21)]{irredmaps}, where we obtained it via
substitution. To my knowledge, it still awaits a bijective
interpretation.

So far we have enumerated $d$-irreducible maps with boundaries. Note
that, by Definition~\ref{Def:sliceirred}, a $d$-irreducible map with a
boundary of length $d$ is reduced to either a tree or a single
$d$-gon. But, if we do not regard the outer face as a boundary
anymore, there may exist nontrivial rooted $d$-irreducible maps whose
outer face has degree $d$, and we denote by $H_d$ their generating
function. In~\citemy[Section~9.1]{irredmaps}, we computed $H_d$
through substitution. We may also characterize it combinatorially as
follows.  Observe that $\frac{\partial H_d}{\partial g_d}$ counts
$d$-irreducible maps with two marked faces of degree $d$, which are
closely related with $(d,d)$-quasi-irreducible maps whose generating
function is $(R^{(d)})^d$. Indeed, the latter are essentially
``sequences'' formed by concatenating the former. There are small
subtleties to be taken care of, and we obtain the relation
\begin{equation}
  \label{eq:sliceirrHd}
  \frac{\partial H_d}{\partial g_d} = 2 + d g_d -
  \frac{d}{2} \cdot \frac{2 g_d + g_d^2}{1+2g_d+g_d^2} - (R^{(d)})^{-d},
\end{equation}
see also~\citemy[Equation~(9.21)]{irredmaps}.

Again, all our discussion may be extended to the nonbipartite case, at
the price of having to introduce another generating function $S^{(d)}$
counting $d$-irreducible elementary slices of tilt $0$. The first part
of Theorem~\ref{thm:sliceannularirr} is unchanged, but the generating
functions are now polynomials in $R^{(d)}$ and $S^{(d)}$.

Finally, let us observe that we may recover the enumerative results
of~\cite{BeFu12b} for maps with prescribed girth, simply by setting
the weight $g_d$ to $0$: as mentioned above a $d$-irreducible map
without $d$-gons is nothing but a map of girth $d+1$ (or $d+2$ in the
bipartite case).

\section{Conclusion and perspectives}
\label{sec:sliceconc}

We now conclude this chapter by evoking some directions for future
research.

A first immediate question is the connection between slice
decomposition and the bijective canvas. There is such a close
connection between the enumerative results of
Section~\ref{sec:sliceirr} and those of~\cite{BeFu12b} that we may
wonder whether the two approaches are the two sides of the same
coin. At the fundamental level it is however not clear how to relate
orientations and geodesics, I currently do not understand the
Bernardi-Fusy bijection between bioriented maps and generalized
mobiles well enough to tell. Superficially, the fact that it is
essential to consider orientations that are \emph{accessible} suggests
that one might associate with it a suitable generalized distance.
Incidentally, we may wonder whether the bijection of
Section~\ref{sec:distmobiles} ---that produces mobiles with several
faces--- might be extended in the Bernardi-Fusy framework to treat
certain ``partially accessible'' orientations.

Another direction concerns the extension of slice decomposition to
Eulerian maps, where we may distinguish two types of faces hence
weight them differently. In view of the Eulerian version of
the~\cite{mobiles} bijection, it is natural not to consider true
geodesics anymore, but instead shortest oriented paths with respect to
the canonical orientation that alternates around each vertex. The
fundamental ideas of slice decomposition seem to adapt to this
setting, see~\citemy[Section~4]{constfpsac} for the case of
constellations and~\cite{DPS16,theseDervieux} for the case of corner
triangulations, which are Eulerian triangulations subject to certain
irreducibility constraints arising from their connection with corner
polyhedra~\cite{EpMu14}.  With Marie Albenque, we are currently
investigating the general theory. One of our goals is to provide
bijective proofs of the several beautiful formulas for bicolored
(Ising) maps which may be found in~\cite[Chapter~8]{Eynard16}. We
already obtained several results which seemed out of reach using the
Eulerian mobiles of~\cite{mobiles} or the blossoming trees
of~\cite{BoSc00}. Our next challenge is the treatment of maps with
``Dobrushin'' boundary conditions, which seem naturally connected with
bidimensional walks by keeping track of the boundary lengths of each
color in some suitable (e.g.\ peeling) decomposition. See also the
recent preprint~\cite{ChTu18} of my former student Linxiao
Chen\footnote{and the corresponding chapter of his
  thesis~\cite[Chapter~V]{theseChen}} and Joonas Turunen, which
enumerates Ising triangulations with Dobrushin boundary conditions via
Tutte equations.  We note that Eulerian maps with girth ---and more
generally cycle-length--- constraints have been considered by Bernardi
and Fusy in~\cite{BeFu14} under the name of hypermaps. Once again one
may attempt to reproduce their results using slices. We speculate that
the choice for the decomposition of a quasi-slice used in
Section~\ref{sec:sliceirr}, where we may choose to either remove the
base edge or cut as in Figure~\ref{fig:slicequasi}, might be made
dependent on some varying parameter to accommodate for cycle-length
constraints.

A third direction which I would like to investigate is the possible
connection with topological recursion, see~\cite[Chapter~7]{Eynard16}
for an account of its map-related aspects. In topological recursion, a
key role is played by the ``Bergman kernel'' which, in maps, is
essentially the generating function of cylinders/annular maps. As we
gave in Section~\ref{sec:slicegen} a new bijective derivation of this
quantity, we might try to look for a slice decomposition of maps with
more involved topologies. A first case to understand is that of
``pants'' or more generally planar maps with several boundaries, where
we may for instance attempt to reinterpret the Collet-Fusy bijective
derivation using slices.

Finally, let me mention some asymptotic questions. Since irreducible
quadrangulations are in bijection with three-connected planar graphs
---whose planar embedding is unique by Whitney's theorem--- it is
tempting to try studying their scaling limits, starting from
quadrangulations using slices or the Fusy-Poulalhon-Schaeffer
bijection~\cite{FPS08}, and going to three-connected graphs possibly
through an irreducible analogue of the Ambj{\o}rn-Budd
bijection~\cite{AmBu13,BJM14}. Then, one might try to deduce results
for general planar graphs~\cite{CFGN15}. Less speculatively,
substitution schemes for planar maps lead to very interesting
asymptotic phenomena~\cite{BFSS01} and it might be worthwhile to
investigate the case of irreducible maps.


\chapter{The $O(n)$ loop model on random planar maps}
\label{chap:on}

In this chapter we come closer to physics, by considering a model of
statistical mechanics on random maps.  The publications in this topic
are~\citemy{recuron,moreloops,pottsloop,treeloop}.  Our initial
interest came from the work of Le Gall and Miermont
on scaling limits of maps with large faces, which differ from
the Brownian map. We were particularly puzzled by the discussion
in~\cite[Section~8]{LGMi11} which suggests a connection with the
so-called $O(n)$ loop model. After some investigation we realized that the
connection becomes much simpler when ``loops are drawn on the dual'',
as we will explain in more detail below. This led us to introduce the
gasket decomposition, which is an exact combinatorial relation between
loop-decorated maps and Boltzmann maps (as defined in
Section~\ref{sec:distCF}). The gasket decomposition allows to rederive
and generalize in a transparent manner some functional equations for
the partition function of the model, which were first derived using
matrix models.

To simplify the presentation, I will focus on the specific case of the
rigid $O(n)$ loop model on quadrangulations, whose analysis is
easier. After recalling some context in Section~\ref{sec:oncont}, I
define the model and present the gasket decomposition in
Section~\ref{sec:ongasket}. I then discuss the phase diagram of the
model in Section~\ref{sec:onsol}. The material for these two sections
comes mostly from the paper~\citemy{recuron} written with Gaëtan Borot
and Emmanuel Guitter. Section~\ref{sec:onnesting} is devoted to the
statistics of nestings between loops, and is based on the more recent
preprint~\citemy{treeloop} written with G.~Borot and Bertrand
Duplantier. Concluding remarks and perspectives are gathered in
Section~\ref{sec:onconc}.

\section{Context}
\label{sec:oncont}

Let me attempt to review the literature on models of statistical
mechanics on random maps. Arguably, the most emblematic model is the
Ising model, which was solved\footnote{In the usual sense in
  statistical mechanics: computation of the partition function.}  by
Kazakov~\cite{Kazakov86} via a two-matrix integral in the case of
``dynamical'' random planar quartic maps. By dynamical it is meant
that the model is annealed: the map and its \emph{decorations} ---here
$\pm 1$ spins--- are drawn together at random, so that the marginal
probability of drawing a given map is proportional to the partition
function of the Ising model defined on it. This is the situation which
is most relevant for 2D quantum gravity, where we want the geometry to
be affected by the presence of the spins (``matter''). Quenched models
have also been studied through scaling arguments and numerical
simulations, see e.g.\ \cite{JaJo00} and references therein, but the
results are much scarcer than in the annealed case which we consider
from now on.

The remarkable feature of Kazakov's solution is that it shows the
existence of a phase transition and a critical point, as in the Ising
model on regular bidimensional lattices. The universal critical
exponents are however different from their regular lattice
counterparts~\cite{BoKa87}, but related via the
Knizhnik-Polyakov-Zamolodchikov (KPZ) relations~\cite{KPZ88}. In fact,
the Ising model was used as a testbed for the KPZ relations, see for
instance~\cite{Duplantier04} for a review of this fascinating
topic. What is most relevant for our discussion is that the ``string
susceptibility exponent'' at the Ising critical point differs from
that of pure gravity, which is why it is expected that the large-scale
geometric properties of critical Ising maps are fundamentally
different from those of usual undecorated maps.

Another, actually older, solvable model is that of maps decorated with
a spanning tree~\cite{Mullin67}. It was argued in~\cite{BKKM86} that
this is equivalent to considering Gaussian embeddings of maps in
``$-2$'' dimensions (note that Boulatov \emph{et al.}\ consider cubic
maps, and do not seem to know about Mullin's formula). This yields yet
another check of the KPZ relations, and yet another large-scale
geometry.

A common generalization of these two models is the $q$-state Potts
model~\cite{Kazakov88}: Ising corresponds to the case $q=2$ and spanning trees
(or forests) to the $q \to 0$ limit. By the Fortuin-Kasteleyn (FK)
random cluster representation, it makes sense to consider noninteger
values of $q$. The case $q=1$ corresponds to bond percolation, for
which the geometry of maps is the same as in the pure gravity case
(since the ``partition function'' of percolation is trivial). It is
however very interesting to study the percolation transition and the
geometry of interfaces, see e.g.\ \cite{Angel03,AnCu15,CuKo15}. For
any $q \in [0,4]$, there is a continuous phase transition whose
critical exponents vary with $q$. Consequently we expect a different
large-scale geometry of the maps for each $q$. Much effort has been
devoted to analyzing the Potts model on maps for general $q$,
sometimes in different languages.

\begin{enumerate}
\item In its original formulation, the Potts model amounts to
  considering $q$-colorings of maps. This topic was considered by
  Tutte himself~\cite{Tutte71}. In particular, he worked for more than
  ten years to solve some functional equations arising in the case of
  ``chromatic sums'' (i.e.\ of properly $q$-colored maps, obtained in
  the antiferromagnatic zero-temperature limit of the Potts
  model). See~\cite{Tutte95} and references therein, and
  also~\cite{Baxter01}. The combinatorial approach of Tutte, based on
  functional equations, has been recently generalized by Bernardi and
  Bousquet-Mélou to nonzero temperatures~\cite{BBM11,BBM17}. The main
  outcome is a characterization of the partition function of the
  $q$-state Potts model on random planar maps: it satisfies an
  explicit algebraic equation for $q$ of the form $2+2\cos(\pi j/m)$
  with $j,m$ integers (and $q\neq 0,4$), and it obeys an explicit
  nonlinear differential equation for general $q$. See also
  \cite{BMC15} for the already nontrivial case $q \to 0$.

  Meanwhile, in theoretical physics, the matrix integral
  representation of the Potts model was studied in several
  papers~\cite{Daul1995,Bonnet99,EyBo99,ZinnJustin00,GJSZ12}. We note
  that there is not so far much connection between the two approaches:
  for instance the existence of ``algebraic points'' is known to
  theoretical physicists, but for a slightly different notion of
  algebraicity (i.e.\ with respect to the parameter coupled to the
  boundary length in the disk generating function, while Tutte,
  Bernardi and Bousquet-Mélou consider the parameter coupled to the
  number of edges).
\item By a well-known construction, which we review
  in~\citemy[Section~2.1]{pottsloop}, the $q$-state Potts model on a
  planar map may be reformulated in terms of loops on a derived
  map. In particular, at a ``self-dual'' point, we obtain precisely
  the fully-packed version of the $O(n)$ loop model with the relation
  $q=n^2$, and the general $O(n)$ loop model corresponds to the dilute
  Potts model. Here, the natural range for observing new geometries is
  $n \in [0,2]$, or actually $[-2,2]$ if we dare consider negative
  Boltzmann weights~\cite{CaSp05}. The $O(n)$ loop model on cubic maps
  turns out to be exactly solvable using matrix integrals and loop
  equations~\cite{Kostov89,GaKo89,KoSt92,EyZJ92,EyKr95,EyKr96} or the
  gasket decomposition~\citemy{recuron,moreloops} presented in this
  chapter. The limit $n \to 0$, which describes polymers or
  self-avoiding walks, was first considered in~\cite{DuKo88}.
  The exact solvability of the $O(n)$ loop model allows to analyze
  completely its critical behavior and exponents, which again are
  consistent with the KPZ relations. More recently, the model was put
  into the framework of topological recursion~\cite{BoEy11}.
\item The self-dual $q$-state Potts model/fully-packed $O(n)$ loop
  model appears under yet another language in the
  paper~\cite{Sheffield16}, which we already mentioned in
  Section~\ref{sec:slicecontext} and which led to many further
  developments \cite{GMS15,BLR17,Chen17,GS17,GS15,GKMW18} including
  the abelian sandpile model on random maps~\cite{SW15} and bipolar
  orientations~\cite{KMSW15}. Most of these papers deal with the
  scaling limits at critical points, using the ``mating-of-trees'' or
  ``peanosphere'' construction.
\item On the square lattice, there is a well-known correspondence
  between the Potts model and the ice or ``six-vertex'' model. This
  correspondence breaks down on a random quartic map, but still there
  is a three-to-one correspondence between $3$-colorings of a planar
  quadrangulation and ice configurations/Eulerian orientations of its
  dual quartic map. The matrix integral representation of the
  six-vertex model on random quartic maps was studied
  in~\cite{Kostov00,ZJ00_6V}. More recently, Eulerian orientations of
  both quartic and general Eulerian maps were considered in
  \cite{BBMDP17,EP18,BMP18}. Here the results are mostly enumerative.
\end{enumerate}

After this long list one may wonder if a model other than the Potts
model has ever been considered on random maps. In 1989,
Kazakov~\cite{Kazakov89} realized that the one-matrix model,
corresponding to the Boltzmann maps considered in
Section~\ref{sec:distCF}, already allows to reach multicritical points
different from pure gravity by fine-tuning the weight sequences
$(g_k)_{k \geq 1}$. Unfortunately some of these weights are negative,
which makes the probabilistic interpretation difficult and the
conformal field theory ``nonunitary'', as realized by Staudacher who
noted a connection with the dimer model on random quartic
maps~\cite{Staudacher90}. More recently, Le Gall and Miermont proposed
another way of escaping from pure gravity with Boltzmann maps, by
considering a nonnegative weight sequence $(g_k)_{k \geq 1}$ with
unbounded support and fine-tuned
asymptotics~\cite{LGMi11}. Interestingly, it appears that there exists
a one-parameter family of weight sequences which interpolates between
the cases considered by Kazakov and by Le Gall and
Miermont~\cite{ABM16}. As we will explain in this chapter,
Boltzmann maps with weight sequences satisfying the assumptions
of~\cite{LGMi11} appear in the gasket decomposition of $O(n)$ loop
configurations on random maps. So, the Potts model is still lurking in
the background!

Still, a few models of statistical mechanics on random maps which are
not (immediately) related to the Potts model have been considered,
such as ADE height models \cite{Kostov92,Kostov96} and hard
particles~\cite{objetsdurs}, which are both exactly solvable.

\section{The gasket decomposition}
\label{sec:ongasket}

The $O(n)$ loop model aims at describing a ``gas of loops'' on a
lattice. By loop, we here mean a cycle (simple closed path), and not a
loop in the graph theoretical sense. As usual in statistical
mechanics, the model admits several different ``microscopic''
definitions, which are expected not to matter much at a macroscopic
level. Their common feature is that $n$ plays the role of a nonlocal
weight per loop, which arises through high-temperature graphical
expansions of models with $O(n)$ symmetry~\cite{DMNS81}. We note that
the term ``$O(n)$ loop model'' is a misnomer since there is no action
of the orthogonal group on the loops themselves, but this term is
traditionally used in the literature so we will stick with it. We now
provide the precise combinatorial (microscopic) definition on which we
will concentrate.

\begin{figure}
  \centering
  \fig{.7}{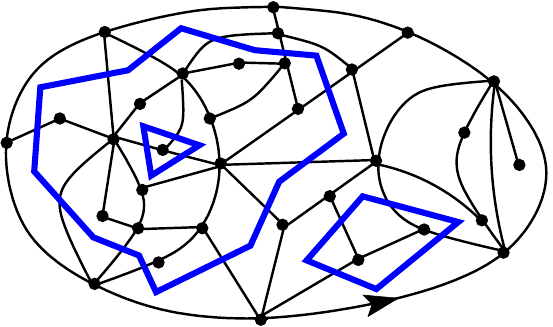}
  \caption{A (non-rigid) loop configuration on a quadrangulation with
    a boundary.}
  \label{fig:loopconfig}
\end{figure}

A \emph{loop configuration} is a collection of disjoint cycles (i.e.\
the loops are both self- and mutually avoiding). On planar maps, we
find it convenient to think of loops as living on the dual map: they
go through the faces and edges of the primal map. For simplicity, we
restrict to the case of a quadrangulation with a boundary: the loops
then live on the dual almost-quartic map, and we assume that no loop
visits the outer face.  See Figure~\ref{fig:loopconfig} for an
example.

\begin{figure}
  \centering
  \includegraphics{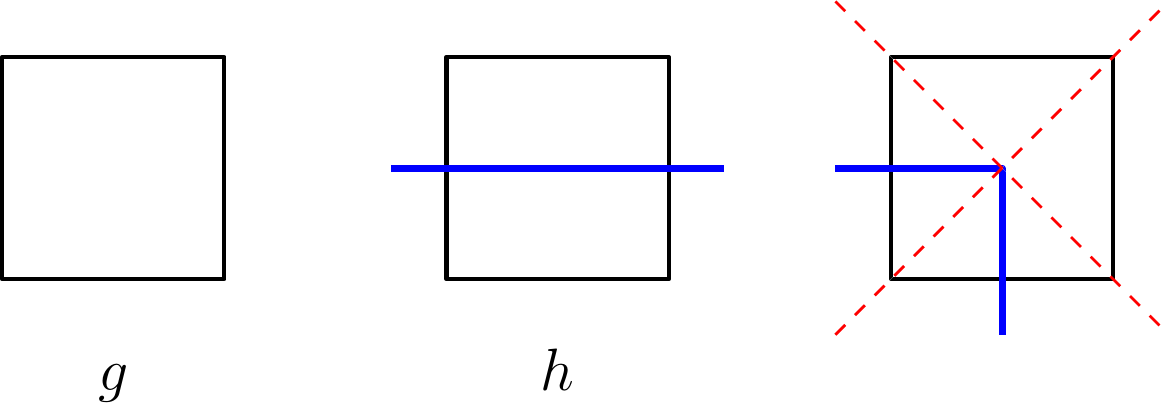}
  \caption{The possible types of faces in the rigid $O(n)$ loop model,
    and their weights.}
  \label{fig:rigidfaces}
\end{figure}

A loop configuration on a quadrangulation with a boundary is said
\emph{rigid} if, in each visited face, the loop enters and exists
through opposite edges. In other word, ``turns'' are forbidden, see
Figure~\ref{fig:rigidfaces}.  The \emph{rigid $O(n)$ loop model} is a
measure on the set of quadrangulations with a boundary which are
endowed with a rigid loop configuration. It is defined by assigning to
such a map $m$ a weight
\begin{equation}
  \label{eq:onweight}
  w(m) = n^{L(m)} g^{f_u(m)} h^{f_v(m)}
\end{equation}
where $n,g,h$ are nonnegative real parameters, and
$L(m),f_u(m),f_v(m)$ denote respectively the number of loops, the
number of unvisited inner faces and the number of visited faces of
$m$. Note that taking $n=0$ or $h=0$ amounts to forbidding the loops,
and we recover the natural weight function of quadrangulations
considered in Section~\ref{sec:dist3P}.

For $p$ a positive even integer, we denote by $Z_p$ the sum of the
weights of all loop-decorated quadrangulations with a boundary of
length $p$. We say that the triplet $(n,g,h)$ is \emph{admissible} if
$Z_p$ is finite. It is not difficult to check that the property of
being admissible does not depend on $p$ and that, for all $n$, it is
satisfied for $g$ and $h$ small enough. We set by convention $Z_0=1$.

We now describe the gasket decomposition, which makes a link with the
face-weighted Boltzmann maps defined in Section~\ref{sec:distCF}. The
general idea of viewing Boltzmann maps as ``gaskets'' of $O(n)$ loop
models actually appears already in~\cite{LGMi11}, but the precise
connection proposed by Le Gall and Miermont poses some technical
difficulties which we were able to circumvent with the idea of having
the loops on the dual map, and introducing the rigid $O(n)$ loop model
which makes the correspondence both exact and manageable.

\begin{figure}
  \centering
  \fig{.7}{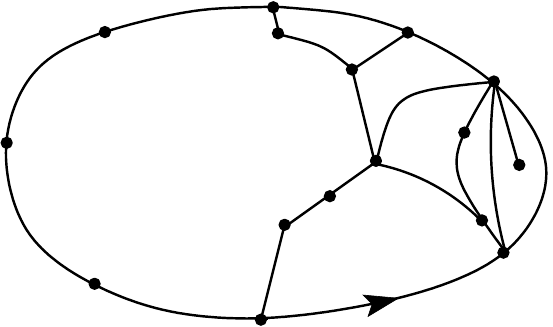}
  \caption{The gasket of the loop-decorated map of
    Figure~\ref{fig:loopconfig}.}
  \label{fig:loopconfig_gasket}
\end{figure}

Starting with a loop-decorated quadrangulation with a boundary, let us
erase all the loops and their interiors (vertices and edges) as well
as the edges they cross. See Figure~\ref{fig:loopconfig_gasket}. We
call \emph{gasket} the resulting object, it is a planar bipartite map
without loops, but possibly with inner faces of high degree. In
particular, it is not necessarily a quadrangulation. Note that the
degree of the outer face is unchanged.

Of course, we lose information in the process so this correspondence
is not bijective. But we can fix this by keeping track of the contents
of the outermost loops, which turn into faces in the gasket. Let us
consider such an outermost loop of length $\ell$. Since we are
considering a rigid loop configuration, it is not difficult to check
that $\ell$ is necessarily even, and that the gasket face associated
with the loop has degree $\ell$. We now consider the \emph{internal
  map} formed by the vertices, edges and loops inside the loop at
hand: it is nothing but a loop-decorated quadrangulation with a
boundary of length $\ell$ (as counted by $Z_\ell$), with a rooting
inherited from that of the original map. Therefore, to make the
decomposition reversible, we need to attach to each face of the gasket
coming from an outermost loop an internal map with matching outer
degree. Note that the gasket may also have \emph{regular} faces of
degree $4$, corresponding to the unvisited faces of the original
quadrangulations which are not surrounded by loops. It is not
difficult to check that the decomposition is now reversible and can be
turned into a recursive bijective decomposition, which has the
following consequences.

\begin{thm}[see {\citemy{recuron}}]
  \label{thm:onfixpt}
  For $p\geq 2$ even, denote by $\mathcal{F}_p(g_2,g_4,g_6,\ldots)$
  the generating function of rooted bipartite maps with a boundary of
  length $p$, with the Boltzmann weight~\eqref{eq:distboltz}. Then,
  the sequence $(Z_p)_{p \geq 2}$ is solution of the fixed-point
  equation
  \begin{equation}
    \label{eq:onfixpt}
    \begin{split}
      Z_p &= \mathcal{F}_p(z_2,z_4,z_6,\ldots) \\
      z_p &= g \delta_{p,4} + n h^p Z_p.
    \end{split}
  \end{equation}
  When $(n,g,h)$ is admissible, the gasket is distributed as a
  Boltzmann bipartite map with the weight sequence $(z_k)_{k \geq 2}$.
\end{thm}

Note that $\mathcal{F}_{2n}(g_2,g_4,g_6,\ldots)$ is the same quantity
as that denoted by $w_n$ in Section~\ref{sec:distCF}, we use here a
different notation to emphasize its dependency on the weight sequence.
It is clear that the fixed-point equation~\eqref{eq:onfixpt} uniquely
determines $Z_p$ and $z_p$ as formal power series in $n$, $g$ and
$h$. The quantity $z_p$ accounts for the possible ways to ``fill'' a
face of degree $p$ in the gasket: it may either be a regular face of
weight $g$ if $p=4$, or arise from an outermost loop (weight $n$)
covering $p$ visited faces ($h^p$) and containing an internal map of
outer degree $p$ ($Z_p$).

\begin{rem}
  The gasket decomposition may be adapted to situations which are more general
  than the rigid case presented here, at the price of making the
  fixed-point equation more complicated (as we have to correctly
  account for the different possible types of ``rings'' formed by the
  faces visited by a given loop). We treated the case of
  quadrangulations (not necessarily rigid) in~\citemy{recuron}, then
  extended to general degrees in~\citemy{moreloops}. This allows to
  recover in particular the $O(n)$ loop on cubic maps, previously
  studied in the references given in Section~\ref{sec:oncont}. We may
  also incorporate an extra ``bending energy'' parameter. Finally
  in~\citemy{pottsloop} we considered ``twofold'' loop models that
  arise from the correspondence with the Potts model outside the
  self-dual situation.
\end{rem}

\section{Exact solution and phase diagram}
\label{sec:onsol}

It turns out that the fixed-point equation~\eqref{eq:onfixpt} may be
``solved'', in the sense that we may rewrite it as a functional
equation whose solution may be written in parametric form using the
techniques from~\cite{Kostov89,GaKo89,EyKr95}. But, before, it is instructive
to have a qualitative discussion of the possible asymptotic behaviors
of $Z_p$ and $z_p$ as $p \to \infty$, as it makes the connection with
the work of Le Gall and Miermont~\cite{LGMi11}.

Consider a sequence $\underline{g}=(g_{2k})_{k \geq 1}$ of nonnegative
weights and set
\begin{equation}
  \label{eq:onphi}
  \varphi(x) := \sum_{k=1}^\infty \binom{2k-1}{k} g_{2k} x^k.
\end{equation}
We saw in Section~\ref{sec:distCF}, precisely at~\eqref{eq:distCFReq},
that the generating function $R$ of pointed rooted bipartite maps with
the Boltzmann weight~\eqref{eq:distboltz} satisfies $R=1+\varphi(R)$.
The weight sequence $\underline{g}$ is said \emph{admissible} if $R$
is finite and, in this case, we necessarily have $\varphi'(R) \leq 1$.
We say that $\underline{g}$ is \emph{subcritical} if $\varphi'(R)<1$,
and \emph{critical} if $\varphi'(R)=1$. This distinction plays a role
when considering the probability that the Boltzmann map has a large
number of vertices: it decays exponentially in the subcritical case,
and as a power law in the critical case. Furthermore, a critical
sequence $\underline{g}$ is said \emph{generic} if $R$ is strictly
smaller than the radius of convergence of $\varphi$, and
\emph{nongeneric} otherwise. Le Gall and Miermont exhibited examples
of nongeneric critical sequences for which
\begin{equation}
  \label{eq:ongknongen}
  g_{2k} \sim c \frac{(4R)^{-k}}{k^a}, \qquad k \to \infty
\end{equation}
where $c$ is a fine-tuned positive constant, and the exponent $a$ is
comprised between $3/2$ and $5/2$. Note that the exponential decay
factor $(4R)^{-k}$ is fixed by the requirement that $\varphi$ has
radius of convergence $R$. For such sequences, the law of the degree
of a typical face has a heavy tail (while it has an exponential tail
in the generic case), and we observe scaling limits different from the
Brownian map~\cite{LGMi11}.

\begin{figure}
  \centering
  \fig{.6}{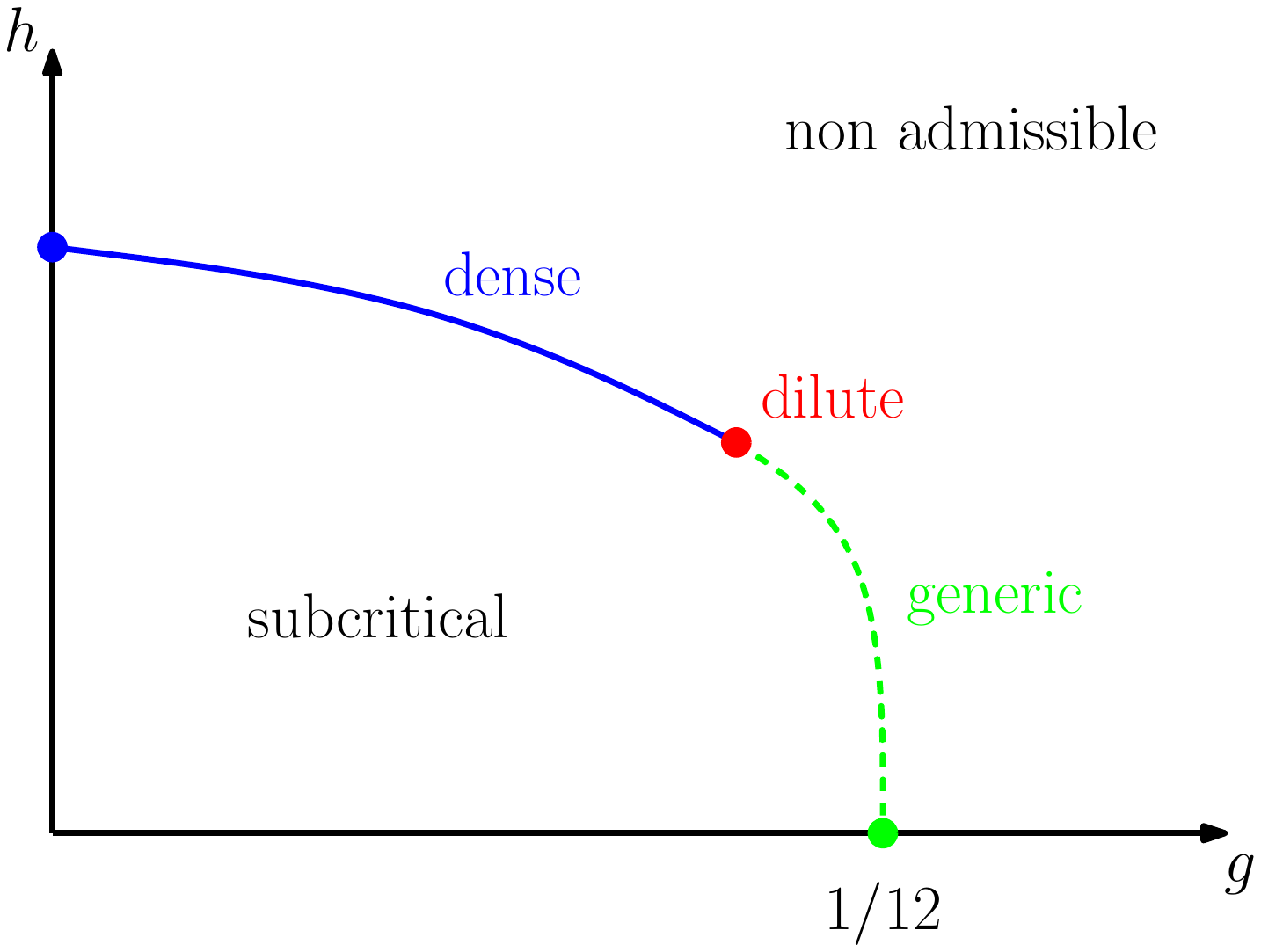}
  \caption{Qualitative phase diagram of the rigid $O(n)$ model for
    fixed $n<2$.}
  \label{fig:qualiphasediag4}
\end{figure}

How do the weight sequence $\underline{z}$ solving the fixed point
equation~\eqref{eq:onfixpt} fit into this classification? We cannot
rule out the possibility that $\underline{z}$ is subcritical or
generic critical. But, more interestingly, an elementary computation
---see~\citemy[Section~3.3]{recuron} for details--- shows that for a
nongeneric critical weight sequence satisfying~\eqref{eq:ongknongen}
we have
\begin{equation}
  \mathcal{F}_{2k}(g_2,g_4,g_6,\ldots) \sim \frac{c}{2 \cos \pi a} \frac{(4R)^{k}}{k^a},
  \qquad k \to \infty.
\end{equation}
The only way to make such asymptotics compatible
with the fixed-point equation~\eqref{eq:onfixpt} is to have the relations
\begin{equation}
  \label{eq:onngcond}
  n = 2 \cos \pi a
\end{equation}
and
\begin{equation}
  \label{eq:onngcondbis}
  h=
  \frac{1}{4R}.
\end{equation}
Thus, for $n>2$ we cannot obtain a nongeneric critical
sequence in the rigid $O(n)$ model. For $n<2$, there are two possible
values of the exponent $a$ in the allowed interval $[3/2,5/2]$, namely
\begin{equation}
  \label{eq:onasol}
  a = 2 \pm b, \qquad b:= \frac{1}{\pi} \arccos(n/2) \in [0,1/2].
\end{equation}
Following the physics terminology, the value $a=2-b \in [3/2,2]$ is
called \emph{dense} and the value $a=2+b \in [2,5/2]$ \emph{dilute}.
These two values coincide for $n=2$. To summarize, for an admissible
triplet of parameters $(n,g,h)$, the weight sequence $\underline{z}$
governing the structure of the gasket in the rigid $O(n)$ model may
either be subcritical, generic critical or nongeneric critical with
the dense or dilute value of the exponent $a$. Correspondingly, we say
that the model is at a \emph{subcritical point}, a \emph{generic
  critical point} or a dense or dilute \emph{nongeneric critical
  point}. Which situation occurs in practice is encoded in the
\emph{phase diagram} of the model. Figure~\ref{fig:qualiphasediag4}
displays the qualitative phase diagram for a fixed value of $n<2$,
which is expected from physical grounds and is confirmed by the exact
solution, as we will explain now.

We start by recalling certain well-known analytical properties of the
generating function of Boltzmann maps with a boundary, see for
instance~\citemy[Section~6]{moreloops} for more details. Let us
introduce the \emph{resolvent}
\begin{equation}
  \label{eq:onboltzres}
  \mathcal{W}(\xi) := \frac{1}{\xi} + \sum_{k \geq 1} \frac{\mathcal{F}_{2k}(g_2,g_4,g_6,\ldots)}{\xi^{2k+1}}
\end{equation}
which is closely related with the quantity $W_0$ considered in
Section~\ref{sec:distCF}. The one-cut lemma states that, when the
weight sequence $\underline{g}$ is admissible, the function
$\xi \mapsto \mathcal{W}(\xi)$ is analytic in
$\mathbb{C} \setminus [-\gamma,\gamma]$ with $\gamma:=2\sqrt{R}$
(which depends on $\underline{g}$), has singularities at $\pm \gamma$
and a cut on $[-\gamma,\gamma]$, with finite limits when one
approaches the cut from the upper or the lower half-plane. For
$\xi \in [-\gamma,\gamma]$, we denote these limits by
$\mathcal{W}(\xi+i0)$ and $\mathcal{W}(\xi-i0)$ respectively.
Note that, in
the bipartite setting considered here, $\mathcal{W}$ is an odd
function of $\xi$; in the general setting the cut may no longer be
symmetric about $0$. Furthermore, the \emph{spectral density}
\begin{equation}
  \rho(\xi) := \frac{\mathcal{W}(\xi-i0)-\mathcal{W}(\xi+i0)}{2i\pi}
\end{equation}
is a nonnegative even function on $[-\gamma,\gamma]$, vanishing at
$\pm \gamma$, and we have the functional relation
\begin{equation}
  \mathcal{W}(\xi-i0)+\mathcal{W}(\xi+i0)= \xi - \sum_{k \geq 1} g_{2k} \xi^{2k-1}, \qquad
  \xi \in [-\gamma,\gamma].
\end{equation}

Let us now substitute $\underline{g}=\underline{z}$ the solution of
the fixed point equation~\eqref{eq:onfixpt}. Then, we get that
\begin{equation}
  \label{eq:Wseries}
  W(\xi) := \frac{1}{\xi} + \sum_{k \geq 1} \frac{Z_{2k}}{\xi^{2k+1}}
\end{equation}
has a cut $[-\gamma,\gamma]$ depending on $(n,g,h)$, and the
functional relation becomes
\begin{equation}
  \label{eq:onfuneq}
  W(\xi-i0)+W(\xi+i0)=\xi - g \xi^3 - \frac{n}{h \xi^2} W\left(\frac{1}{h \xi}\right) + \frac{n}{\xi}, \qquad \xi \in [-\gamma,\gamma].
\end{equation}
This equation is very similar to that for the $O(n)$ loop model on
cubic maps, which was originally derived using matrix integrals, and
which we recovered using the gasket decomposition
in~\citemy{moreloops}. 

The functional equation~\eqref{eq:onfuneq} relates the values of $W$
on the cut $[-\gamma,\gamma]$ to that on its image through the
inversion $\xi \mapsto \frac{1}{h \xi}$. Admissibility entails that
the cut and its image have disjoint interiors, but they can possibly
meet at their endpoints when $\gamma$ is equal to the fixed point
$h^{-1/2}$ of the inversion. From the relation $\gamma=2\sqrt{R}$,
this happens precisely when the second
condition~\eqref{eq:onngcondbis} for nongeneric criticality is
satisfied.  As we shall see, the first condition~\eqref{eq:onngcond}
is then a consequence of the functional equation. In other words, we
have $\gamma \leq h^{-1/2}$ with equality if and only if the model is
a nongeneric critical point.

The key observation for solving the functional
equation~\eqref{eq:onfuneq} is that, given the value of $\gamma$, it
is an inhomogeneous \emph{linear} equation in $W$. Therefore, we have
\begin{equation}
  \label{eq:Wparthom}
  W(\xi) = W_{\mathrm{part}}(\xi)+W_{\mathrm{hom}}(\xi)
\end{equation}
where
\begin{equation}
  W_{\mathrm{part}}(\xi) := \frac{2(\xi-g \xi^3)-n\left(\frac{1}{h^2\xi^3}-\frac{g}{h^4\xi^5}\right)}{4-n^2}+\frac{n}{(2+n)\xi}
\end{equation}
is the particular solution of~\eqref{eq:onfuneq} which is a Laurent polynomial in
$\xi$, and $W_{\mathrm{hom}}$ is an odd solution of the homogeneous linear
equation
\begin{equation}
  \label{eq:onfuneqhom}
  W_{\mathrm{hom}}(\xi-i0)+W_{\mathrm{hom}}(\xi+i0) + \frac{n}{h \xi^2} W_{\mathrm{hom}}\left(\frac{1}{h \xi}\right) = 0, \qquad \xi \in [-\gamma,\gamma].
\end{equation}
The set of meromorphic solutions of this equation is a certain
infinite-dimensional vector space, and the precise solution
$W_{\mathrm{hom}}$ which we should pick is determined by the
analyticity properties of $W$. Indeed, we know that $W$ cannot have
poles, but $W_{\mathrm{part}}$ has poles at $0$ and $\infty$: these
should be therefore cancelled by $W_{\mathrm{hom}}$. Also,
by~\eqref{eq:Wseries}, we should moreover impose that
$W(\xi) \sim \frac{1}{\xi}$ for $\xi \to \infty$. Such constraints
turn out to be sufficient to fix completely $W_{\mathrm{hom}}$ given
$\gamma$.

The actual value of $\gamma$ is fixed by the
\emph{consistency condition} that the spectral density $\rho$ is
nonnegative on $[-\gamma,\gamma]$ and vanishes at $\pm \gamma$. All
the nonlinearity of the problem is hidden in this condition, as the
space of solutions of~\eqref{eq:onfuneqhom} depends on $\gamma$ in an
essential way.

Let us write the solution in the nongeneric critical case
$\gamma = h^{-1/2}$, which is both easier and more interesting. In this case the
solution of~\eqref{eq:onfuneqhom} takes the form
\begin{equation}
  \label{eq:whomcrit}
  W_{\mathrm{hom}}(\xi) =
  \left( B(\xi) - {\textstyle \frac{\gamma^2}{\xi^2} B\left(\frac{\gamma^2}{\xi} \right) }\right)
  \left(\frac{\xi - \gamma}{\xi + \gamma}\right)^b -
  \left( B(-\xi) - {\textstyle \frac{\gamma^2}{\xi^2} B\left(- \frac{\gamma^2}{\xi} \right) } \right)
  \left(\frac{\xi + \gamma}{\xi - \gamma}\right)^b
\end{equation}
where the exponent $b$ is as in~\eqref{eq:onasol}, and $B$ is a
meromorphic function which, from the analyticity requirements on $W$,
is equal to
\begin{equation}
  \label{eq:Bexpres}
  B(\xi) = \frac{g}{4-n^2} \left( \xi^3 + 2 b \gamma \xi^2 + 2 b^2 \gamma^2 \xi +
    \frac{2}{3} (b+2 b^3) \gamma^3 \right) - \frac{1}{4-n^2} (\xi + 2 b \gamma).
\end{equation}
Finally, the consistency condition yields a relation between the
parameters $n$, $g$ and $h$ of the model, which is nothing but the
equation for the nongeneric critical line.

\begin{thm}[{see \citemy[Section~6.2]{recuron} and \cite[Section~1.2.1]{Budd18}}]
  \label{thm:oncritng}
  For $n \in [0,2]$, the nongeneric critical points of the rigid
  $O(n)$ loop model are obtained for parameters $(g,h)$ such that
  \begin{equation}
    g = \frac{3}{2+b^2} \left(h - \frac{2-n}{2b^2} h^2 \right)
  \end{equation}
  and $g \leq g^*$, $h \geq h^*$, with
  \begin{equation}
    g^* := \frac{3b^2(2-b)^2}{2(2-n)(b^2-2b+3)^2}, \qquad
    h^* := \frac{b^2(2-b)^2}{(2-n)(b^2-2b+3)}
  \end{equation}
  and $b$ as in~\eqref{eq:onasol}. The point $(g^*,h^*)$ is a dilute
  critical point, and all the other points are dense critical points,
  including the ``full-packed'' critical point $(0,\frac{2b^2}{2-n})$.
\end{thm}

Identifying the dilute and dense critical exponents is rather
straightforward from~\eqref{eq:whomcrit}, as it yields the expansion
of $W$ around the singularities at $\pm \gamma$, which are related
with the asymptotic behavior of $Z_{2k}$ for $k \to \infty$.

It is also possible to obtain a parametrization for the generic
critical line displayed on Figure~\ref{fig:qualiphasediag4}, which
connects the point $(g,h)=(1/12,0)$ to the dilute critical point. It
is more complicated since, for $\gamma<h^{-1/2}$, the solution of the
functional equation~\eqref{eq:onfuneqhom} passes through the use of an
elliptic parametrization, see~\citemy[Sections~6.3 and 6.4]{recuron}
for details.

Let me mention that the derivation of Theorem~\ref{thm:oncritng} which
we gave in~\citemy{recuron} was based on numerical evidence which we
did not fully justify. More precisely, we assumed that the consistency
condition, which we use to identify the correct value of $\gamma$, is
not only necessary but also sufficient.  This has been proved
rigorously by Timothy Budd and Linxiao Chen in~\cite[Theorems~1 and
2]{Budd18}. In a nutshell, the first theorem, due to T.~Budd, asserts
that the model is admissible if the fixed-point
equation~\eqref{eq:onfixpt} admits a solution in the space of
nonnegative real sequences (which is then unique), while the second
theorem, due to L.~Chen, asserts that such solution exists if the
functional equation~\eqref{eq:onfuneq} admits a solution satisfying
the consistency condition\footnote{In fact, L.~Chen proved that it is
  sufficient to check the positivity of the spectral density in the
  vicinity of~$\pm \gamma$, which is the actual criterion we used
  in~\citemy{recuron}.} (which is then unique). See also the
discussion in Linxiao's thesis~\cite[Chapter~II]{theseChen}.

In the papers~\citemy{moreloops} and \citemy{pottsloop}, we follow a
similar approach to analyze the phase diagrams of the $O(n)$ loop
model on cubic maps with bending energy and of the ``twofold'' loop
models, respectively. These are the most general models that we may
solve by the techniques presented in this section. New insights are
needed to solve models where the loops may visit faces of higher
degree.

By combining Theorems~\ref{thm:onfixpt} and \ref{thm:oncritng} with
the results from~\cite{LGMi11}, we obtain some interesting information
about the geometry of the gasket in the rigid $O(n)$ loop model for
$n<2$: at the nongeneric critical points, whose existence has been
shown, the gasket is distributed as a Boltzmann bipartite map with a
nongeneric critical weight sequence $\underline{z}$ such that
$z_{2k} \sim c \frac{\gamma^{-2k}}{k^{2 \pm b}}$ where $b$ is as in
\eqref{eq:onasol}, and where the $+$ and $-$ sign is taken in the
dilute and dense case respectively. Therefore, by the results of Le
Gall and Miermont, if we condition the gasket to have a large number
$U$ of vertices, the typical distance between gasket vertices is of
order $U^{3 \pm 2b}$, and if we divide by this factor the gasket
converges (at least along subsequences) in the Gromov-Hausdorff sense
as $U \to \infty$ towards a random compact metric space of Hausdorff
dimension $3 \pm b \in (2,4)$. This limit is different from the
Brownian map which has dimension $4$.

\section{Nesting statistics}
\label{sec:onnesting}

So far we have obtained a relatively precise information about the
structure of the gasket, but it is natural to ask also for geometric
information about the full map with loops. It seems difficult to
handle distances in the full map by our approach but, interestingly,
we may obtain information about the \emph{nestings} between loops.

\begin{figure}[t]
  \centering
  \fig{}{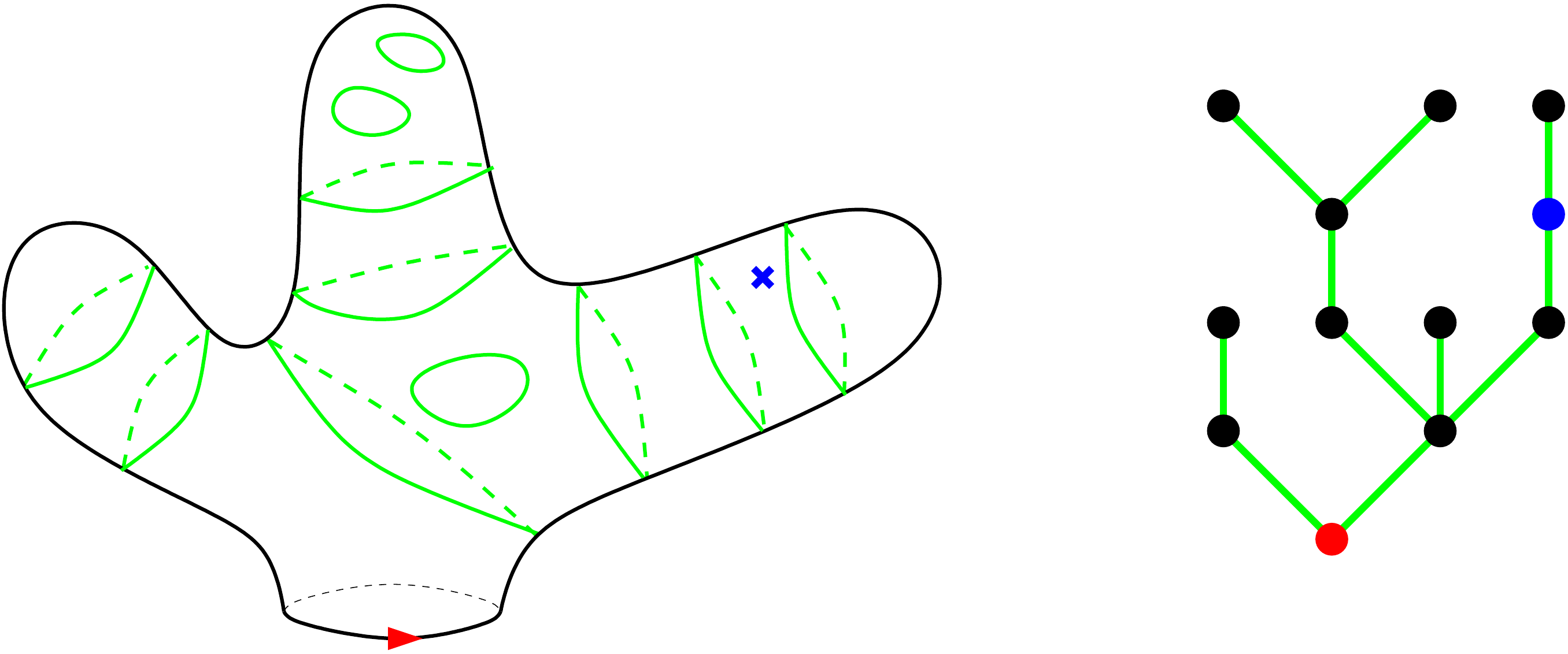}
  \caption{Left: schematic representation of a pointed loop-decorated
    quadrangulation with a boundary (the red arrow and the blue cross
    represent respectively the root edge and the distinguished
    vertex). Right: the corresponding nesting tree.}
  \label{fig:nestingtreedepth}
\end{figure}

Let us consider a loop-decorated quadrangulation with a boundary. The
associated \emph{nesting tree} is the graph whose edges and vertices
(hereafter called \emph{nodes}) correspond respectively to the loops
and to the regions which they delimit on the underlying surface of the
quadrangulation; the incidence relations in the nesting tree keep
track of those in the surface. See Figure~\ref{fig:nestingtreedepth}
for an illustration. The nesting tree is indeed a tree by the
planarity of the surface\footnote{The higher genus case, where the
  nesting graph is not necessarily a tree, has been considered
  in~\cite{BoGF16}.} and by the fact the loops are disjoint
cycles. The \emph{root node} of the nesting tree corresponds to the
region incident to the boundary, which may be identified with the
gasket. We are interested in understanding the structure of the
nesting tree, especially at a nongeneric critical point where we
expect it to have an nontrivial structure.

In the paper~\citemy{treeloop} we considered the situation where the
loop-decorated map has an extra marking, such as a distinguished
vertex or another boundary. The extra marking turns into a
distinguished node in the nesting tree, which may differ from the root
vertex. The \emph{depth} is defined as the distance between the root
and distinguished nodes in the nesting tree; it corresponds to the
minimal number of loops that one must cross to go from the root to the
extra marking in the loop-decorated map.

For simplicity I will restrict to the case where the extra marking is
a vertex, which we pick uniformly at random in the loop-decorated
map. Note that the corresponding node is in general \emph{not} uniform
in the nesting tree. If the loop-decorated map is itself picked at
random, according to the rigid $O(n)$ loop model measure with a
prescribed boundary length $2k$, the depth is a random variable which
we denote by $P_k$.

\begin{thm}[{see \citemy{treeloop}}]
  \label{thm:treeloop}
  At a nongeneric critical point of the rigid $O(n)$ loop model with
  $n \in (0,2)$, the depth $P_k$ of a uniformly chosen vertex in a
  loop-decorated quadrangulation of boundary length $2k$ grows
  logarithmically with $k$ as $k \to \infty$.
  More precisely, we have the convergence in distribution
  \begin{equation}
    \label{eq:treeloopCLT}
    \frac{P_k- \frac{n}{\pi \sqrt{4-n^2}} \ln k}{\sqrt{\ln k}}
    \overset{d}{\to} N(0,\sigma^2) \qquad (k \to \infty)
  \end{equation}
  where $N(0,\sigma^2)$ is the normal distribution with mean $0$ and
  variance $\sigma^2 := \frac{4n}{\pi (4-n^2)^{3/2}}$, and we have the
  large deviation estimate
  \begin{equation}
    \label{eq:treeloopLDP}
    \ln \Prob\left( P_k = \left\lfloor \frac{p}{\pi} \ln k \right\rfloor \right)
    \sim -\frac{J(p)}{\pi} \ln k \qquad (k \to \infty, \ p \geq 0)
  \end{equation}
  where the rate function $J$ reads
  \begin{equation}
    \label{eq:Jpdef}
    J(p) := p \ln \left( \frac{2 p}{n \sqrt{1+p^2}} \right) + \arccot(p)-\arccos(n/2).
  \end{equation}
\end{thm}

\begin{figure}
  \centering
  \fig{.6}{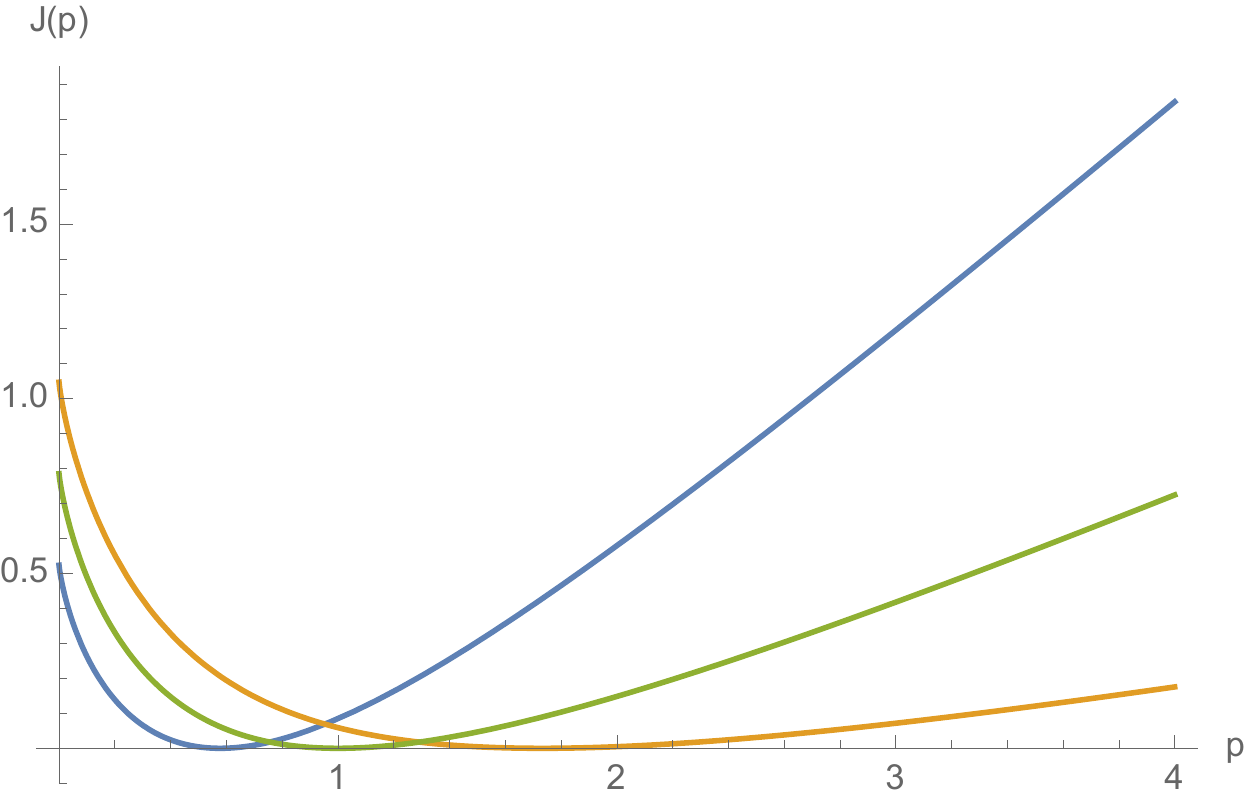}
  \caption{A plot of the rate function $J$ for $n=1$
    (blue), $n=\sqrt{2}$ (green) and $n=\sqrt{3}$ (yellow).}
  \label{fig:Jplot}
\end{figure}

The rate function $J$ is plotted on Figure~\ref{fig:Jplot} for
a few values of $n$. It is strictly convex---we have
$J''(p)=\frac{1}{p(p^2+1)}>0$---and attains it minimum at
$p=\frac{n}{\sqrt{4-n^2}}$ consistently
with~\eqref{eq:treeloopCLT}. At $p=0$ it takes the value
$\arcsin(n/2)$. Note that, remarkably, the statement of the theorem
does not depend on the dilute/dense nature of the nongeneric critical
point.

Theorem~\ref{thm:treeloop} differs slightly from the results stated
in~\citemy{treeloop}. First, we consider here the rigid $O(n)$ loop
model instead of the $O(n)$ loop model on cubic maps with bending
energy. This change is rather innocent since the computations are
entirely similar. Second, we consider the asymptotics for a large
perimeter $2k$ while, in~\citemy{treeloop}, we condition the
loop-decorated map to contain a fixed number $V$ of vertices and
consider the limit $V \to \infty$ (keeping either the perimeter $2k$
finite, or having it grow with $V$ in an appropriate ``crossover''
scaling). This change somewhat simplifies the proof of
Theorem~\ref{thm:treeloop} which we will sketch in the remainder of
this section (see Remark~\ref{rem:Vcond} for a discussion of the
conditioning).

The key ingredient in the proof of Theorem~\ref{thm:treeloop} is an
exact expression for a refined generating function keeping track of
the depth of the marked vertex. More precisely, for $k$ a nonnegative
integer and $s$ a formal or nonnegative real variable, we set
\begin{equation}
  \label{eq:Zpdef}
  Z_{2k}^\bullet[s] := \sum_m w(m) s^{P(m)}
\end{equation}
where the sum runs over all \emph{pointed} loop-decorated
quadrangulations $m$ with a boundary of length $2k$, the weight $w(m)$
is as in~\eqref{eq:onweight}, and $P(m)$ denotes the depth of the
distinguished vertex. The probability generating function of $P_k$ is
then given by
\begin{equation}
  \label{eq:Pkpgf}
  \E\left(s^{P_k}\right) = \frac{Z_{2k}^\bullet[s]}{Z_{2k}^\bullet}
\end{equation}
where the unrefined generating function
$Z_{2k}^\bullet=Z_{2k}^\bullet[1]$ is related to the generating
function $Z_{2k}$ considered in the previous sections by
\begin{equation}
  Z_{2k}^\bullet = \left( g \frac{\partial}{\partial g} + h \frac{\partial}{\partial h} + k + 1 \right) Z_{2k}.
\end{equation}
Indeed, by Euler's relation, the number of vertices of a
loop-decorated quadrangulation $m$ with a boundary of length $2k$ is
equal to $f_u(m)+f_v(m)+k+1$.

Now, by differentiating the fixed point equation~\eqref{eq:onfixpt},
we find that $Z_{2k}^\bullet$ satisfies the relation
\begin{equation}
  \label{eq:Zbulleteq}
  Z_{2k}^\bullet = \mathcal{F}_{2k}^\bullet(z_2,z_4,z_6,\ldots) + n \sum_{\ell=0}^\infty
  h^{2\ell} \mathcal{F}^\circ_{2k,2\ell}(z_2,z_4,z_6,\ldots) Z_{2\ell}^\bullet
\end{equation}
where $\mathcal{F}_{2k}^\bullet(g_2,g_4,g_6,\ldots)$ denotes the
generating function of pointed rooted bipartite maps with a boundary
of length $2k$, and
$\mathcal{F}^\circ_{2k,2\ell}(g_2,g_4,g_6,\ldots)=\frac{\partial}{\partial
  g_{2\ell}} \mathcal{F}_{2k}(g_2,g_4,g_6,\ldots)$ that of annular
bipartite maps with boundaries of lengths $2k$ and $2\ell$, the
boundary of length $2k$ being rooted\footnote{If we choose to also
  root the other boundary for symmetry, then we have to divide the
  term of index $\ell$ in~\eqref{eq:Zbulleteq} by $2\ell$. This is the
  convention that we follow in~\citemy{treeloop}, where the cylinder
  generating function $\mathcal{F}^{(2)}$ which we consider is related
  to the current $\mathcal{F}^\circ$ by
  $\mathcal{F}^{(2)}_{2k,2\ell}=2\ell
  \mathcal{F}^\circ_{2k,2\ell}$.}. The combinatorial meaning
of~\eqref{eq:Zbulleteq} is the following. The first term in the
right-hand side corresponds to the case where the marked vertex
belongs to the gasket. The second term corresponds to the opposite
case: then, we consider the outermost loop that separates the marked
vertex from the boundary. Denoting the length of this loop by $2\ell$,
we find that, in the gasket decomposition, the gasket has a
distinguished face of degree $2\ell$, inside which we find a pointed
internal map with outer degree $2\ell$. This explains the different
factors appearing in the sum over $\ell$.

We then claim that the refined generating function $Z_{2k}^\bullet[s]$
satisfied the modified equation
\begin{equation}
  \label{eq:Zbulletseq}
  Z_{2k}^\bullet[s] = \mathcal{F}_{2k}^\bullet(z_2,z_4,z_6,\ldots) + n s \sum_{\ell=0}^\infty
  h^{2\ell} \mathcal{F}^\circ_{2k,2\ell}(z_2,z_4,z_6,\ldots) Z_{2\ell}^\bullet[s]
\end{equation}
where the weight sequence $\underline{z}$ does not depend on $s$, and
is still given by the fixed point equation~\eqref{eq:onfixpt}. Indeed,
if we consider a map $m$ contributing to $Z_{2k}^\bullet[s]$, it
contributes to the first term of the right-hand side if $P(m)=0$, and
to the second term if $P(m) \geq 1$. In this latter case, we have
$P(m)=P(m')+1$, where $m'$ is the internal map containing the marked
vertex.

We now form the series
\begin{equation}
  W^\bullet_s(\xi) := \frac{1}{\xi} + \sum_{k \geq 1} \frac{Z_{2k}^\bullet[s]}{\xi^{2k+1}}.
\end{equation}
For $|s| \leq 1$, it is analytic in the same domain as the series
$W(\xi)$ considered in Section~\ref{sec:onsol}, namely
$\C \setminus [-\gamma,\gamma]$ ($\gamma$ being independent of $s$).
The equation~\eqref{eq:Zbulletseq} then translates into the functional equation
\begin{equation}
  \label{eq:onfuneqs}
  W^\bullet_s(\xi-i0)+W^\bullet_s(\xi+i0)=- \frac{n s}{h \xi^2} W^\bullet_s\left(\frac{1}{h \xi}\right) + \frac{n s}{\xi}, \qquad \xi \in [-\gamma,\gamma],
\end{equation}
see~\citemy[Section~4]{treeloop} for details. This functional equation
may be solved along the same lines as~\eqref{eq:onfuneq}. In
particular, the homogeneous linear equation is the same
as~\eqref{eq:onfuneqhom}, except that $n$ is replaced by $n s$.
At a nongeneric critical point, the solution takes the form
\begin{equation}
  \label{eq:Wbulletsol}
  W^\bullet_s(\xi) = \frac{n s}{(2 + n s) \xi} + \frac{1}{2 (2 + n s) b(s) \gamma}
  \left( \left(\frac{\xi + \gamma}{\xi - \gamma}\right)^{b(s)} -
    \left(\frac{\xi - \gamma}{\xi + \gamma}\right)^{b(s)} \right)
\end{equation}
with
\begin{equation}
  \label{eq:bsol}
  b(s) := \frac{1}{\pi} \arccos(n s/2).
\end{equation}
Note that the expression for $W^\bullet_s(\xi)$ is much more compact
than that for $W(\xi)$, cf~\eqref{eq:whomcrit}
and~\eqref{eq:Bexpres}. This is again a manifestation of the general
principle that the generating functions of pointed rooted maps are
simpler than those of maps which are just rooted\footnote{In
  particular, for $s=0$---which amounts to forcing the marked vertex
  to be in the gasket---we have
  $W^\bullet_0(\xi)=\frac{1}{\xi \sqrt{1-(\gamma/\xi)^2}}$ consistently with
  the universal form of the series $\mathcal{W}^\bullet(\xi)$ of
  pointed rooted Boltzmann maps.}. Close to the endpoints of the
cut, we have the asymptotic equivalent
\begin{equation}
  W^\bullet_s(\xi) \sim \pm 
  \frac{2^{b(s)-1}}{(2+n s)b(s)\gamma}
  \cdot (1 \mp \gamma/\xi)^{-b(s)} \qquad (\xi \to \pm \gamma)
\end{equation}
which implies (using the analyticity
of $W^\bullet_s$ in $\C \setminus [-\gamma,\gamma]$) the asymptotics
\begin{equation}
  Z_{2k}^\bullet[s] \sim \frac{4^{b(s)}}{2(2+n s)\Gamma(1+b(s))} \cdot \frac{\gamma^{2k}}{k^{1+b(s)}} \qquad (k \to \infty).
\end{equation}
Note that this asymptotic behavior does not depend on the dense/dilute
nature of the nongeneric critical point, and that the subleading
exponent $1+b(s)$ takes the value $3/2$ at $s=0$ regardless of $n$.
By~\eqref{eq:Pkpgf}, we find that the probability generating function
of $P_k$ satisfies
\begin{equation}
  \E\left(s^{P_k}\right) \sim \text{cst} \cdot k^{b-b(s)}.
\end{equation}
Theorem~\ref{thm:treeloop} follows by standard arguments of analytic
combinatorics: the asymptotic normality~\eqref{eq:treeloopCLT} follows
from the quasi-powers theorem~\cite[Theorem~IX.8]{FlSc09}, and the
large deviation estimate~\eqref{eq:treeloopLDP} is given by
considering the Legendre transform of the function
$t \mapsto b-b(e^t)$, which is nothing but
\begin{equation}
  \frac{J(p)}{\pi} = \inf_{t \in (-\infty,\ln(2/n))} \left( b-b(e^t) - \frac{p}{\pi} t \right).
\end{equation}

\begin{rem}
  \label{rem:outsideng}
  Outside the nongeneric critical situation, it should be possible to
  obtain an expression for $W^\bullet_s(\xi)$ using the same elliptic
  parametrization as for $W(\xi)$. We expect to find an asymptotic
  behavior of the form
  $Z_{2k}^\bullet[s] \sim C(s) \frac{\gamma^{2k}}{k^{3/2}}$ for
  $k \to \infty$, with only the constant depending on $s$. This would
  imply that $P_k$ has a discrete limit law as $k \to \infty$.
\end{rem}

\begin{rem}
  \label{rem:Vcond}
  Conditioning on having a fixed number $V$ of vertices, as we do
  in~\citemy{treeloop}, is more involved since we need to extract from
  the refined generating function $Z_{2k}^\bullet[s]$ the contribution
  of maps with $V$ vertices, which may be done by introducing an
  auxiliary weight per vertex and performing a contour integral. The
  limit $V \to \infty$ may then be studied by a bivariate saddle-point
  analysis. Remarkably, we find that Theorem~\ref{thm:treeloop} holds
  with very little modification in this setting: set $c=1$ if we are
  at a dilute critical point, and $c=1/(1-b)$ otherwise. Then, in the
  regime $V \to \infty$ with $k$ fixed, we simply have to replace
  $\ln k$ by $c \ln V$ in \eqref{eq:treeloopCLT} and
  \eqref{eq:treeloopLDP}. And in the crossover regime $V \to \infty$
  with $k \propto V^{c/2}$, the statements hold without modification.
\end{rem}

Interestingly, we found that the rate function $J$ of
Theorem~\ref{thm:treeloop} also appears when considering the so-called
Conformal Loop Ensembles (CLE) coupled to Liouville quantum
gravity. More precisely, our starting point was the expression of the
\emph{multifractal spectrum of extreme nesting in CLE} which has been
computed by Miller, Watson and Wilson~\cite{MWW16}. In a nutshell, it
is the function that associates to $\nu>0$ the Hausdorff dimension of
the set of points $z \in \C$ such that the number of CLE loops
surrounding the ball $B(z,e^{-r})$ grows as $\nu r$ for
$r \to \infty$. To obtain our $J$, we need to replace the Euclidean
radius $e^{-r}$ of the ball by its \emph{Liouville quantum
  measure}. This is done by applying a functional version of the KPZ
relations, see~\citemy[Section~2.6]{treeloop} for details.

To conclude this section, let me mention that our results have been
complemented by Chen, Curien and Maillard~\cite{CCM17}---see
also~\cite[Chapter~III]{theseChen}---who showed that, when the
perimeter $2k$ of the loop-decorated map tends to infinity, the
nesting tree may be described by a certain~\emph{multiplicative
  cascade}. Their approach consists in keeping track of the lengths of
the longest outermost loops, and showing that they scale linearly with
$k$, with random proportionality constants whose law is explicit.

\section{Conclusion and perspectives}
\label{sec:onconc}

In this chapter we have presented the gasket decomposition and some of
its applications to the study of the $O(n)$ loop model on random
maps. Many other results on this model have been obtained recently by
Timothy Budd~\cite{Budd17talk,Budd17,Budd18}. In particular, while
studying the peeling process of loop-decorated random maps, he
uncovered a surprising connection with walks on the square lattice,
which allows to study their winding angles around the origin.

Let me now list a few directions for future research. First, it would be
interesting to make a connection between our approach and that
developed by Bernardi and Bousquet-Mélou (recall that the $O(n)$ loop
model is intimately connected with the Potts model, as mentioned in
Section~\ref{sec:oncont}). Do the techniques discussed in this chapter
allow to recover the same algebraicity results as those~\cite{BBM11}?
And how does differential algebraicity~\cite{BBM17} arise in this
context? Another challenging question is to extend the gasket
decomposition in the presence of degree constraints on the
vertices. In particular, the results from~\cite{EyBo99} still await a
combinatorial explanation.

The topic of maps with large faces, which was my initial motivation
for considering the $O(n)$ loop model, is still a very active domain
of research~\cite{BuCu17,Richier18,CR18,Mar18b}. From the point of
view of statistical physics, it is still not completely understood how
the phase transition between the dense and dilute regimes manifests
itself at the geometric level. More precisely, in the qualitative
discussion at the beginning of Section~\ref{sec:onsol}, the phase
transition occurs when the exponent $a$ appearing
in~\eqref{eq:ongknongen} takes the value $2$. The dense and dilute
phases corresponds to the respective cases $a<2$ and $a>2$. By analogy
with the $O(n)$ loop model on regular lattices (whose scaling limit at
a critical point is believed to be a conformal loop ensemble), we
expect the contours of the faces to be simple in the scaling limit for
$a \geq 2$, and nonsimple for $a<2$. Some results in that direction were
obtained in~\cite{Richier18} but, to my knowledge, the question is not
settled. A very interesting duality between the dense and dilute
regimes has been studied in~\cite{CR18}.

Finally, I would like to extend the gasket decomposition to other
models of statistical physics on random maps such as the six-vertex
model---see the references given in Section~\ref{sec:oncont} and the
recent preprint~\cite{BMEPZJ19}---and the ADE height
model~\cite{Kostov92,Kostov96}. Some preliminary investigation shows
that our approach may indeed be adapted to this setting, with new
technical challenges to be overcome. For instance, loop
representations of the six-vertex model typically involve complex
weights, whose probabilistic interpretation is unclear. The primary
motivation is to gain geometric insight on these models.


\chapter{Schur processes}
\label{chap:dominos}

In this chapter, we leave the the realm of planar maps and enter that
of Schur processes. They are certain measures over sequences of
partitions, with deep connections with models of statistical mechanics
such as lozenge or domino tilings, last passage percolation (LPP) and
the totally asymmetric simple exclusion process (TASEP). The publications
closest to this topic
are~\citemy{pyramids,dimerstat,sampling,freeboundaries,cylindricschur,BBNVfpsac2019}.
In this chapter I will present my contributions from a personal
perspective.

After recalling some context in Section~\ref{sec:dominoscont}, I
discuss in Section~\ref{sec:dominosschur} how Schur processes arise in
connection with a family of domino tilings called \emph{steep
  tilings}~\citemy{pyramids}, that are equivalent to dimer coverings
of \emph{rail yard graphs}~\citemy{dimerstat}. In
Section~\ref{sec:dominosbc}, I explain how to compute the correlation
functions in the presence of periodic or free boundary conditions
using the free fermion formalism. Then, in
Section~\ref{sec:dominosasymp}, I discuss the corresponding
asymptotics. The material for these two sections comes
from~\citemy{freeboundaries,cylindricschur}. I conclude with some
perspectives in Section~\ref{sec:dominosconc}.

\section{Context}
\label{sec:dominoscont}

Let me start by telling how I became interested in the subject: in
2011-2012 I was spending the academic year at
LIAFA\footnote{Laboratoire d'informatique algorithmique : fondements
  et applications, now merged into IRIF (Institut de recherche en
  informatique fondamentale)}, and participating in their \emph{groupe
  de lecture de combinatoire} which that year was on the topic of
integrable hierarchies, following the book by Miwa, Jimbo and
Date~\cite{MJD00}. The initial motivation was to understand their
connection with map generating functions, but I actually did not
pursue further work in that direction\footnote{Unlike my colleague
  Guillaume Chapuy\cite{CaCh15,ACEH18,ChapuyHDR}}. However, in Spring
2012 I also visited the Mathematical Sciences Research Institute in
Berkeley during the program \emph{Random spatial processes}, where I
learnt about the work of Benjamin Young on pyramid
partitions~\cite{Young09}. Back in Paris, together with Guillaume
Chapuy and Sylvie Corteel, we were struck that the fact that the
\emph{vertex operator formalism}, that plays a key role in integrable
hierarchies, is also useful for the study of pyramid
partitions~\cite{Young10}. Also, we were puzzled by the similarity
between the expression of the generating function of pyramid
partitions and that of domino tilings of the Aztec
diamond~\cite{EKLP92a,EKLP92b}. This led us to introduce the notion of
steep tilings~\citemy{pyramids}, which encompasses both families and
may be enumerated by the vertex operator formalism.  After Sylvie
presented our work at Aléa, we were joined by Cédric Boutillier and
Sanjay Ramassamy, with whom we introduced rail yard
graphs~\citemy{dimerstat}. Dimer models on rail yard graphs are
essentially equivalent to steep tilings, as we shall see in
Section~\ref{sec:dominosschur}, but they come with a coordinate system
which is more convenient for expressing the so-called correlation
functions. These models are, in some sense, the most general
combinatorial realizations of Schur processes.

Schur processes were introduced by Okounkov and
Reshetikhin~\cite{OkRe03,OkRe07} as a generalization of Okounkov's
Schur measures~\cite{Okounkov01}. They also appear, more or less
implicitly, in the works of
Johansson~\cite{Johansson02,Johansson03,Johansson05}. Together with
their nondeterminantal generalization called Macdonald
processes~\cite{BoCo14}, they are central models in the field known as
\emph{integrable probability}~\cite{Borodin14,BoGo16}. I will not
attempt to review the abundant literature in this field but, besides
the references cited in the previous papers, point to the book by
Romik~\cite{Romik15}. It gives a nice account of the story of the
\emph{Ulam--Hammersley problem}, which consists in studying the
asymptotic distribution of the length of a longest increasing
subsequence of a random permutation, and which is intimately connected
with the \emph{Plancherel measure on integer
  partitions}~\cite{BaDeJo99,BOO00,Johansson01}, the simplest instance of a
Schur measure.

The vertex operator formalism, also known as the free fermion
formalism, the (semi) infinite wedge, the fermionic Fock space, the
boson-fermion correspondence, etc, is a classical topic in
mathematical physics, see e.g.~\cite{JiMi83,MJD00,AlZa13},
\cite[Chapter~14]{Kac90} or \cite[Chapter~1]{ZinnHDR}. Its
probabilistic application for the study of Schur measures and
processes was pioneered by Okounkov~\cite[Appendix~A]{Okounkov01}, see
also \cite{PrSp02}. In contrast, other authors such as
Johansson~\cite{Johansson01} and
Borodin--Rains~\cite{Rains00,BoRa05,Borodin07} used variants of the
orthogonal polynomial method developed in random matrix theory.

Since my own entry in the subject was through the vertex operator
formalism, I found it natural to continue using it. In our work on
steep tilings~\citemy{pyramids}, we understood how to apply it to
enumerate tilings with periodic or free boundary conditions. But it
was unclear how to extend the approach to handle the correlation
functions. This is the problem that we solved in the
papers~\citemy{freeboundaries,cylindricschur}, which will be reviewed
in Section~\ref{sec:dominosbc}.

\section{From domino tilings to Schur processes}
\label{sec:dominosschur}

\begin{figure}[t]
  \centering
  \includegraphics[scale=0.6]{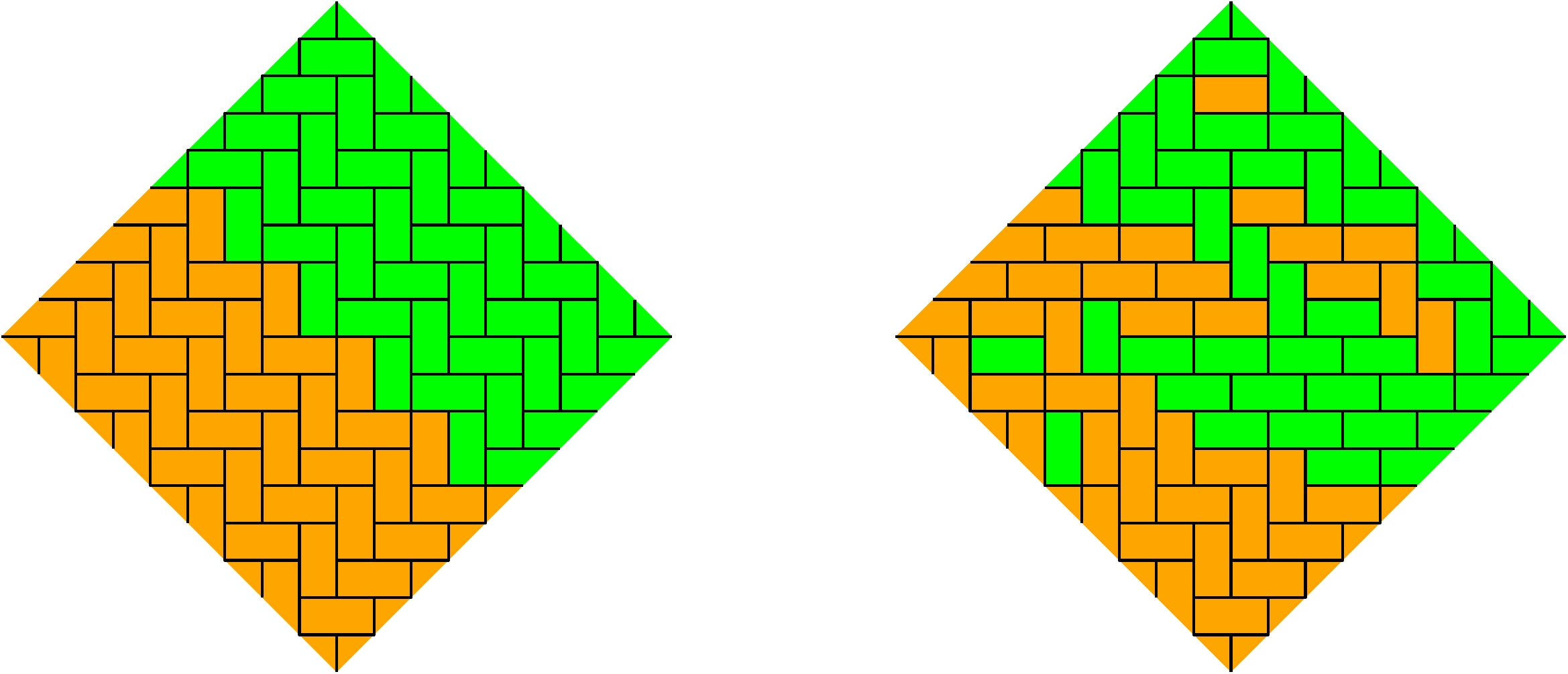}
  \caption{A portion of the fundamental tiling (left), and a pure
    steep tiling (right) corresponding to a perturbation of it by
    finitely many ``flips''. North and east (resp.\ south and west)
    dominoes are represented in green (resp.\ orange). The fundamental
    tiling is periodic along the north-west to south-east direction,
    along the perpendicular direction it is almost periodic (notice
    the ``zig-zag'' pattern gets shifted at the green-orange
    interface, which is the only place where we can perform initial
    flips). A general steep tiling coincides with the fundamental
    tiling far away from the interface.}
  \label{fig:universalSteepTiling}
\end{figure}

In this section we explain how Schur processes arise in connection
with a certain family of domino tilings, that we call \emph{steep
  tilings}, and that are in bijection with \emph{dimer coverings of
  rail yard graphs}. The definitions in this section differ slightly
from those in~\citemy{pyramids} and~\citemy{dimerstat}, see
Remark~\ref{rem:dominosconnect}. The term ``steep'' comes from the
fact that the associated height function, which we will not discuss
here for brevity, has eventually the maximal possible slope in two
opposite directions.

Consider the domino tiling represented on the left of
Figure~\ref{fig:universalSteepTiling}, which we call the
\emph{fundamental tiling}.
We choose the Cartesian coordinate system $(x,y)$ such that the
corners of dominoes belong to the lattice $\Z^2$ and, by convention,
the origin $(0,0)$ is the point at the center of the figure (hence is
a corner common to four different dominoes). A point of $\Z^2$ is said
\emph{odd} or \emph{even} depending on the parity of the sum of its
coordinates. Parity allows to distinguish four types of dominoes: a
horizontal (resp.\ vertical) domino is said \emph{north} (resp.\
\emph{east}) if its top-left corner is odd, and \emph{south} (resp.\
\emph{west}) otherwise. In the fundamental tiling, the region
$x+y \geq 1$ contains only north and east dominoes, while the region
$x+y \geq -1$ contains only south and west dominoes.

A general \emph{steep tiling} is obtained from the fundamental tiling
by performing flips. A~\emph{flip} is the elementary operation which
consists in replacing a pair of parallel dominoes forming a
$2 \times 2$ block by a pair of dominoes with the opposite
orientation. Note that, initially in the fundamental tiling, the only
flippable blocks are centered at the positions $(n,n)$ with $n$ an odd
integer. But, after such flips have been performed, new flippable
blocks appear. An example of steep tiling is represented on the right
of Figure~\ref{fig:universalSteepTiling}. A steep tiling is
said~\emph{pure} if it may be obtained from the fundamental tiling by
finitely many flips, hence it coincides with it outside a finite
region. For a general steep tiling, we require that the number of
flips centered on the line $x-y=n$ be finite for each $n$, but the
total number of flips may be infinite. Informally this says that a
steep tiling coincides with the fundamental tiling sufficiently far
away from the line $x+y=0$. Such more general setting is necessary
when considering periodic steep tilings, for instance.
Keeping track of the number of flips on each line amounts to putting
certain weights on dominoes, but these weights are easier to define in
the language of rail yard graphs, which we now present.

\begin{figure}[t]
  \centering
  \includegraphics[scale=0.6]{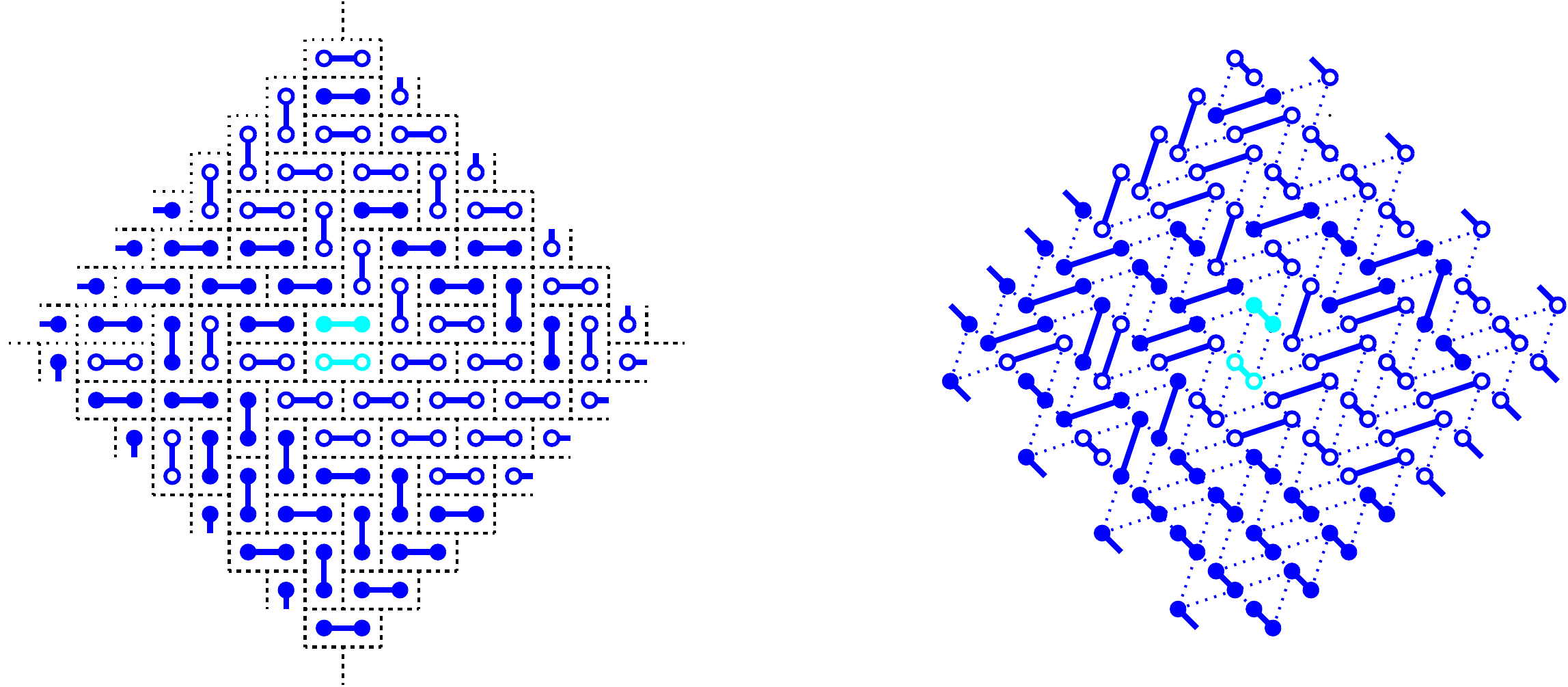}
  \caption{Left: the dimer covering on $(\Z')^2$ corresponding to the
    steep tiling displayed on the right of
    Figure~\ref{fig:universalSteepTiling}. To facilitate the further
    identification with a sequence of interlaced partitions, south and
    west (resp.\ north and east) dimers are represented with solid
    (resp.\ hollow) endpoints, that indicate the presence of a
    particle (resp.\ of a hole).  Right: the same dimer covering, now
    represented on the ``universal'' rail yard graph. To help
    visualize the deformation, we represent in lighter color the
    dimers close to the origin. There are finitely many dimers on the
    longer edges, and the weights are associated with them.}
  \label{fig:universalSteepTilingRYG}
\end{figure}

Steep tilings are in bijection with certain \emph{dimer coverings} of
the (universal) \emph{rail yard graph}, obtained as follows. Given a
steep tiling, the first step consists in passing to the dual picture,
namely to the corresponding dimer covering of the dual graph, as
displayed on the left of Figure~\ref{fig:universalSteepTilingRYG}. The
vertex set of the dual graph may be thought as $(\Z')^2$, where
$\Z':=\Z+\frac{1}{2}$ is the set of half-integers. Next, we slightly
deform the dual graph as follows: given a vertex $(x,y) \in (\Z')^2$
\begin{itemize}
\item we leave it at the same position if $x-y$ is even,
\item we translate it by $(\frac{1}{2},\frac{1}{2})$ if $x-y \equiv 1 \mod 4$,
\item we translate it by $(-\frac{1}{2},-\frac{1}{2})$ if $x-y \equiv -1 \mod 4$.
\end{itemize}
See the right of Figure~\ref{fig:universalSteepTilingRYG} for an
illustration. A more convenient choice of
coordinates~\citemy{dimerstat} for the vertices of the deformed graph,
which we call the \emph{universal rail yard graph} (URYG), is to set
\begin{equation}
  x = \frac{X}{2} + Y, \quad y = - \frac{X}{2} + Y
\end{equation}
where $X \in \Z$ and $Y \in \Z'$. Thus, the vertex set of the URYG may
be thought as $\Z \times \Z'$ in terms of the $(X,Y)$
coordinates. There are two types of edges: \emph{shorter} edges
connecting vertices with the same value of $Y$, and \emph{longer}
edges connecting vertices whose values of $Y$ differ by $\pm 1$. We do
not detail here the full characterization of dimer coverings
corresponding to steep tilings,
but simply note that the finiteness condition on the number of flips
along a line $x-y=n$ implies that the number of longer edges covered
by dimers, hereafter called \emph{covered longer edges}, is finite
along each such line. In particular \emph{pure} dimer coverings,
corresponding to pure steep tilings, have finitely many covered longer
edges, but the converse is not true\footnote{Dimer coverings with
  finitely many covered longer edges may have different boundary
  conditions for $X \to \pm \infty$. A boundary condition corresponds
  to a partition, and the pure case corresponds to having the empty
  partition as boundary condition for both $X \to +\infty$ and
  $X \to -\infty$.  Boundary conditions are analogous to the
  ``asymptotics'' for ``3d partitions'' (plane partitions) considered
  in \cite{ORV06}.\label{fn:asymp}}. To a dimer covering $C$ with
finitely many covered longer edges, we assign a weight
\begin{equation}
  \label{eq:dimerwC}
  w(C) := \prod_{i \in \Z'} u_i^{n_i(C)} 
\end{equation}
where $n_i(C)$ is the number of covered longer edges connecting
vertices with abscissas $i-\frac{1}{2}$ and $i+\frac{1}{2}$ in the URYG
coordinates (so their middles have abscissa $i \in \Z'$), and the
$u_i$'s are arbitrary variables. By working out the correspondence
backwards, we see that the weight $w(C)$ corresponds to putting
certain weights on dominoes in the steep tiling picture. Details are
left to the reader. The only pure tiling with weight $1$ (i.e.\
containing no covered longer edge in the dimer picture) is the
fundamental tiling.

\begin{rem}
  \label{rem:dominosconnect}
  The original definition of steep tilings in~\citemy{pyramids}
  involves the so-called asymptotic data, which is a binary word on
  the alphabet $\{+,-\}$. Here we consider only steep tilings with
  asymptotic data $\cdots++--++--++--\cdots$. There is no loss of
  generality in doing so, as arbitrary asymptotic data may be
  ``emulated'' by setting appropriate variables $u_i$
  in~\eqref{eq:dimerwC} to zero. Similarly the original definition of
  rail yard graphs in~\citemy{dimerstat} involves the so-called $LR$
  and sign sequences, and here we consider the case where these are
  respectively $\cdots LRLRLRLR \cdots$ and $\cdots++--++--++--\cdots$
  (again, without loss of generality, which is why we call this case
  ``universal''). The URYG which we consider here corresponds
  actually to the contracted graph $\tilde{R}$
  of~\citemy[Section~5.2]{dimerstat}.
\end{rem}

The main result of~\citemy{pyramids} concerns the weighted enumeration
of steep tilings subject to various kinds of boundary conditions
(pure, periodic, free, etc), and the main result of~\citemy{dimerstat}
is an explicit expression for their so-called correlation functions in
the \emph{pure} case. So, let us first review the results for the pure
case. We start with the enumerative result.

\begin{thm}[see {\citemy[Theorem~3]{pyramids} and
    \citemy[Theorem~1]{dimerstat}}]
  \label{thm:steepZ}
  The \emph{pure partition function} $Z$, defined as the sum of the
  weights~\eqref{eq:dimerwC} of all pure dimer coverings, is equal to
  \begin{equation}
    \label{eq:steepZ}
    Z = \prod_{\substack{(i,j) \in \Z^2 \\ i < j}} \frac{
      \left(1+u_{4i+\frac{1}{2}} u_{4j-\frac{1}{2}}\right)
      \left(1+u_{4i+\frac{3}{2}} u_{4j-\frac{3}{2}}\right)}{
      \left(1-u_{4i+\frac{1}{2}} u_{4j-\frac{3}{2}}\right)
      \left(1-u_{4i+\frac{3}{2}} u_{4j-\frac{1}{2}}\right)}.
  \end{equation}
\end{thm}

It is clear that $Z$ is a well-defined formal power series in the
variables $(u_i)_{i \in \Z'}$, with nonnegative integer
coefficients. Furthermore, it is analytically convergent when the
$u_i$'s are small enough and decay fast enough (e.g.\ exponentially)
at $\pm \infty$.

We now discuss some examples, that are obtained by \emph{reduction},
i.e.\ by setting some $u_i$'s to zero, cf.\
Remark~\ref{rem:dominosconnect}. Reduction amounts to forbidding
certain types of dominoes/dimers at certain positions, such that the
resulting subfamily is in bijection with another family of
tilings. For brevity, we only write down the reduced formulas, and
refer to the original papers for the combinatorial details of
reduction.

\begin{exmp}[Domino tilings of the Aztec diamond]
  Take $u_{4i+\frac{3}{2}}=u_{4i-\frac{3}{2}}=0$ for all $i \in \Z$,
  and set $u_{4i+\frac{1}{2}}=v_{2i+1}$,
  $u_{4i-\frac{1}{2}}=v_{2i}$. We get
  \begin{equation}
    Z_{\text{Aztec}} = \prod_{i<j} (1 + v_{2i+1} v_{2j})
  \end{equation}
  which corresponds to Stanley's weighting scheme for the Aztec
  diamond, see~\citemy[Remark~2]{pyramids}. Taking further $v_i=1$ for
  $1 \leq i \leq 2\ell$ and $v_i=0$ otherwise, we recover the
  well-known fact that there are $2^{\binom{\ell+1}{2}}$ domino
  tilings of the Aztec diamond of order $\ell$. Taking more generally
  $v_i=q^{(-1)^i i}$ for $1 \leq i \leq 2\ell$, we recover the
  $q$-series
  $\prod_{i=1}^\ell (1+q^{2i-1})^{\ell+1-i}$~\cite{EKLP92a}.
\end{exmp}

\begin{exmp}[Pyramid partitions]
  For $i \geq 0$, take $u_{4i+\frac{1}{2}}=u_{4i+\frac{3}{2}}=0$,
  $u_{4i+\frac{5}{2}}=v_{2i}$ and
  $u_{4i+\frac{7}{2}}=v_{2i+1}$. Similarly for $i \leq 0$, take
  $u_{4i-\frac{1}{2}}=u_{4i-\frac{3}{2}}=0$,
  $u_{4i-\frac{5}{2}}=v'_{2\norm{i}}$ and $u_{4i-\frac{7}{2}}=v'_{2\norm{i}+1}$. We get
  \begin{equation}
    Z_{\text{pyramids}} = \prod_{i,j=0}^\infty \frac{(1+v_{2i} v'_{2j})(1+v_{2i+1} v'_{2j+1})}{
      (1-v_{2i} v'_{2j+1})(1-v_{2i+1} v'_{2j})}.
  \end{equation}
  By taking $v_i=v'_i=q^{i+\frac{1}{2}}$ we obtain the $q$-series
  $\prod_{k\geq 1}\frac{(1+q^{2k-1})^{2k-1}}{(1-q^{2k})^{2k}}$ which
  is the generating function of pyramid
  partitions~\cite{Young09,Young10}.
\end{exmp}

\begin{exmp}[Plane partitions]
  For $i \geq 0$, take
  $u_{4i+\frac{1}{2}}=u_{4i+\frac{3}{2}}=u_{4i+\frac{7}{2}}=0$ and
  $u_{4i+\frac{5}{2}}=v_i$. For $i \leq 0$, take
  $u_{4i-\frac{1}{2}}=u_{4i-\frac{3}{2}}=u_{4i-\frac{5}{2}}=0$ and
  $u_{4i-\frac{7}{2}}=v'_{\norm{i}}$. We get
  \begin{equation*}
    Z_{\text{pp}} = \prod_{i,j=0}^\infty \frac{1}{1-v_i v'_i}.
  \end{equation*}
  By taking $v_i=v'_i=q^{i+\frac{1}{2}}$ we obtain the $q$-series
  $\prod_{k \geq 1} \frac{1}{(1-q^k)^k}$ which is the well-known
  MacMahon generating function of unboxed plane partitions.
\end{exmp}

We now discuss the correlation functions. Here it is slightly easier
to work in the dimer picture: suppose that the $u_i$'s are nonnegative
and such that the partition function $Z$ is finite. Then, we may
interpret $w(C)/Z$ as the probability of the pure dimer covering $C$
in the~\emph{URYG dimer model}. For $E$ a finite set of edges of the
URYG, the \emph{correlation function} $P(E)$ is defined as the
probability that all the edges in $E$ are covered by dimers.
Since the URYG is bipartite, we
expect\footnote{Let us point out that we only ``expect'' and do not
  ``know'' a priori: since the URYG is infinite there are technical
  difficulties in applying the Kasteleyn theory. To my knowledge, the
  general theorems of Kenyon \emph{et al.} ~\cite{KOS06} need certain
  periodicity assumptions that are not satisfied for the URYG with
  general, non periodic, weights.} by the general Kasteleyn theory
(see e.g.\ \cite{Tiliere15} and references therein) that $P(E)$ may be
expressed in the form of a determinant of size $|E|$, which is a minor
of the so-called inverse Kasteleyn matrix. This turns out to be the
case, and we are furthermore able to give an explicit expression for
the entries of the inverse Kasteleyn matrix.

To state this expression, we need some further definitions and
notations: for $X \in \Z$, we set
\begin{equation}
  F_X(z) := \prod_{\substack{i \in 4\Z+\frac{1}{2} \\ i<X}} (1 - u_i z)^{-1}
  \prod_{\substack{i \in 4\Z+\frac{3}{2} \\ i<X}} (1 + u_i z)
  \prod_{\substack{i \in 4\Z+\frac{5}{2} \\ i>X}} (1 - u_i z^{-1})
  \prod_{\substack{i \in 4\Z+\frac{7}{2} \\ i>X}} (1 + u_i z^{-1})^{-1}.
\end{equation}
This quantity, and its inverse $F_X(z)^{-1}$, are to be interpreted as
Laurent series in $z$, where the first two factors should be seen as
power series in $z$, and the last two as power series in $z^{-1}$. If
we assume, for simplicity, that all the $u_i$'s are smaller than $1$
and decay exponentially as $i \to \pm \infty$, then these Laurent
series are convergent in an open annulus containing the unit circle
(other cases may be treated by analytic continuation). Let $(X,Y)$ and
$(X',Y')$ be the URYG coordinates of two vertices, with $X$ odd and
$X'$ even. Then, we set
\begin{equation}
  \mathcal{C}(X,Y;X',Y') :=
  \begin{cases}
    \displaystyle [z^Y w^{-Y'}] \frac{F_X(z)}{F_{X'}(w)} \sum_{k=0}^\infty
    \left(\frac{w}{z}\right)^{k+\frac{1}{2}} & \text{if $X<X'$,} \\
    \displaystyle [z^Y w^{-Y'}] \frac{F_X(z)}{F_{X'}(w)} \sum_{k=0}^\infty
    \left(\frac{z}{w}\right)^{k+\frac{1}{2}} & \text{if $X>X'$,} \\
  \end{cases}
\end{equation}
where $[z^Y w^{-Y'}]$ means extraction of the coefficient of
$z^Y w^{-Y'}$ in the bivariate Laurent series on the right (which may
be represented as a double contour integral).

\begin{thm}[see {\citemy[Theorem~5]{dimerstat}}]
  \label{thm:dimerstat}
  $\mathcal{C}$ is an inverse Kasteleyn matrix for the URYG dimer
  model, in the sense that, for any finite set $E=\{e_1,\ldots,e_n\}$
  of edges, we have
  \begin{equation}
    P(E) = \pm u_{i_1} \cdots u_{i_m}
    \det_{1 \leq k,\ell \leq n} \mathcal{C}(X_k,Y_k;X'_\ell,Y'_\ell).
  \end{equation}
  where the edge $e_k$ has endpoints $(X_k,Y_k)$ and $(X'_k,Y'_k)$
  (with $X_k$ assumed odd and $X'_k$ even), and $i_1,\ldots,i_m$
  ($m \leq n$) denote the abscissas of the longer edges in $E$.
\end{thm}

The proofs of Theorems~\ref{thm:steepZ} and \ref{thm:dimerstat} rely
on two ingredients:
\begin{itemize}
\item a bijection between steep tilings (or URYG dimer coverings) and
  \emph{sequences of interlaced partitions} that form a \emph{Schur
    process};
\item the vertex operator/free fermion formalism (plus some
  combinatorial tricks to make the connection with Kasteleyn theory).
\end{itemize}
In the remainder of this section, we will give the first ingredient,
for the second we refer to the original papers.

Consider again Figure~\ref{fig:universalSteepTilingRYG} and pay now
attention to the endpoints of the dimers, which are represented either
solid ($\bullet)$ or hollow ($\circ$). These symbols form a
\emph{particle configuration}: the symbol $\bullet$ indicates the
presence of a particle, and the symbol $\circ$ the absence of a
particle, hence the presence of a \emph{hole}. In detail, we put
particles (resp.\ holes) on the endpoints of dimers corresponding to
south or west (resp.\ north or east) dominoes. In terms of URYG
coordinates, there are particles on the endpoints of a dimer covering
an edge between $(X,Y)$ and $(X+1,Y')$ if and only if $X$ is
even. Therefore, to each dimer covering we may associate a particle
configuration on the URYG vertex set.

\begin{figure}
  \centering
  \fig{.7}{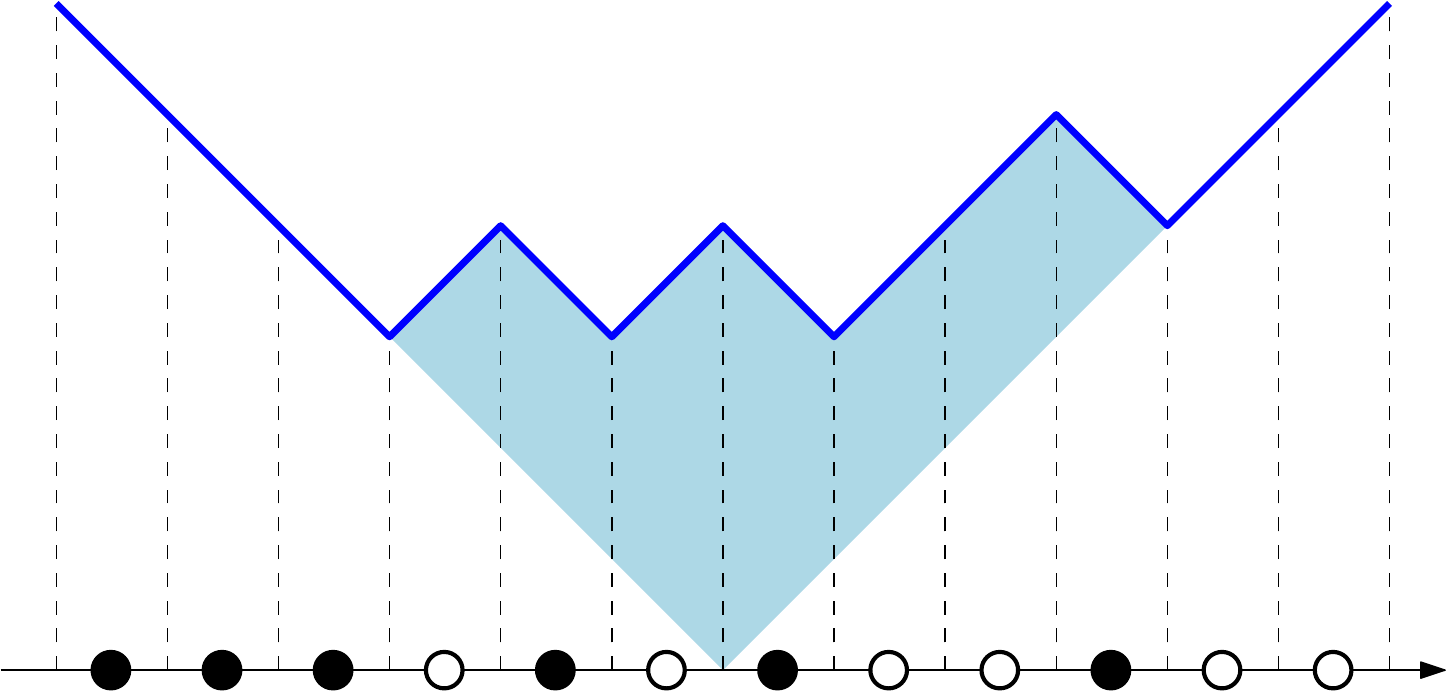}
  \caption{The correspondence between Maya diagram and
    partitions. Here we consider the partition $\lambda=(4,2,1)$,
    whose parts give the displacements of the particles ($\bullet$)
    with respect to their ``ground state''. The parts of the conjugate
    partition $\lambda'=(3,2,1,1)$ give the displacements of the holes
    ($\circ$).}
  \label{fig:russianNoVAxis}
\end{figure}

If we restrict to the vertices having a given abscissa $X \in \Z$, then
we obtain a 1D particle configuration that is a \emph{Maya
  diagram}~\cite{MJD00}. Such diagram is in bijection with a partition
$\lambda(X)$: again we trust that Figure~\ref{fig:russianNoVAxis} is worth a
thousand words. Therefore, a URYG dimer covering corresponds to a
sequence of partitions $(\lambda(X))_{X \in \Z}$. This sequence
satisfies the \emph{interlacing property}
\begin{equation}
  \label{eq:schurint}
  \lambda(4k) \prec \lambda(4k+1) \prec' \lambda(4k+2) \succ \lambda(4k+3) \succ
  \lambda(4k+4), \qquad' k \in \Z.
\end{equation}
This is a straightforward consequence of the URYG structure, of the
coding of partitions by Maya diagrams, and of the
characterization~\eqref{eq:vertint} of interlacing. For instance, the
relation $\lambda(4k+1) \prec' \lambda(4k+2)$ is obtained by
considering the dimers between vertices with abscissas $4k+1$ and
$4k+2$: for each such dimer there are particles on the endpoints,
whose ordinates differ by $0$ or $1$, so we conclude
by~\eqref{eq:vertint} that $\lambda(4k+2)/\lambda(4k+1)$ is a vertical
strip. Note that its size is equal to the number of covered longer
edges with middle abscissas $4k+\frac{3}{2}$. Other cases are obtained
by symmetry and ``particle-hole duality''.

The fundamental tiling corresponds to the constant sequence equal to
the empty partition, and pure steep tilings map to sequences with
finite support (i.e.\ equal to the empty partition except for finitely
many terms). The weight $w(C)$ of a dimer covering may be rewritten in
terms of partitions as
\begin{equation}
  \label{eq:schurw}
  w(C) = \prod_{i \in \Z'} u_i^{\lvert \norm{\lambda(i+\frac{1}{2})}-\norm{\lambda(i-\frac{1}{2})} \rvert}.
\end{equation}
Interestingly, both the interlacing property~\eqref{eq:schurint} and
the weight~\eqref{eq:schurw} can be rewritten in terms of \emph{Schur
  functions} in one variable. More precisely, by
\eqref{eq:schursingle}, we have 
\begin{multline}
  \label{eq:schurweight}
  w(\underline{\lambda}) := \prod_{k \in \Z} \left(s_{\lambda(4k+1)/\lambda(4k)}\left(u_{4k+\frac{1}{2}}\right)
    s_{\lambda'(4k+2)/\lambda'(4k+1)}\left(u_{4k+\frac{3}{2}}\right) \right. \times \\
  \left. s_{\lambda(4k+2)/\lambda(4k+3)}\left(u_{4k+\frac{5}{2}}\right)
    s_{\lambda'(4k+3)/\lambda'(4k+4)}\left(u_{4k+\frac{7}{2}}\right) \right) \\
  = \begin{cases}
    \prod_{i \in \Z'} u_i^{\lvert \norm{\lambda(i+\frac{1}{2})}-\norm{\lambda(i-\frac{1}{2})} \rvert} &\text{if \eqref{eq:schurint} is satisfied,}\\
    0 & \text{otherwise.}
  \end{cases}
\end{multline}
for any sequence of partitions
$\underline{\lambda}=(\lambda(X))_{X \in \Z}$ which is eventually
constant for $X \to \pm \infty$. If we restrict to sequences with
finite support, corresponding to pure steep tilings, the
form~\eqref{eq:schurweight} tells that $\underline{\lambda}$ is a
\emph{Schur process}~\cite{OkRe03,OkRe07}. We then may reuse the
results and methods of Okounkov and Reshetikhin, that are based on
free fermions, to prove Theorem~\ref{thm:steepZ} and to characterize
the correlation functions of the particle configuration (i.e., for any
finite set $F$ of vertices, the probability that there is a particle
on each element of $F$). By relating the correlation functions of
particles to those of dimers, we obtain Theorem~\ref{thm:dimerstat}.

\section{Periodic and free boundary conditions: fermionic approach}
\label{sec:dominosbc}

In the paper~\citemy{pyramids}, we did not restrict to pure steep
tilings, which coincide with the fundamental tiling outside a finite
region (see again Figure~\ref{fig:universalSteepTiling}), but we also
considered other types of ``boundary conditions''.

\begin{figure}
  \centering
  \includegraphics[scale=0.6]{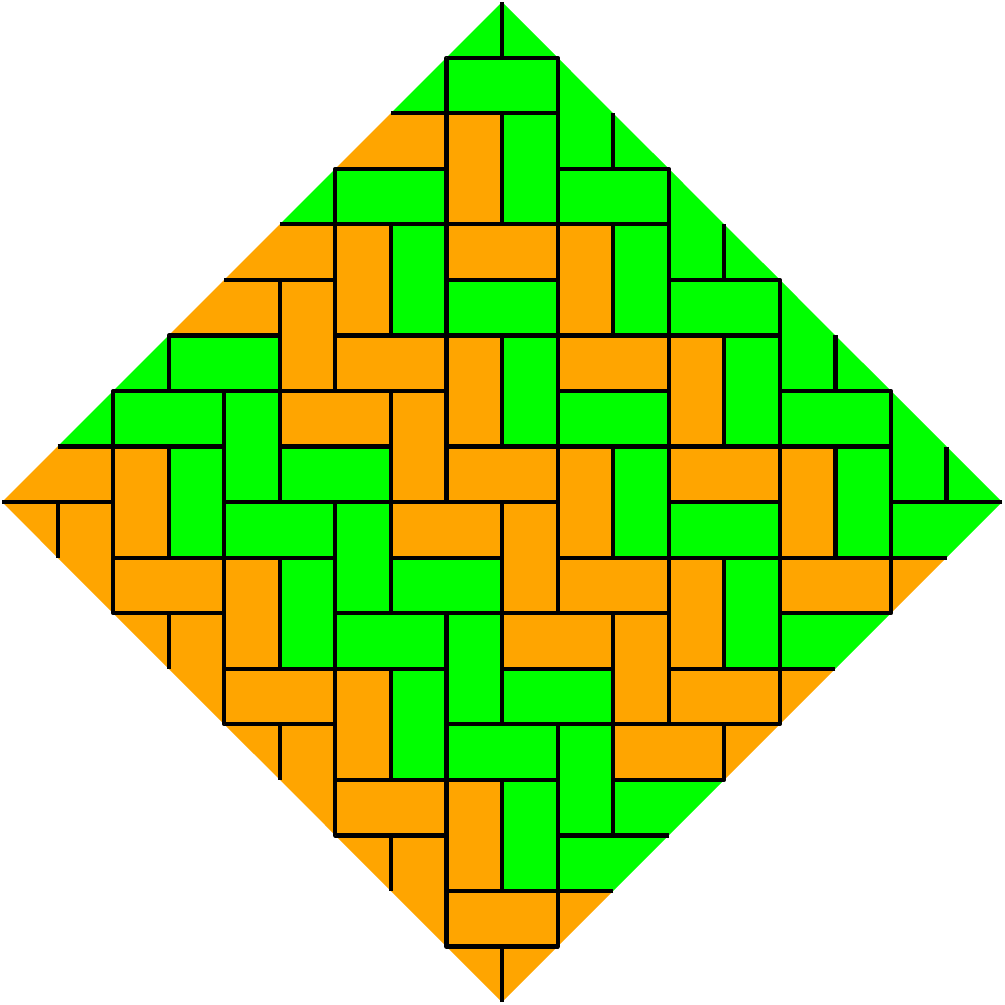}
  \caption{The periodic steep tiling corresponding to the constant
    sequence of partitions equal to $(3,2,2)$.}
  \label{fig:steepTilingPeriodic}
\end{figure}

For instance, we may consider~\emph{periodic} steep tilings, that have
a period $(2T,-2T)$ in the $(x,y)$ coordinates for some positive
integer $T$ (note that the fundamental tiling has period $(2,-2)$). In
terms of sequences of interlaced partitions, such periodic tilings
correspond to sequences of the form~\eqref{eq:schurint} which have a
period $4T$. A first caveat is that the sum of the
weights~\eqref{eq:dimerwC} of all periodic steep tilings is
ill-defined, since there are infinitely many periodic tilings $C$ such
that $w(C)=1$, for instance the tiling displayed on
Figure~\ref{fig:steepTilingPeriodic}. Such tilings are in bijection
with constant sequences of partitions, for which the product of Schur
functions $w(\underline{\lambda})$ of~\eqref{eq:schurweight} evaluates
to $1$. To fix this problem, we introduce the modified partition
function
\begin{equation}
  \label{eq:Zperdef}
  Z_{\mathrm{per}} := \sum_{C \text{ periodic}} q^{\norm{\lambda(0)}} w(C)
\end{equation}
where $q$ is an extra variable, which should be of modulus smaller
than $1$ to make the sum convergent, and where we restrict the range
of $i$'s in~\eqref{eq:dimerwC} to the interval $\Z' \cap [0,4T]$ by
periodicity. The variables $q$ and $u_i$'s may be related to the
number of \emph{periodic flips} that are needed to obtain the tiling
$C$ from the fundamental tiling (a periodic flip is the operation that
consists in performing simultaneous flips at all translates of a given
position, so as to preserve periodicity).

\begin{thm}[see {\citemy[Theorems~5 and 12]{pyramids}}]
  \label{thm:Zper}
  The periodic partition function is given by
  \begin{equation}
    Z_{\mathrm{per}} = \prod_{k = 1}^\infty \left(\frac{1}{1-q^k} \prod_{i,j=1}^T \frac{
        \left(1+q^{k-\mathbbm{1}_{i<j}} u_{4i+\frac{1}{2}} u_{4j-\frac{1}{2}}\right)
        \left(1+q^{k-\mathbbm{1}_{i<j}} u_{4i+\frac{3}{2}} u_{4j-\frac{3}{2}}\right)}{
        \left(1-q^{k-\mathbbm{1}_{i<j}} u_{4i+\frac{1}{2}} u_{4j-\frac{3}{2}}\right)
        \left(1-q^{k-\mathbbm{1}_{i<j}} u_{4i+\frac{3}{2}} u_{4j-\frac{3}{2}}\right)} \right)
  \end{equation}
  where $\mathbbm{1}_{i<j}=1$ if $i<j$ and $0$ otherwise.
\end{thm}

In the same way that the sequence of partitions corresponding to a
random pure steep tiling is a Schur process as defined
in~\cite{OkRe03}, the sequence $\underline{\lambda}$ corresponding to
a random periodic steep tiling (with the weight
$q^{\norm{\lambda(0)}} w(\underline{\lambda})$) forms a \emph{periodic
  Schur process}, as defined in~\cite{Borodin07}.  In fact,
Theorem~\ref{thm:Zper} follows from Borodin's results.
In~\citemy{pyramids} we gave another derivation based on the vertex
operator formalism. It might be possible to derive the periodic
analogue of Theorem~\ref{thm:dimerstat} using Borodin's expression for
the correlation functions, but the main difficulty is that the
periodic Schur process is \emph{not} determinantal as soon as $q>0$,
i.e.\ the particle correlation function is not directly given by a
determinant.

Borodin found that the process becomes determinantal if one performs
``shift-mixing'': in colloquial terms this corresponds to choosing
some integer $c \in \Z$ independent of the tiling, with probability
proportional to $q^{\frac{c^2}{2}} t^c$ where $q$ is the same
parameter as before and $t$ is another arbitrary positive real
parameter, and shifting the tiling by the vector $(c,c)$. Then, for
the shift-mixed tiling, the particle correlation function has
determinantal form.

I became quite puzzled by Borodin's result. His proof was based on the
so-called $L$-ensembles, and not the free fermion formalism (though he
mentions that his original unpublished derivation
was based on Fock space techniques). I therefore took it as a
challenge to rederive his result using fermions. We solved this
problem with Dan Betea in~\citemy{cylindricschur}, using \emph{finite
  temperature} ensembles of free fermions. The idea is of course
rather natural in quantum statistical mechanics, where inverse
temperature may be thought of as an imaginary period, but some details
are nontrivial. In particular, we understood that shift-mixing amounts
in physical terms to passing to the grand canonical ensemble.

In parallel, in our work on steep tilings we also found expressions
for the partition function of tilings with \emph{free boundary
  conditions}, see~\citemy[Theorem~4]{pyramids}. Intuitively this
corresponds to not prescribing the boundary conditions\footnote{cf.\
  footnote~\ref{fn:asymp}} of the URYG dimer coverings at
$X \to \pm \infty$. When there is only one free boundary, i.e.\ when
the asymptotics at $X \to +\infty$ is free but the one at
$X \to -\infty$ corresponds to the empty partition, the sequence of
interlaced partitions forms a \emph{pfaffian Schur process}, as
defined in~\cite{BoRa05}. For such a process, there is an analogue of
Theorem~\ref{thm:dimerstat} where the determinant is replaced by a
pfaffian. When there are two free boundaries, we obtain a variant of
the Schur process that, to my knowledge, had not been considered
before~\citemy{pyramids}. By analogy with the periodic case, I
conjectured that the correlation functions are not pfaffian in
general, but become so after performing a suitable shift-mixing. We
proved this conjecture with Dan Betea, Peter Nejjar and Mirjana
Vuleti\'c in~\citemy{freeboundaries}, using again fermionic
techniques. The basic idea here is to perform a
certain~\emph{Bogoliubov transformation}, which is another classical
method in mathematical physics~\cite{BlRi86}.

In the remainder of this section, I will sketch the basic ideas
of~\citemy{cylindricschur} and \citemy{freeboundaries}. I mostly focus
on the partition picture, but the reader may keep in the back of
her mind the tiling/dimer picture.

Let me start with the most elementary setting, which corresponds to
plain random partitions.  Consider the periodic partition function
$Z_{\text{per}}$ of Theorem~\ref{thm:Zper}, in which we set all
$u_i$'s to zero. It then reduces to the well-known generating function
of partitions
\begin{equation}
  \label{eq:Zeuler}
  Z_{\text{par}} = \prod_{k=1}^\infty \frac{1}{1-q^k}.
\end{equation}
Indeed, URYG dimer coverings without covered longer edges are in
bijection with constant sequences of partitions. In probabilistic
terms, we are considering the \emph{canonical measure} on partitions,
in which a given partition $\lambda$ appears with probability
$\frac{1}{Z_{\text{par}}} q^{\norm{\lambda}}$. To such a random
partition we associate a random Maya diagram, via the bijection of
Figure~\ref{fig:russianNoVAxis}. In explicit terms, the positions of
the particles form the set
\begin{equation}
  S(\lambda) = \left\{ \lambda_1 - \frac{1}{2}, \lambda_2 - \frac{3}{2},
    \lambda_3 - \frac{5}{2}, \ldots \right\} \subset \Z'
\end{equation}
where the origin is by convention at the ``bottom corner'' of the
Young diagram. It is clear that, for different $s \in \Z'$, the events
$\{ s \in S(\lambda) \}$ cannot be independent, because there are as
many positive elements in the set $S(\lambda)$ as negative elements in
its complementary (in other words, the Maya diagram has \emph{charge}
zero). There is, however, a surprisingly simple way to obtain
independence: consider a random integer taking value $c$ with
probability proportional to $q^{\frac{c^2}{2}} t^c$ (it is nothing but
Borodin's shift-mixing parameter) and independent of $\lambda$. Then,
consider the shifted set
\begin{equation}
  S(\lambda) + c = \left\{ \lambda_1 - \frac{1}{2} + c, \lambda_2 - \frac{3}{2} + c,
    \lambda_3 - \frac{5}{2} + c, \ldots \right\} \subset \Z'.
\end{equation}
It is not difficult\footnote{Consider first the case where
  $\lambda=\emptyset$, where the formula~\eqref{eq:qtdiag} is
  easy. Then, consider the effect of ``creating'' the parts of
  $\lambda$ one at a time: to add the part $\lambda_k$ we remove the
  element $-k+\frac{1}{2}+c$ from $S$, then add the element
  $\lambda_k-k+\frac{1}{2}+c$ to it. Observe that both sides
  of~\eqref{eq:qtdiag} get multiplied by $q^{\lambda_k}$. This
  establishes the formula for any partition $\lambda$.}  to see that the
probability that this set is equal to a given $S \subset \Z'$ is
proportional to
\begin{equation}
  \label{eq:qtdiag}
  q^{\norm{\lambda} + \frac{c^2}{2}} t^c = \prod_{s \in S \cap (0,\infty)} (q^s t)
  \prod_{s \in (\Z' \setminus S) \cap (-\infty,0)} (q^{-s} t^{-1}).
\end{equation}
Here, it is understood that the right-hand side vanishes as soon as
one of the sets of indices is infinite (since
$0<q<1$). Summing~\eqref{eq:qtdiag} over all pairs $(\lambda,c)$ or
subsets $S \subset \Z'$ yields a classical combinatorial proof of the
well-known Jacobi triple product identity. Moreover, it also implies
that
\begin{equation}
  \label{eq:fermidirac}
  \Prob\left(s \in S(\lambda) + c\right) = \frac{q^s t}{1 + q^s t}, \qquad s \in \Z'
\end{equation}
and that these events are independent for different $s$'s.
We recognize the \emph{Fermi-Dirac distribution} describing a grand
canonical ensemble of fermions, where $q$ is related to the physical
temperature ($q=0$ is zero temperature, $q\to 1$ is the
high-temperature limit) and $t$ to the chemical potential. Asymptotics
will be discussed in Section~\ref{sec:dominosasymp}.

We now turn to the general setting of the periodic Schur process,
corresponding to taking nonzero $u_i$'s in the partition function
$Z_{\text{per}}$ of Theorem~\ref{thm:Zper}. In other words, we are
consider a random $4T$-periodic sequence of integer partitions, which
is equal to $\underline{\lambda}=(\lambda(X))_{X \in \Z}$ with
probability proportional to $q^{\lambda(0)}$ times the Schur
weight~\eqref{eq:schurweight} (with the range of indices in the
product restricted to a period). At this stage, we make some use of
the free fermion formalism: for brevity we will not include the
definitions here, but refer instead to the relevant sections of the
original papers, namely~\citemy[Section~3]{freeboundaries} and
\citemy[Appendix~B]{cylindricschur}, and references therein.

We are interested in the particle correlation function which, as
mentioned at the end of the previous section, is defined as the
probability that there are particles on all sites of a given finite
set $F$. More precisely, let
$F=\{(X_1,Y_1),\ldots,(X_n,Y_n)\} \subset \Z \times \Z'$ be such a
finite set, written in URYG coordinates. We assume that the abscissas
are ordered ($X_1 \leq \cdots \leq X_n$) and, by periodicity, that all
of them belong to a same period, say $\{1,\ldots,4T\}$. The
correlation function $\rho(F)$ is defined explicitly in terms of
$\underline{\lambda}$ as
\begin{equation}
  \rho(F) := \Prob\left( Y_k \in S(\lambda(X_k)) \text{ for all $k=1,\ldots,n$}\right).
\end{equation}
The free fermion formalism allows, after some manipulations, to write
this correlation function in the form
\begin{equation}
  \label{eq:rhoferm}
  \rho(F) = \left\langle \Psi(X_1,Y_1) \Psi^*(X_1,Y_1) \cdots \Psi(X_n,Y_n) \Psi^*(X_n,Y_n) \right\rangle
\end{equation}
where the $\Psi(X_i,Y_i)$ are certain \emph{fermionic creation
  operators} acting on the \emph{fermionic Fock space}, the
$\Psi^*(X_1,Y_1)$ are their dual \emph{annihilation operators}, and
$\langle \cdot \rangle$ denotes a certain \emph{expectation value} (or
state) on the corresponding operator
algebra. See~\citemy[Section~4.3]{dimerstat} for a relatively detailed
discussion of the way to arrive at the form~\eqref{eq:rhoferm} in the
nonperiodic case. In the periodic case, the expectation value of an
operator $\mathcal{O}$ reads
\begin{equation}
  \label{eq:finiteTexp}
  \langle \mathcal{O} \rangle = \frac{\Tr( \mathcal{O} q^H \Pi_0 )}{\Tr(q^H \Pi_0 )}
\end{equation}
where $H$ is the energy operator, and $\Pi_0$ the projector on the
subspace of charge $0$.

For $q=0$ (zero temperature), the expectation value reduces to the
\emph{vacuum expectation value}. In this case, \emph{Wick's lemma}
entails that the correlation function reads
\begin{equation}
  \label{eq:Wick}
  \begin{split}
  \rho(F) &= \det_{1 \leq k,\ell \leq n} 
  K(X_k,Y_k;X_\ell,Y_\ell), \\
  K(X,Y;X',Y') :&=
  \begin{cases}
    \left\langle \Psi(X,Y) \Psi^*(X',Y') \right\rangle & \text{if $X \leq X'$,}\\
    - \left\langle \Psi^*(X',Y') \Psi(X,Y) \right\rangle & \text{if $X > X'$.}
  \end{cases}
\end{split}
\end{equation}
This says that the particle configuration is a \emph{determinantal
  point process}, and $K$ is its correlation kernel. In fact, in the
case $q=0$ the correlation kernel has essentially the same expression
as the inverse Kasteleyn matrix $\mathcal{C}$ of the previous section.

In the case $q>0$, the difficulty is that Wick's lemma does not hold
anymore for the expectation value~\eqref{eq:finiteTexp}. Hence, the
process is nondeterminantal. But the solution to this problem is the
same as for the canonical measure on partitions, namely we shall pass
to the \emph{grand canonical ensemble} for which the expectation value
reads
\begin{equation}
  \label{eq:finiteTexpG}
  \langle \mathcal{O} \rangle = \frac{\Tr( \mathcal{O} q^H t^C )}{\Tr(q^H t^C )}
\end{equation}
where $C$ is the charge operator. This is nothing but the operator
version of the Fermi-Dirac distribution~\eqref{eq:fermidirac}.
Passing to it amounts to considering the ``shift-mixed'' correlation
function
\begin{equation}
  \label{eq:tilderho}
  \tilde{\rho}(F) := \Prob\left( Y_k - c \in S(\lambda(X_k)) \text{ for all $k=1,\ldots,n$}\right)
\end{equation}
with $c$ independent of the Schur process and distributed as
above. The correlation function $\rho(F)$ may be recovered through a
simple contour integration on $t$. For the expectation
value~\eqref{eq:finiteTexpG}, Wick's lemma holds for any $q$ and $t$
(the fundamental reason being that $q^H t^C$ is the exponential of a
``free Hamiltonian'', namely a quadratic form in the
creation/annihilation operators), so we find that $\tilde{\rho}(F)$
admits the same determinantal expression as in~\eqref{eq:Wick}, with
$\langle \cdot \rangle$ the grand canonical expectation value. This
was, in a nutshell, the idea of our rederivation of Borodin's
correlation function, see~\citemy[Sections~3 and 4]{cylindricschur}.

We now discuss the free boundary case. We again find an expression of
the form~\eqref{eq:rhoferm} for the correlation function $\rho(F)$,
where the expectation value is now
\begin{equation}
  \langle \mathcal{O} \rangle = \frac{\langle \underline{u} | \mathcal{O} | \underline{v} \rangle}{\langle \underline{u} | \underline{v} \rangle}
\end{equation}
with $\langle \underline{u} |$, $| \underline{v} \rangle$ certain
\emph{free boundary states} depending on positive parameters $u,v$
such that $uv<1$ (the product $uv$ plays a role similar to the $q$ of
the periodic case). We have the same issue that Wick's lemma does not
hold for this expectation value. In view of the periodic case, it is
natural to try to fix this problem by modifying the expectation value.
After some efforts, we were able to identify the appropriate
modification, which consists in replacing the free boundary states
$\langle \underline{u} |$, $| \underline{v} \rangle$ by some
\emph{extended free boundary states} $\langle \underline{u,t} |$,
$| \underline{v,t} \rangle$ that are obtained by performing a certain
Bogoliubov transformation on the vacuum,
see~\citemy[Section~3.2.1]{freeboundaries} for the details. Here, $t$
is a positive real parameter playing a role similar to the periodic
case.  In terms of correlation functions, passing to the extended free
boundary states amounts to replacing the original correlation function
$\rho(F)$ by its shift-mixed variant $\tilde{\rho}(F)$ as
in~\eqref{eq:tilderho}, with $c$ now an \emph{even} random integer
independent of the Schur process (its law is obtained by conditioning
the $c$ of the periodic case to be even, with $q=uv$).

A peculiar property of the extended free boundary states is that they
are \emph{not} eigenvectors of the charge operator. As a consequence,
for the corresponding expectation value we have
\begin{equation}
  \left\langle \Psi(X,Y) \Psi(X',Y') \right\rangle \neq 0.
\end{equation}
This says that, when we apply Wick's lemma to evaluate the correlation
function~\eqref{eq:rhoferm}, we shall not ignore the ``contractions''
between two $\Psi$'s, or between two $\Psi^*$'s. The outcome is that
the shift-mixed correlation function takes the pfaffian form
\begin{equation}
  \label{eq:Wickpf}
  \tilde{\rho}(F) = \pf_{1 \leq k,\ell \leq n}  K(X_k,Y_k;X_\ell,Y_\ell)
\end{equation}
where $K$ is now a \emph{$2 \times 2$ matrix-valued kernel}
\begin{equation}
  K(X,Y;X',Y') :=
  \begin{pmatrix}
    K_{1,1}(X,Y;X',Y') & K_{1,2}(X,Y;X',Y') \\
    K_{2,1}(X,Y;X',Y') & K_{2,2}(X,Y;X',Y') \\
  \end{pmatrix}
\end{equation}
with
\begin{equation}
  \begin{split}
    K_{1,1}(X,Y;X',Y') &:= \left\langle \Psi(X,Y) \Psi(X',Y') \right\rangle \\
    K_{2,2}(X,Y;X',Y') &:= \left\langle \Psi^*(X,Y) \Psi^*(X',Y') \right\rangle \\
    K_{1,2}(X,Y;X',Y') &:= 
    \begin{cases}
      \left\langle \Psi(X,Y) \Psi^*(X',Y') \right\rangle & \text{if $X \leq X'$}\\
      - \left\langle \Psi^*(X',Y') \Psi(X,Y) \right\rangle & \text{if $X > X'$}
    \end{cases} \\
    &=: - K_{2,1}(X',Y';X,Y).
  \end{split}
\end{equation}
Here, the meaning of \eqref{eq:Wickpf} is that we form the
$2n \times 2n$ matrix made of the blocks $K(X_k,Y_k;X_\ell,Y_\ell)$, and we
take its pfaffian (the matrix is indeed antisymmetric by properties of
the kernel). The conclusion is that the shift-mixed process is a
\emph{pfaffian point process}. In the case of one free boundary ($u=0$
or $v=0$), the shift-mixed process is identical to the original
process, and we recover the result of Borodin and Rains~\cite{BoRa05}
regarding the correlation functions of the pfaffian Schur process. For
two free boundaries ($uv>0$), our result was new,
see~\citemy[Theorem~2.5]{freeboundaries}.

\section{Periodic and free boundary conditions: asymptotics}
\label{sec:dominosasymp}

Beyond the technical challenge of adapting the free fermion formalism,
our motivation for studying the Schur process with periodic and free
boundary conditions is that it allows to observe different large-scale
behaviors from the usual Schur process with empty boundary conditions.

In random partitions, one distinguishes two types of scaling limits,
called the \emph{bulk} and \emph{edge} scaling
limits. Following~\citemy[Section~2]{cylindricschur}, let us
illustrate the distinction in the case of a random partition $\lambda$
distributed according to the canonical measure. From the
expression~\eqref{eq:Zeuler} of the partition function, a second
moment calculation shows that we have the convergence in probability
\begin{equation}
  r^2 \norm{\lambda} \overset{\mathbb{P}}{\to} \frac{\pi^2}{6}, \qquad \text{where } q=e^{-r},\ r \to 0^+.
\end{equation}
In particular, $r^{-1}$ plays not only the role of a temperature, but
also of a length scale for the Young diagram of $\lambda$ (since its
area $\norm{\lambda}$ is of order $r^{-2}$), hence for its associated
Maya diagram. This suggests to take $s=\lfloor x r^{-1} \rfloor$
in~\eqref{eq:fermidirac}, for some fixed $x \in \R$, with the
immediate result
\begin{equation}
  \Prob\left(s \in S(\lambda) + c\right) \to \frac{e^{-x+\ln t}}{1 + e^{-x+\ln t}}.
\end{equation}
In fact, the same result holds without the $+c$ shift, with $t=1$. The
right-hand side gives the slope of Vershik's limit shape for uniform
partitions~\cite{Vershik96}\footnote{To recover Vershik's result, we
  should prove concentration, and that the limit of canonical
  partitions for $q \to 1$ is the same as the limit of uniform random
  partitions of size $n \to \infty$ (``equivalence of
  ensembles'').}. Also, for $i \in \Z$, the indicators
$\mathbbm{1}_{s+i \in S(\lambda)}$ become asymptotically i.i.d.\
Bernoulli random variables: this is Okounkov's observation that ``a
typical (canonical) partition is locally a random
walk''~\cite[Section~3.4]{Okounkov01}. Such a result describes
the~\emph{bulk scaling limit} of canonical partitions. We now turn to
the~\emph{edge scaling limit} which corresponds to studying the
distribution of the first part(s) of $\lambda$, or equivalently the
position of the rightmost particle(s) in the Maya diagram. By the
independence property of the Fermi-Dirac
distribution~\eqref{eq:fermidirac}, it is not difficult to check that,
for any $x \in \R$, we have
\begin{equation}
  \label{eq:canongum}
  \Prob\left(\lambda_1 \leq r^{-1} \ln r^{-1} + x r^{-1} \right) \to e^{-e^{-x}}. 
\end{equation}
In other words, the rescaled fluctuations of the first part $\lambda_1$
are governed by the Gumbel distribution, consistent with the result
of Erd\H{o}s and Lehner~\cite{ErLe41}. More generally, the rescaled
positions of the $k$ first parts $\lambda_1,\ldots,\lambda_k$ converge
to the positions of the $k$ rightmost points in a Poisson point
process with intensity $e^{-x}dx$.

Let us now discuss the periodic Schur process. The bulk scaling limit
has been studied in great generality by Borodin~\cite{Borodin07}, but
he did not consider the edge scaling limit. We addressed this question
with Dan Betea~\citemy{cylindricschur}, in the context of
the~\emph{cylindric Plancherel measure/process} (CPM/CPP). The CPP
corresponds to some sort of ``poissonian'' limit of periodic steep
tilings, but we will not enter into details here. The CPM is a
marginal of the CPP, and it may be defined as the measure on
partitions
\begin{equation}
  \label{eq:CPM}
  p(\lambda) = \frac{1}{Z_{\mathrm{CPM}}} \sum_{\mu \subset \lambda} q^{\norm{\mu}}
  (s_{\lambda/\mu}(\ex_y))^2
\end{equation}
where $q$ and $y$ are nonnegative real parameters with $q<1$,
$s_{\lambda/\mu}(\ex_y)$ is the exponential specialization of the skew
Schur function as defined in \eqref{eq:expspec}, and the partition
function $Z_{\mathrm{CPM}}$ is equal to
$e^{y^2/(1-q)} \prod_{k=1}^\infty (1-q^k)^{-1}$. The CPM interpolates
between the canonical measure, recovered for $y=0$, and the Plancherel
measure, recovered for $q=0$.

Let $\lambda$ denote a random partition distributed according to the
CPM. We are interested in the limit $q \to 1$ and/or $y \to \infty$
where $\norm{\lambda}$ gets large, and we consider the largest part
$\lambda_1$. We set as before $q=e^{-r}$. Since the CPM interpolates
between canonical and Plancherel, we intuitively expect to observe at
least two different regimes:
\begin{itemize}
\item when we let $r \to 0$ keeping $y$ ``small'', $\lambda_1$ should
  have Gumbel-type fluctuations as in the canonical measure,
\item when we let $y \to \infty$ keeping $r$ finite, the fluctuations
  of $\lambda_1$ should be of the same nature as in the Plancherel
  measure which, as shown by Baik, Deift and
  Johansson~\cite{BaDeJo99}, are governed by the Tracy--Widom GUE
  distribution.
\end{itemize}
In fact, there exists a \emph{crossover} regime where the rescaled
fluctuations of $\lambda_1$ follow a distribution that interpolates
between Gumbel and Tracy--Widom. This interpolating distribution was
first encountered by Johansson~\cite{Johansson07} in a random matrix
model. It depends on a positive parameter $\alpha$ and may be defined
as the Fredholm determinant
\begin{equation}
  \label{eq:Falpha}
  F_\alpha(x) := \det (I-M_\alpha)_{L^2(x,\infty)}
\end{equation}
where $M_\alpha$ is the \emph{finite-temperature Airy kernel}
\begin{equation}
  \label{eq:ftAk}
  M_\alpha(x,x') := \int_{-\infty}^\infty \frac{e^{\alpha u}}{1+e^{\alpha u}} \Ai(x+u) \Ai(x'+u) du
\end{equation}
with $\Ai$ the Airy function.  For $\alpha=\infty$, $M_\alpha$ reduces
to the usual Airy kernel, 
hence~\eqref{eq:Falpha} reduces to the classical expression for the
Tracy--Widom GUE distribution as a Fredholm determinant. We may
now state our main result.

\begin{thm}[see {\citemy[Theorems~1 and 2]{cylindricschur}}]
  \label{thm:CPMedge}
  Consider the cylindric Plancherel measure~\eqref{eq:CPM} with
  $q=e^{-r}$, and suppose that $r \to 0$ and/or $y \to \infty$ in such
  a way that $r^2 y \to \alpha$, where $\alpha \in [0,\infty]$ is fixed.
  \begin{itemize}
  \item In the \emph{low-temperature} regime $\alpha=\infty$ or the
    crossover regime $0<\alpha<\infty$, we have for all $x \in \R$
    \begin{equation}
      \label{eq:CPMcross}
      \Prob\left( \frac{\lambda_1 - 2 L}{L^{1/3}} \leq x \right) \to
      F_\alpha(x), \qquad L:=\frac{y}{1-q}.
    \end{equation}
  \item In the \emph{high-temperature} regime $\alpha=0$, we have for all $x \in \R$
    \begin{equation}
      \Prob\left( \frac{\lambda_1 - r^{-1} \ln (r^{-1} I_0(2y+ry))}{r^{-1}} \leq x \right) \to e^{-e^{-x}}
    \end{equation}
    where
    $I_0(z):=\frac{1}{2\pi} \int_{-\pi}^\pi e^{z \cos \phi} d\phi$ is
    the modified Bessel function of the first kind and order zero.
  \end{itemize}
\end{thm}

Our theorem encompasses the canonical case $y=0$, for which we
recover~\eqref{eq:canongum}, and the Plancherel case $q=0$ for which
we recover the result of Baik, Deift and
Johansson~\cite{BaDeJo99}. The physical intuition behind our theorem
is the following. In the CPM the fluctuations of $\lambda_1$ have two
origins: \emph{thermal} fluctuations of order $r^{-1}$, governed by
the Gumbel distribution, and \emph{quantum} fluctuations of order
$L^{1/3}$, governed by the Tracy--Widom GUE distribution. In the
high-temperature regime ($r^{-1} \gg L^{1/3}$) the former wins, hence
we observe the Gumbel distribution. In the low-temperature regime
($r^{-1} \ll L^{1/3}$) the latter wins, hence we observe the
Tracy--Widom GUE distribution. Finally in the crossover regime
($r^{-1} \propto L^{1/3}$) both have the same order of magnitude,
hence we observe the interpolating distribution. Note that, in the
high-temperature regime, the parameter $y$ actually affects the
deterministic first-order asymptotics of $\lambda_1$. This may be
related to the bulk asymptotics of the CPM, see the discussion
in~\citemy[Section~5]{cylindricschur}. We have also studied the
crossover edge asymptotics of the cylindric Plancherel process, and
shown that it involves the \emph{finite-temperature extended Airy
  kernel} previously introduced by Le Doussal, Majumdar and
Schehr~\cite{LDMaSc17} in the context of the equilibrium dynamics of
noninteracting fermions in a harmonic
trap. See~\citemy[Section~6]{cylindricschur} for more details.

While we have not considered the edge asymptotics of periodic steep
tilings in our paper, we believe that, by universality,
Theorem~\ref{thm:CPMedge} should hold with little modifications for
generic edge points. In fact, our approach based on saddle-point
computations should be easy to adapt to this situation.

Finally, let us briefly discuss the free boundary Schur process. In
the paper~\citemy{freeboundaries}, we considered the asymptotics in
the case of one free boundary, corresponding to the pfaffian Schur
process of Borodin and Rains~\cite{BoRa05}. We discussed the bulk
asymptotics for two instances of the process, namely symmetric plane
partitions and plane overpartitions. It should not be difficult to
extend our results to the case of steep tilings. Regarding the edge
asymptotics, we considered its application to \emph{last passage
  percolation}, and encountered a limiting distribution that
interpolates between the three Tracy--Widom GOE/GUE/GSE distributions.

The case of two free boundaries is still pretty much work in
progress. We have announced some of our results in the extended
abstract~\citemy{BBNVfpsac2019}, where we establish a convergence in
distribution similar to~\eqref{eq:CPMcross} for some free boundary
analogues of the CPM. The corresponding limiting distributions are
expressed in terms of certain \emph{Fredholm pfaffians}, and appear
not to have been considered before. Together with Dan Betea, Peter
Nejjar and Mirjana Vuleti\'c, we are currently writing the long
journal paper that will be the sequel of~\citemy{freeboundaries}, and
which will treat both bulk and edge asymptotics in the presence of two
free boundaries.

\section{Conclusion and perspectives}
\label{sec:dominosconc}

In this chapter I have explained how, starting from the question of
finding a general family of tilings encompassing both domino tilings
of the Aztec diamond and pyramid partitions, I was led to study
various aspects of Schur processes and join the very active field of
integrable probability.

One paper left out of this chapter is~\citemy{sampling}, where we
revisit the Robinson-Schensted-Knuth correspondence and its growth
diagram formulation~\cite{Fomin95}, from the point of view of its
applications to the sampling of Schur processes. In particular, we
found an unexpected connection with the famous domino shuffling
algorithm~\cite{EKLP92b}.

Let me now list some of the questions I would like to address from now
on. In the immediate future, I am eager to explore further the
applications of the periodic and free boundary Schur processes to last
passage percolation and exclusion processes. A particularly
tantalizing question is the connection with the $(1+1)$-dimensional
Kardar-Parisi-Zhang (KPZ) equation: as noted for instance
in~\cite{DLDMS15}, the finite-temperature Airy kernel which we
encountered in~\eqref{eq:ftAk} also appears in the KPZ context. To our
knowledge, the fundamental origin of this coincidence is yet to be
found. Also, we would like to investigate whether a similar connection
exists for the kernels which we encounter in~\citemy{BBNVfpsac2019}.

The approach of Baik, Deift and Johansson~\cite{BaDeJo99} for studying
the edge asymptotics of the Plancherel measure on partitions was based
on Riemann-Hilbert techniques. It would be interesting to extend this
approach to the CPM, as our derivation of Theorem~\ref{thm:CPMedge}
was based rather different methods, namely determinantal point
processes in the spirit of~\cite{BOO00,Johansson01}. The
Riemann-Hilbert approach seems better suited to analyze the
asymptotics of the distribution $F_\alpha$, as was done in the very
recent preprint~\cite{CaCl19} which complements the probabilistic
approach of~\cite{CoGh18}. Also, this might shed light on the free
boundary case for which everything is open.

The next direction of research is the extension to nondeterminantal
models such as Macdonald processes~\cite{BoCo14}. These processes have
been recently related to the six vertex model by graphical
methods~\cite{BBW16} and the approach was extended to the case of one
free boundary~\cite{BBCW18}. It is then natural to ask whether
periodic and (two) free boundary versions of the processes may be
considered.

Finally, it seems that we left random maps out of this chapter, but a
connection with them was made in~\cite{Okounkov00}: in this paper
Okounkov uses an interplay between maps on surfaces and ramified
coverings of the sphere, to give the first proof of a conjecture left
by Baik, Deift and Johansson regarding the asymptotic joint
distribution of the first parts $\lambda_1,\lambda_2,\lambda_3,\ldots$
of a Plancherel random partition (subsequent proofs were given
in~\cite{BOO00,Johansson01} by determinantal point process
techniques). I have just started supervising the M2 internship of
Alejandro Caicedo who will study Okounkov's paper, and look for possible
extensions to related models of random partitions.


\renewcommand{\bibname}{General bibliography}
\bibliographystyle{hdr_myhalpha}
\bibliography{HDR}

\end{document}